\documentclass[11pt]{article}
\usepackage{amsmath,amssymb,amsthm}
\usepackage{amsfonts,amssymb}
\usepackage{dsfont}
\usepackage[colorlinks=true, linkcolor=red, urlcolor=blue, citecolor=gray]{hyperref}
\usepackage{color,soul}
\usepackage{xcolor}
\usepackage{mdframed}
\usepackage{comment}
\usepackage{graphicx}
\usepackage{mathtools}
\usepackage[export]{adjustbox}
\usepackage{algorithm}
\usepackage[noend]{algpseudocode}
\usepackage[normalem]{ulem}

\usepackage{tikz}

\definecolor{green2}{RGB}{34,139,34}
\definecolor{green}{RGB}{34,139,34}

	\newcommand{\ignore}[1]{}
	\newcommand{\B}{{\bf B}}

\newcommand{\Cam}[1]{}
\newcommand{\Navid}[1]{}

\newcommand{\R}{\mathbb{R}}
\newcommand{\TT}{\mathcal{T}}

\DeclareMathOperator{\polylog}{polylog}
\DeclareMathOperator{\Vol}{Vol}
\DeclarePairedDelimiter{\ceil}{\lceil}{\rceil}

\let\Pr\relax
\DeclareMathOperator*{\Pr}{\mathbb{P}}

\newcommand{\e}{\epsilon}

\newtheorem{theorem}{Theorem}
\newtheorem{corollary}{Corollary}
\newtheorem{remark}{Remark}
\newtheorem{lemma}{Lemma}

\theoremstyle{definition}
\newtheorem{definition}{Definition}
\newtheorem{construction}{Construction}

\newtheorem{claim}{Claim}

\newtheorem{fact}{Fact}

\newcommand{\poly}{\text{poly}}
\newcommand{\wt}{\widetilde}
\newcommand{\T}{\top}
\newcommand{\ape}{\approx_\e}

\newenvironment{proofof}[1]{\noindent{\bf Proof of #1:}}{$\qed$\par}

\advance \topmargin by -\headheight
\advance \topmargin by -\headsep
\evensidemargin \oddsidemargin
{\makeatletter
 \gdef\xxxmark{%
   \expandafter\ifx\csname @mpargs\endcsname\relax 
     \expandafter\ifx\csname @captype\endcsname\relax 
       \marginpar{xxx}
     \else
       xxx 
     \fi
   \else
     xxx 
   \fi}
 \gdef\xxx{\@ifnextchar[\xxx@lab\xxx@nolab}
 \long\gdef\xxx@lab[#1]#2{{\bf [\xxxmark #2 ---{\sc #1}]}}
 \long\gdef\xxx@nolab#1{{\bf [\xxxmark #1]}}
 \long\gdef\xxx@lab[#1]#2{}\long\gdef\xxx@nolab#1{}%
}

\newcommand{\todo}[1]{\textcolor{red}{TODO: #1}}

\newcommand{\plog}{\mathop\mathrm{polylog}}

\newcommand{\bv}[1]{\mathbf{#1}}

\title{Faster Spectral Sparsification in Dynamic Streams\thanks{The algorithmic part of this paper has recently been improved by a subset of the authors~\cite{KNST19}.}}
\author{Michael Kapralov\\EPFL \and Aida Mousavifar\\EPFL \and Cameron Musco\\Microsoft Research \and Christopher Musco\\Princeton University\and Navid Nouri \\EPFL}

  \usepackage{nth}
  \usepackage{intcalc}

  \newcommand{\cSTOC}[1]{\nth{\intcalcSub{#1}{1968}}\ Annual\ ACM\ Symposium\ on\ Theory\ of\ Computing\ (STOC)}

  \newcommand{\cFOCS}[1]{\nth{\intcalcSub{#1}{1959}}\ Annual\ IEEE\ Symposium\ on\ Foundations\ of\ Computer\ Science\ (FOCS)}

  \newcommand{\cICALP}[1]{\nth{\intcalcSub{#1}{1973}}\ International\ Colloquium\ on\ Automata,\ Languages and\ Programming\ (ICALP)}

  \newcommand{\cPODS}[1]{\nth{\intcalcSub{#1}{1981}}\ Symposium\ on\ Principles\ of\ Database\ Systems\ (PODS)}
  \newcommand{\cSODA}[1]{\nth{\intcalcSub{#1}{1989}}\ Annual\ ACM-SIAM\ Symposium\ on\ Discrete\ Algorithms\ (SODA)}

  \newcommand{\pSTOC}[1]{Preliminary\ version\ in\ the\ \cSTOC{#1}, #1}

  \newcommand{\pFOCS}[1]{Preliminary\ version\ in\ the\ \cFOCS{#1}, #1}

  \newcommand{\pICALP}[1]{Preliminary\ version\ in\ the\ \cICALP{#1}, #1}

  \newcommand{\STOC}[1]{Proceedings\ of\ the\ \cSTOC{#1}}

  \newcommand{\PODS}[1]{Proceedings\ of\ the\ \cPODS{#1}}
  \newcommand{\SODA}[1]{Proceedings\ of\ the\ \cSODA{#1}}

  \newcommand{\arXiv}[1]{\href{http://arxiv.org/abs/#1}{arXiv:#1}}

\begin{document}

\maketitle

\begin{abstract}
Graph sketching has emerged as a powerful technique for processing massive graphs that change over time (i.e., are presented as a dynamic stream of edge updates) over the past few years, starting with the work of Ahn, Guha and McGregor (SODA'12) on graph connectivity via sketching. In this paper we consider the problem of designing {\em spectral} approximations to graphs, or spectral sparsifiers, using a small number of linear measurements, with the additional constraint that the sketches admit {\em an efficient recovery scheme}.

Prior to our work, sketching algorithms were known with near optimal $\tilde  O(n)$ space complexity, but $\Omega(n^2)$ time decoding (brute-force over all potential edges of the input graph), or with subquadratic time, but rather large $\Omega(n^{5/3})$ space complexity (due to their reliance on a rather weak relation between connectivity and effective resistances).
In this paper we first show how a simple relation between effective resistances and edge connectivity leads to an improved $\widetilde O(n^{3/2})$ space and time algorithm, which we show is a natural barrier for connectivity based approaches. Our main result then gives the first algorithm that achieves subquadratic recovery time, i.e. avoids brute-force decoding, and at the same time nontrivially uses the effective resistance metric, achieving $n^{1.4+o(1)}$ space and recovery time.

Our main technical contribution is a novel method for `bucketing' vertices of the input graph into clusters that allows fast recovery of edges of high effective resistance: the buckets are formed by performing ball-carving on the input graph using (an approximation to) its effective resistance metric. We feel that this technique is likely to be of independent interest.

 Our second technical contribution is a new PRG for graph sketching applications that allows stretching a seed vector of random bits of length $n\log^{O(1)} n$ to polynomial length pseudorandom strings with only $\log^{O(1)} n$ cost per evaluation.  In fact, one notes that the aforementioned runtime bounds for graph sketches formally only hold under the assumption of free perfect randomness, and deteriorate by another factor of $n\log^{O(1)} n$ if Nisan's PRG is used, as is standard. Our PRG is the first efficient PRG for graph sketching applications, allowing us to remove the free randomness assumption at only a polylogarithmic factor loss in runtime.
\end{abstract}

\newpage
\tableofcontents

\newpage

\section{Introduction}
A surprising and extremely useful algorithmic fact is that any graph can be approximated, in a strong sense, by a very sparse graph. In particular, given a graph $G$ on $n$ nodes that has possibly $O(n^2)$ edges, it is possible to find a graph $\tilde{G}$ with just $O\left(\frac{n}{\epsilon^2}\right)$ edges such that, for any vector $x \in \mathbb{R}^n$, 
\begin{align}
\label{eq:intro_guarantee}
(1-\epsilon)x^TL_Gx \leq x^TL_{\tilde G}x \leq (1+\epsilon)x^TL_Gx
\end{align}
Here $L_G$ and $L_{\tilde G}$ are the Laplacian matrices of $G$ and $\tilde{G}$, respectively. Any 
$\tilde{G}$ satisfying \eqref{eq:intro_guarantee} is called a \emph{spectral sparsifier} of $G$.  A spectral sparsifier preserves many important structural properties of $G$: the total weight of edges crossing any cut in $\tilde G$ is within a $(1\pm\epsilon)$ factor of the weight crossing the same cut in $G$, each eigenvalue of $L_{\tilde G}$ is within a $(1\pm\epsilon)$ factor of each eigenvalue of $L_{G}$, and electrical flows in $\tilde G$ well approximate those in $G$.

These properties and more allow $\tilde G$ to be used as a surrogate for $G$ in many algorithmic applications. Since it can be stored in less space and operated on more efficiently, using the spectral sparsifier can generically reduce computational costs associated with processing large graphs \cite{BatsonSurvey}.

\subsection{Spectral sparsifiers via linear sketching}
\label{sec:main_questions}
The first algorithm for computing spectral sparsifiers was introduced by Spielman and Teng \cite{SpielmanTengSparsification}, with improvements offered in a number of subsequent papers. \cite{spielman2011graph} gives a simple algorithm based on randomly sampling $G$'s edges and \cite{BatsonSpielmanSrivastava:2012} gives the first result achieving sparsifiers with an optimal number of edges, $O\left(\frac{n}{\epsilon^2}\right)$. Algorithms in \cite{LeeSun15} and \cite{yinTatSDP} offer faster alternatives. 

Recently, there has been great interest in algorithms that can recover a sparsifier based on a \emph{linear sketch} of $G$ \cite{ahn2012analyzing,ahn2012graph,GoelKapralovPost12,ahn2013spectral,kapralov2017single}. The idea is to compress some representation of $G$ (usually its edge-vertex incidence matrix $B$) by multiplying that representation by a \emph{random sketching matrix}, $S$, with a small  number of rows. We then extract a sparsifier $\tilde{G}$ from $S B$, which ideally does not store much more than $O(n)$ bits of information itself.

This approach can be viewed as reframing the sparsification problem as a highly structured sparse recovery problem. In traditional sparse recovery, the goal is to compress a vector $x$ with a linear sketch $S$. From $S x$, we extract a sparse vector $\tilde{x}$ that approximates $x$. Here, the object we compress is a graph, and we extract a sparse approximation to the graph. As in vector sparse recovery, we are interested in two central questions:
\begin{enumerate}
	\item How small of a linear sketch still allows for recovery of a spectral sparsifier $\tilde{G}$?
	\item How quickly can we extract $\tilde{G}$ from this linear sketch?
\end{enumerate}
We refer to the second cost as the time to ``decode'' our linear sketch.
In traditional sparse recovery, the answer to both of these questions in roughly $O(k \plog N)$. I.e., we can recover a $k$ sparse vector approximation using space and decoding time that are both linear in $k$.

The case is far less clear for graph sketching. Current methods achieve space $\tilde{O}({n})$, which is nearly optimal, but use brute force decoding techniques that run in $\Omega(n^2)$ time. We conjecture that this cost can be improved to $\tilde{O}(n)$. This paper makes the first progress towards that goal.

\subsection{Why study this problem?}

Like algorithms for vector sparse recovery \cite{GilbertIndyk:2010}, linear sketching algorithms for graph sparsification offer powerful tools for distributed or streaming computational environments. In particular, they can be far more flexible than traditional sparsification algorithms.

For example, any linear sketching algorithm for computing a sparsifier immediately yields a \emph{dynamic streaming} algorithm for sparsification. In the dynamic streaming setting, the algorithm receives a stream of edge updates to a changing graph $G$ (i.e. edge insertions or deletions). The goal is to maintain a small space, e.g. $\tilde{O}(n)$ space\footnote{We use $\tilde{O}(X)$ as shorthand for $O(X \log^c X)$, where $c$ is a fixed constant that does not depend on $X$.}, compression of $G$ and to eventually extract a sparsifier from this compression. To apply a linear sketching algorithm to this problem, we simply note that any edge update can be expressed as a \emph{linear} update to the edge-vertex incidence matrix $B$, so $S B$ can be maintained dynamically. A sparsifier $\tilde{G}$ can then be extracted from $S B$ at any time.

The dynamic streaming setting naturally models computational challenges that arise when processing dynamically changing social networks, web topologies, transportation networks, and other large graphs. Not only does linear graph sketching offer a powerful approach to dealing with these computation problems, but it is the \emph{only known approach}: all known dynamic streaming algorithms for graph sparsification, and in fact any other graph problem, are based on linear sketching \cite{McGregor14}\footnote{In fact, it can be shown formally that any dynamic streaming algorithm can be implemented as a linear sketching algorithm \cite{LiYiWoodruff14}.}.

\subsection{Prior work}

For a general survey on linear sketching methods and streaming graph algorithms more generally, we refer the reader to \cite{McGregor14}. We focus on reviewing prior work specifically related to sparsification, which in some sense is the most generic graph compression objective that has been studied.

The idea of linear graph sketching was introduced in a seminal paper by Ahn, Guha, and McGregror \cite{ahn2012analyzing}. They present an algorithm for computing a \emph{cut sparsifier} of $G$, which is a strictly weaker, but still useful, approximation than a spectral sparsifier \cite{BenczurKarger:1996}. Their work was improved in \cite{ahn2012graph} and \cite{GoelKapralovPost12}, which use a linear compression of size $\tilde{O}(\frac{n}{\epsilon^2})$ to compute a cut sparsifier.

The more challenging problem of computing a spectral sparsifier from a linear sketch was addressed in \cite{ahn2013spectral}, who give an $\tilde{O}(\frac{n^{5/3}}{\epsilon^2})$ space solution. An $\tilde{O}(\frac{n}{\epsilon^2})$ space solution was obtained in \cite{kapralov2017single} by more explicitly exploiting the connection between graph sketching and vector sparse recovery.

We also mention that spectral sparsifiers have been studied in the insertion-only streaming model, where edges can only be added to $G$ \cite{KelnerLevin:2013}, and in a dynamic data structure model \cite{dynamicSparsifiers,AndoniCKQWZ16,jambulapati2018efficient}, where more space is allowed, but the algorithm must quickly output a sparsifier at every step of the stream. While these models are superficially similar to the dynamic streaming model, they seem to allow for different techniques, and in particular do not require linear sketching.

\paragraph{Effective resistance, spectral sparsification, and random spanning trees.} 
The \emph{effective resistance metric} or \emph{effective resistance distances} induced by an undirected graph plays a central role in spectral graph theory and has been at the heart of numerous algorithmic breakthroughs over the past decade. They are central to the to obtaining fast algorithms for constructing spectral sparsifiers \cite{spielman2011graph, KoutisLP16}, spectral vertex sparsifiers \cite{KyngLPSS16}, sparsifiers of the random walk Laplacian \cite{ChengCLPT15, JindalKPS17}, and subspace sparsifiers \cite{LiS18}. They have played a key role in many advances in solving Laplacian systems \cite{SpielmanT04,KoutisMP10,KoutisMP11,PengS14,CohenKMPPRX14,KoutisLP16,KyngLPSS16,KyngS16} and are critical to the current fastest (weakly)-polynomial time algorithms for maximum flow and minimum cost flow in certain parameter regimes \cite{LeeS14}. Given their utility, the computation of effective resistances has itself become an area of active research \cite{JambulapatiS18,ChuGPSSW18}.

In a line of work particularly relevant to this paper, the effective resistance metric has played an important role in obtaining faster algorithms for generating random spanning trees~\cite{KelnerM09,MadryST15,Schild18}. The result of~\cite{MadryST15} partitions the graph into clusters with bounded diameter in the effective resistance metric in order to speed up simulation of a random walk, whereas~\cite{Schild18} proposed a more advanced version of this approach to achieve a nearly linear time simulation. While these results seem superficially related to ours, there does not seem to be any way of using spanning tree generation techniques for our purpose. The main reason is that the objective in spanning tree generation results is quite different from ours: there one would like to find a partition of the graph that in a sense minimizes the number times a random walk  crosses cluster boundaries, which does not correspond to a way of recovering `heavy' effective resistance edges in the graph. In particular, while in spanning tree generation algorithms the important parameter is the number of edges crossing the cuts generated by the partitioning, whereas it is easily seen that heavy effective resistance edges cannot be recovered from small cuts. 

\subsection{Our results}
In general, we cannot hope to improve on the $\tilde{O}({n})$ space complexity of the solution in \cite{kapralov2017single} because any spectral sparsifier extracted from $S B$ takes at least $O(n)$ space to represent (for constant $\epsilon$). However, there still remains a major gap in addressing our second question of decoding time. The algorithm in \cite{kapralov2017single} uses $\tilde{O}(n^2)$ decoding time. The method in \cite{ahn2013spectral} is faster, running in $\tilde{O}(n^{5/3})$ time, but it requires space $\tilde{O}({n^{5/3}})$, which is far from optimal.

We present two results that improve on these bounds. The first, summarized in Theorem \ref{thm:intro_n32}, gives a simple  algorithm that runs in space and time $\tilde{O}(n^{3/2})$. This second, summarized in Theorem \ref{thm:main_thm}, gives a more involved method that runs in $\tilde{O}(n^{1.4 + \delta})$ space and time for any constant $\delta$. Both of these algorithms are based on \emph{effective resistance sampling}, which is a powerful way of constructing spectral sparsifiers in the offline setting \cite{spielman2011graph}.

We give a detailed technical overview of both methods in Section \ref{sec:overview}. At a high level, our second algorithm can be viewed as the first successful attempt to apply ``bucketing'' methods to the graph sparse recovery problems. The most naive way to recover a sparse approximation to a vector $x$ from a sketch $S x$ is to use $S x$ to check whether or not an individual entry in $x$ is large in comparison to $\|x\|$. This ``brute-force'' approach leads to algorithms that run in $O(N)$ time for an $N$ length vector. To achieve $O(k\plog N)$ decoding time, it is necessary to check multiple entries at once, which can be done with hashing or bucketing schemes that divide $x$ into intervals of different sizes, checking the mass of entire intervals at once. 

Similarly, the graph sketching algorithm of \cite{kapralov2017single} recovers a sparsifier $\tilde{G}$ by using $SB$ to find edges with high effective resistance. It does so by checking all possible $O(n^2)$ edges in $G$, leading to a $\Omega(n^2)$ runtime. Our improvement hinges on a method for bucketing $G$ into node clusters that effectively allow for many edges to be checked simultaneous.

\subsection{Fast pseudorandomness for linear sketching algorithms}

Finally, we mention that, besides a better understanding of bucketing methods for graphs, obtaining faster sketching methods for sparsification requires solving a largely orthogonal issue, which we discuss in Section \ref{sec:reducing_cost_rand}. In particular, like many streaming algorithms, our methods are developed with the assumption that we have access to a large number of fully random hash functions. To ensure that the algorithms actually run in small space, we need to eliminate this assumption. One potential way of doing so is through the use of a pseudorandom number generator (PRG) for small space computation \cite{indyk2000stable}. However, existing PRGs used in the streaming literature run slowly in our setting, creating another $\Omega(n^2)$ time bottleneck for decoding \cite{nisan1992pseudorandom}. 

We address this issue by describing a much faster, ``locally computable'' pseudorandom generator based on construction of Nisan and Zuckerman \cite{NisanZuckerman96} and a locally computable randomness extractor of De and Vidick \cite{DeVidick10}. We hope this result will be more widely useful in designing faster sketching algorithms for graph problems and other applications.

\section{Preliminaries}
\label{sec:prelim}
Let $G=(V,E)$ be an unweighted undirected graph with $n$ vertices and $m$ edges. Let $B\in \R^{\binom{n}{2}\times n}$ denote the vertex edge incidence matrix of $G=(V,E)$.\footnote{For any distinct pair of vertices $(u,v)\in V\times V$, if $(u,v)\notin E$ then the corresponding row in $B$ is zeroed out. } Also, for any set $E' \subseteq E$ and any vertex $v\in V$ we define $\delta_{E'}(v):=\left(\{v\}\times{V}\right) \cap E'$ as the the set of edges in $E'$ connected to $v$. For any vertex $v\in V$, $\chi_v \in \R^{|V|}$ be the indicator vector of $v$. Let $B_n\in \R^{\binom{n}{2}\times n}$ denote the vertex edge incidence matrix of an unweighted and undirected complete graph, where for any edge $e=(u,v)\in V \times V, u\ne v$, its $e$'th row is equal to $\mathbf{b}_e:=\chi_u-\chi_v$.  In order to avoid repeating trivial conditions, we usually drop the condition $u\ne v$, when we say edge $e=(u,v)\in V\times V$, however, we have this condition implicitly. Also, for any distinct pair of vertices $(u,v)$, let $\mathbf{b}_{uv}:=\chi_u-\chi_v$.

For weighted graph $G=(V,E,w)$, where $w:E\rightarrow \R_+$ denotes the edge weights, let $W\in {\R_{+}}^{\binom{n}{2} \times \binom{n}{2} }$ be the diagonal matrix of weights where $W(e, e) = w(e)$. Note that $L=B^\T W B$, is the laplacian matrix of the graph. Also, let $L^{+}$ denote the
Moore-Penrose pseudoinverse of $L$.
\begin{definition}\label{def:Ggamma}
	For any unweighted graph $G=(V,E)$ and any $\gamma\ge0$, we define $L_{G^\gamma}$, as follows:
	$$L_{G^\gamma}=L_G+\gamma I.$$
	This can be seen in the following way. One can think of $G^{\gamma}$ as graph $G$ plus some regularization term and in order to distinguish between edges of $G$ and regularization term in $G^\gamma$, we let $B_{G^\gamma}=B \oplus \sqrt{\gamma}I$, where $B \oplus \sqrt{\gamma}I$ is the operation of appending rows of $\sqrt{\gamma}I $ to matrix $B$. 
\end{definition}

\subsection{Effective Resistance}
\label{subsec:resistance}
Suppose that we inject a unit current at vertex $u$ and extract from vertex $v$. Let vector $\mathbf{i}\in \R^{m}$ denote the the currents induced in the edges. Thus by Kirchoff's current law, the sum of the currents entering (exiting) any vertex is zero except for the source $u$ and the sink $v$ of electrical network. Hence, we obtain $\chi_u-\chi_v=B^\T\mathbf{i}$. Let vector $\mathbf{\varphi}\in\R^{n}$ denote the potentials induced at the vertices by the mentioned setting. Thus by Ohm's law we have $\mathbf{i}=WB\varphi$. Putting these together we get $$\chi_u-\chi_v=B^\T WB\varphi=L\varphi\text{.}$$
Observe that $(\chi_u-\chi_v) \perp \text{ker}(L)$, hence $\varphi=L^{+}(\chi_u-\chi_v )$. 

The \textit{effective resistance} between vertices $u$ and $v$ in graph $G$, denoted by $R_{uv}^G$ is defined as the voltage difference between vertices $u$ and $v$, when a unit of current  is injected into $u$ and is extracted from $v$. Thus, 
\begin{equation}\label{eq:Eff-Res}
R_{uv}=\mathbf{b}_{uv}^\T L^{+}\mathbf{b}_{uv}
\end{equation}
We also let $R_{uu}:=0$ for any $u\in V$, for convenience.

Also, for any pair $(w_1,w_2)$, the potential difference induced on this pair can be calculated as follows 
\begin{equation}
\varphi(w_1)-\varphi(w_2)=\mathbf{b}_{w_1w_2}^\top L^+\mathbf{b}_{uv}.
\end{equation}
Furthermore, if the graph is unweighted, the flow on edge $f=(u_1,u_2)$ is
\begin{equation}
\mathbf{i}(f)=\mathbf{b}_{f}^\top L^+\mathbf{b}_{uv}.
\end{equation}
 
\begin{lemma}
\label{lem:reff-energy}
Suppose that in a weighted graph $G=(V,E,w)$, we inject $1/R_{uv}$ unit of flow to $u$ and extract it from $v$. 
Let vector $\varphi \in \R^{n}$, denote the potentials induced on the vertices, i.e., $\varphi=\frac{1}{R_{uv}}L^+\mathbf{b}_{uv}$. Then $$\sum_{(a,b)=e \in E} w(e)\left(\varphi(a)-\varphi(b) \right) ^2 = \frac{1}{R_{uv}}\text{.}$$
\end{lemma}
\begin{proof}
Suppose that one injects $\frac{1}{R_{uv}}$ unit of flow to $u$ and removes it from $v$, then for the potential induced on the vertices, $\varphi=\frac{1}{R_{uv}}L^+\mathbf{b}_{uv}$. Thus, 
$$
\sum_{(a,b)=e \in E}w(e) (\varphi(a)-\varphi(b))^2 = \varphi^\top L \varphi = \frac{1}{R_{uv}^2} \cdot \bv{b}_{uv}^\top L^+ L L^+ \bv{b}_{uv} = \frac{1}{R_{uv}^2}  \cdot \bv{b}_{uv}^\top L^+ \bv{b}_{uv} =\frac{1}{R_{uv}}.
$$
Additionally, note that,
\begin{align}
\varphi(u)-\varphi(v)=\frac{1}{R_{uv}}\mathbf{b}_{uv}^\top L^+\mathbf{b}_{uv}=1,
\end{align}
which means that instead of having a current source with $\frac{1}{R_{uv}}$ unit of current, alternatively, we can have a one unit voltage source on $u$ and $v$, setting $\varphi(u)-\varphi(v)=1$.
\end{proof}
For graph $G^\gamma$ (see Definition \ref{def:Ggamma}), where $\gamma\ge 0$ and $G=(V,E)$ is an unweighted graph, we have the following corollary. 
\begin{corollary}\label{cor:Reff-energy}
	For graph $G^\gamma$, where $G=(V,E)$ is an unweighted graph and $\gamma \ge 0$, for any pair of vertices $(u,v)\in V \times V$, if vector $\varphi=\frac{1}{R_{uv}^{G^\gamma}}L_{G^\gamma}^+\mathbf{b}_{uv}$, then
	 $$\sum_{(a,b)=e \in E} \left(\varphi(a)-\varphi(b) \right) ^2+\frac{\gamma}{n} \sum_{\{a,b\}\in \binom{V}{2}} \left(\varphi(a)-\varphi(b) \right) ^2= \frac{1}{R_{uv}^{G^{\gamma}}}\text{.}$$
\end{corollary}
We also have the following characterizations of effective resistance, which we use several times in this paper. 
\begin{fact}\label{fact:eff-pot}
	For every weighted graph $G=(V,E,w)$, the effective resistance can 
	 be characterized as 
\begin{equation*} 
\begin{split}
\frac1{R_{uv}}=&\min_{\substack{\varphi\in \R^V\\\text{s.t.}: \varphi(u)-\varphi(v)=1}} \sum_{e=(u, v)\in E} w(e)(\varphi(u)-\varphi(v))^2.
\end{split}
\end{equation*}

\end{fact}
For regularized graphs, we have the following corollary, for convenience.
\begin{corollary}\label{cor:Ggamma-eff-pot}
	For graph $G^\gamma$, where $G=(V,E)$ is an unweighted graph and $\gamma \ge 0$, for any pair of vertices $(u,v)\in V \times V$, if vector $\varphi=\frac{1}{R_{uv}^{G^\gamma}}L_{G^\gamma}^+\mathbf{b}_{uv}$, then
	$$	\frac1{R_{uv}^{G^\gamma}}=\min_{\substack{\varphi\in \R^V\\\text{s.t.}: \varphi(u)-\varphi(v)=1}} \sum_{(a,b)=e \in E} \left(\varphi(a)-\varphi(b) \right) ^2+\frac{\gamma}{n} \sum_{\{a,b\}\in \binom{V}{2}} \left(\varphi(a)-\varphi(b) \right) ^2.$$
\end{corollary}
\begin{fact}\label{fact:minEnergyFlow}
For every weighted graph $G=(V,E,w)$, the effective resistance can 
be characterized as
\begin{equation*}
\begin{split}
R_{uv}=\min_{\substack{\mathbf{f}\\\text{s.t.~}B^\top \mathbf{f}=\chi_u-\chi_v}}{\sum_{e\in E} \frac{\mathbf{f}(e) ^2}{w(e)}}.
\end{split}
\end{equation*}
\end{fact}
Also, we frequently use the following simple fact.
\begin{fact}[See e.g. \cite{kapralov2017single}, Lemma 3]\label{fact:maxphi}
	For any pair of vertices $(u,v), (u',v')\in V \times V$, we have, 
	\begin{align}
	\mathbf{b}_{uv}^\top L^+\mathbf{b}_{u'v'}=\mathbf{b}_{u'v'}^\top L^+\mathbf{b}_{uv}\le \mathbf{b}_{uv}^\top L^+\mathbf{b}_{uv}=R_{uv}.
	\end{align}
\end{fact} 
We also use Rayleigh's monotonicity law throughout the paper.
\begin{fact}[Rayleigh's monotonicity law]\label{fact:mon}
	For every graph $G=(V, E)$ and every edge $e\in E$ the removal of an edge $e$ from $G$ can only increase effective resistances of other edges.
\end{fact}
\begin{definition}
In any graph $G=(V,E)$, for $u\in V$ and any $r\ge0$, we define 
$$\B_G(u,r)=\{v : v\in V , R_{uv}^G\le r \}.$$
Recall that since we defined $R_{uu}=0$, then $u\in \B_G(u,r)$.
\end{definition}
\begin{lemma}{\label{ball_approx}}
	Suppose that graphs $\wt{G}$ and $G$ are such that $V_G=V_{\wt{G}}$ and for any pair of vertices $(u,v)\in V_G\times V_G$, $R_{uv}^{\wt{G}} \le \Gamma \cdot R_{uv}^G$. We claim that for any $r \in \R_{\ge 0}$ and any vertex $u \in V_G$, we have the following:
	$$\mathbf{B}_G\left(u,\frac{r}{\Gamma}\right) \subseteq \mathbf{B}_{\wt{G}}(u,r).$$
\end{lemma}
\begin{proof}
	For any $v \in \mathbf{B}_G\left(u,\frac{r}{\Gamma}\right)$ we have that $R_{uv}^G\le \frac{r}{\Gamma}$. Therefore, by the assumption of the lemma we get $R_{uv}^{\wt{G}}\le r$, which means $v \in \mathbf{B}_{\wt{G}}(u,r)$. 
\end{proof}

In graph $G=(V,E)$, with $|V|=n$, Suppose that $\sigma \in \R^{n}$ is a demand vector, satisfying the condition $\sigma^\top \bv{1}=0$. A flow vector $\mathbf{f}\in \R^{\binom{n}{2}}$ is called $\sigma$-flow, if $B^\top \mathbf{f}=\sigma$.

In graph $G=(V,E)$, for any set of vertices $V' \subseteq V$, we denote the graph induced on $V'$ by $G_{V'}$. Also, we let $\text{diam}_{\text{eff}}(V'):= \max_{(u,v)\in V'\times V'}R_{uv}^G$ and $\text{diam}_{\text{eff}}^{\text{Ind}}(V'):= \max_{(u,v)\in V'\times V'}R_{uv}^{G_{V'}}$, which indicate the effective resistance diameter of vertices $V'$ in $G$ and $G_{V'}$, respectively. 

For matrices $C,D \in \R^{p \times p}$, we write $C\preceq D$, if $\forall x \in \R^p$, $x^\T C x \leq x^\T D x$. We say that $\wt{C}$ is $(1\pm\e)$-spectral sparsifier of $C$, and we write it as $\wt{C}\approx_\e C$, if $(1-\e)C \preceq \wt{C} \preceq (1+\e)C$.  Graph $\wt{G}$ is $(1\pm\e)$--spectral sparsifier of graph $G$ if, $L_{\wt{G}}\approx_\e L_G$. We also sometimes use a slightly weaker notation $(1-\e)C \preceq_r \wt{C} \preceq_r (1+\e)C$, to indicate that $(1-\e)x^\top C x\le x^\top \wt{C} x \le (1+\e)x^\top C x$, for any $x$ in the row span of $C$.

\section{Technical overview}\label{sec:overview}

Graph sketching, started by the work of Ahn, Guha and McGregor on solving graph connectivity in dynamic streams~\cite{ahn2012analyzing}, is the idea of designing graph algorithms that access the input graph via linear measurements. While graph sketching is a relatively recent development, the idea of linear sketching has been applied to design basic statistical estimation problems on vectors (e.g. norm estimation, heavy hitters) for more than a decade, with many efficient algorithms for fundamental problems known. 

A very successful approach to designing linear sketches for graphs, originally suggested by Ahn, Guha and McGregor, amounts to applying a classical sketching matrix $S$ to the edge incidence matrix $B$ of the graph, and then designing an offline decoding algorithm that reconstructs useful information about the graph from $S\cdot B$.  Such sketches turn out to have a very useful composability property: post-multiplying the sketch by any vector $x\in \R^n$ lets one obtain a sketch $SBx$ for any given vector in the column space of $B$, i.e. in the cut space of the graph. This property has been exploited in the literature \cite{ahn2012analyzing, ahn2012graph, ahn2013spectral, kapralov2017single, KapralovW14} to obtain space efficient sketches for connectivity (where the sketch $S$ is an $\ell_0$ sampler) and spectral sparsification (here $S$ is an  $\ell_2$-heavy hitters sketch).

\subsection{Graph sketching vs classical sparse recovery: small sketch size and efficient decoding}
A graph sketching algorithm consists of two phases. First, one maintains the sketch $S\cdot B$, where $S\in \R^{s\times {n \choose 2}}$ is the sketch and $B\in \R^{{n \choose 2} \times n}$ is the edge incident matrix of the input graphs, under dynamic edge updates (insertions and deletions). Next, at the end of the stream one runs a (usually nonlinear) decoding algorithm on the sketch to produce a sparsifier.  Note that the space complexity of the algorithm is the number of rows in $S$ times $n$, the number of vertices of the graph. We now describe approaches for spectral sparsification through linear sketches that have been developed in the literature. We focus on {\em single-pass} dynamic streaming algorithms, i.e. {\em oblivious sketches}. In this setting, the only known approach for designing space efficient sketches is to implement the effective resistance sampling approach of Spielman and Srivastava~\cite{spielman2011graph}\footnote{If two passes over the stream are allowed, one can use a relation between spectral sparsifiers and spanners~\cite{KapralovP12} and exploit spanner construction algorithms~\cite{KapralovW14}, but this approach is not known to extend to the single pass setting.}.

 In order to construct a spectral sparsifier, as per ~\cite{spielman2011graph}, one samples edges with probabilities proportional to their effective resistances, and gives sampled edges appropriate weights (inverse of the sampling probability) to make the estimate unbiased. Our main problem is to design an oblivious sketch $S\in \R^{s\times {n \choose 2}}$ with a small number of rows $s$ that allows efficient recovery of such a sample. 
 
It is known from prior work~\cite{kapralov2017single} that the main challenge in constructing sketches for spectral sparsification lies in designing a sketch $S$ that allows recovery of `heavy' edges of the graph, or edges with large (e.g., larger than $\approx 1/\log n$ -- such edges need to be included in a sparsifier with probability $1$ as per the sampling approach of Spielman and Srivastava) effective resistance. For simplicity we focus on this question of finding heavy edges for the purposes of our overview, as a reduction introduced in~\cite{kapralov2017single} can be used to convert any such primitive into a full-fledged sparsification routine. We refer to the problem of recovering high effective resistance edges as the \textsc{HeavyEdges} problem. The task is to design a sketch $S\in \R^{s\times {n \choose 2}}$ with a small number of rows $s\ll n$ that allows solving the following problem:

\begin{center}
\fbox{
\parbox{0.9\textwidth}{
\textsc{HeavyEdges}($S\cdot B, \wt{G}$)\vspace{0.1in}\\
\vspace{0.1in}
\noindent{\bf Input:} ~~~~Sketch $S\cdot B$ of graph $G=(V, E)$, a coarse spectral sparsifier $\wt{G}$ of $G$\\
\vspace{0.1in}
\noindent{\bf Output:} List $Q\subseteq E$ of size $s\cdot n^{1+o(1)}$ that contains all edges $e\in E$ with effective resistance $\geq 1/\log n$
}}
\end{center}
In the description above we refer to $\wt{G}$ as a coarse sparsifier of $G$  if for some parameter $\Gamma>1$ one has 
$$
\frac1{\Gamma} K\preceq \wt{K} \preceq K,
$$
where $K$ is the Laplacian of $G$, $\wt{K}$ is the Laplacian of $\wt{G}$ and $\preceq$ stands for the positive semidefinite ordering on matrices.

Two approaches have been designed for this problem in the literature. 

\paragraph{Spectral sparsifiers through inverse connectivity sampling: suboptimal space but subquadratic decoding time.} The first approach, introduced by Ahn, Guha and McGregor~\cite{ahn2013spectral}, is based on relating effective resistances in graphs to edge connectivity: one proves that an edge of large effective resistance   (e.g., $\geq 1/\log n$), must have nontrivially large connectivity, and concludes that sampling with probabilities proportional to (overestimates of) inverse connectivities suffices as long as a large enough number of samples is taken. The latter is possible using a spanning forest sketch~\cite{ahn2012analyzing}. Unfortunately, the relation between effective resistance and inverse connectivities is rather weak in general, and this approach leads to an algorithm with $\wt{O}(n^{5/3})$ space and time complexity. We note that this approach inherently relies on the idea of recovering high effective resistance edges by using the fact that they must cross (reasonably) small cuts.

\paragraph{Spectral sparsifiers through $\ell_2$ heavy hitters: optimal space but quadratic decoding time.} Another approach, introduced in~\cite{kapralov2017single}  uses an $\ell_2$-based characterization of edges with high effective resistance. An edge $e=(u, v)$ has effective resistance at least $\gamma\in (0, 1]$ in the graph $G$ if and only if at least a $\gamma$ fraction of the energy of the electrical flow from $u$ to $v$ is contributed by that edge. This characterization allows one to recover high effective resistance edges $(u, v)$ by applying an $\ell_2$ heavy hitters sketch to the edge incidence matrix $B$. Specifically, if $\varphi\in \R^{|V|}$ is the vector of vertex potentials induced by injecting one unit of flow at $u$ and removing $1$ unit of flow from $v$, then the effective resistance $R_e$ of edge $e=(u, v)$ satisfies
\begin{equation}\label{eq:reff}
R_e=\frac{(B\varphi)_{e}^2}{||B\varphi||_2^2}.
\end{equation}
The relation~\eqref{eq:reff} above implies that if one chooses $S\in \R^{\log^{O(1)} n\times {n \choose 2}}$ to be a $1/\log n$-heavy hitters sketch,  post-multiplies the sketch $SB$ by the vector of potentials $\varphi$ and decodes the resulting sketching using standard $\ell_2$ heavy hitters decoding, the resulting list of nonzero coordinates will contain $e$.  This is a very useful observation, but it does not quite lead to an algorithm, since in order to compute the vector of potentials $\varphi$, one needs to know the entire graph $G$! It turns out, however, that the exact $\varphi$ is not needed. If a {\em coarse (large constant factor) sparsifier} $\wt{G}$ of $G$ is available explicitly, then one can instead compute the corresponding vector of potentials $\wt{\varphi}$ and decode $S\cdot B\wt{\varphi}$ instead. The approximation quality of $G$ by $\wt{G}$ affects the size of the sketch, but the approach still works -- this is the algorithm of~\cite{kapralov2017single}. This approach leads to an optimal $\wt{O}(n)$ space complexity, but suffers from large runtime: the decoding is essentially brute-forcing over all potential edges $\{u, v\}\in {V \choose 2}$, and hence the runtime is quadratic.

\paragraph{Our approach to efficient recovery: `bucketing' vertices by ball carving in effective resistance metric.} It is interesting to contrast the state of the art in graph sketching with classical sketching algorithms for heavy hitters. The problem in heavy hitters is: design a sketch $S\in \R^{s\times n}$ such that for every vector $x\in \R^n$ most of whose mass is in the top $k$ coordinates one can recover a good approximation to those coordinates from the low dimensional vector $Sx$. This problem can be solved in optimal space by hashing into {\bf $\approx k$ buckets} and then using brute force decoding over the universe of size $n$ (e.g. the \textsc{CountSketch} algorithm).  This is similar in flavor to the result of \cite{kapralov2017single}, where space complexity (or, sketching dimension) is optimal, but the decoding is brute-force over the universe of possible edges, i.e. over ${V \choose 2}$. For classical sketching solutions have been proposed that achieve the optimal bounds on sketch size and also run in {\em sublinear} $\wt{O}(k)$ time by careful decoding of the buckets, but no equivalent approach was known prior to our work. The question that we ask is:

\begin{center}
	\fbox{
\parbox{0.9\textwidth}{\begin{center} Can one construct a `bucketing scheme' for spectral sparsification via sketching that will allow fast recovery?\end{center}}}
\end{center}
 Our result is the first to define a notion of `buckets' in graph sketching that admit efficient decoding primitives. Our `bucketing scheme' is based on {\bf ball carving in the effective resistance metric} of the input graph: our algorithm (Algorithm~\ref{alg:main-sparsify} for sparsification and Algorithm~\ref{alg:main-heavy-edge} for the \textsc{HeavyEdges} problem) is a recursive procedure that constructs progressively better approximations to the effective resistance metric of the input graph, and at every point partitions the vertices of the graph by ball carving in the effective resistance metric learnt so far in order to speed up recovery of new important edges.

\subsection{Our techniques}\label{sec:our-tech}

We now present an overview of our techniques. 

\paragraph{Ensuring a lower bound on minimum degree via peeling.} Our development in this paper starts with the observation that for every $d\geq 1$, at the cost of $\wt{O}(nd)$ space and time, a linear sketching algorithm can assume that the input graph $G$ has minimum degree lower bounded by $d$. This is due to the fact that we can store a sketch $S\cdot B$ of the edge incidence matrix of $G$ with $\wt{O}(d)$ rows that allows recovery of all edges incident on any given vertex $v$ of degree at most $d$, with high probability, as well as store all vertex degrees exactly (using a sketch or a simple counter). The algorithm can therefore perform the following decoding operation in $\wt{O}(nd)$ time. Iteratively find the smallest degree vertex in the graph, recover all incident edges, subtract these edges from the sketch and repeat on the residual graph while a vertex of degree at most $d$ exists. Such iterative processes are often hard to implement using sketches due to dependencies that may develop, but in this case this issue does not arise: the algorithm stores degrees of all vertices {\bf exactly}, and therefore the execution path of such a peeling process does not depend on the sketches as long as high probability success events for sparse recovery sketches happen at all intermediate iterations. Since a given edge can be subtracted from any sketch that we use in $\log^{O(1)} n$ time, this results in an algorithm with $\wt{O}(nd)$ runtime. See proof Lemma~\ref{lem:E_j-contains} in Section~\ref{sec:n15} for details. 

In what follows that our input graph has minimum degree lower bounded by a parameter $d\geq 1$, at the cost of a $\wt{O}(nd)$ additive term in space and decoding time. We set $d=\wt{\Theta}(n^{1/2})$ for our warm-up result below (a simple sketch with $\wt{O}(n^{3/2})$ space and decoding time; see below for overview and Section~\ref{sec:n15} for the actual algorithm), and as $d=\wt{\Theta}(n^{0.4})$ for our main result (see below for an overview  and Section~\ref{sec:algo} onwards for details).

\paragraph{Warm-up result: $\wt{O}(n^{3/2})$ space and time via edge connectivities.} We start by noting that if a lower bound on the minimum degree of a graph is assumed, one can prove a stronger relation between edge connectivities and effective resistances than the one used in~\cite{ahn2013spectral}. Specifically, we show that if the minimum degree $d$ in the input graph $G=(V, E)$ is lower bounded by $d\approx \sqrt{n}$, then every edge $e\in E$ with effective resistance $R_e^G\geq 1/\log n$, say, necessarily has connectivity at most $\wt{O}(\sqrt{n})$. More formally, we show:

\begin{lemma}[Informal version of Lemma~\ref{lem:min-cut-R}]
For every graph $G=(V,E), |V|=n$, every integer $d\geq 1$, if for every vertex $v \in V$, we have $\deg(v) \geq \sqrt{n}(\log n)^2$, then for every edge $e\in E$ with edge-connectivity $\lambda_{e} \geq 200\sqrt{n}$ one has $R_e^G \leq 1/\log n$. 
\end{lemma}
To prove the lemma, for every edge $e=(u, v)$ we consider the optimal line embedding of the graph $G$ given the vertex potentials $\varphi$ induced by injecting one unit of flow into $u$ and taking it out at $v$. The sum of squared potential differences over edges of $G$ (i.e. the energy of the  embedding $\varphi$) is the effective conductance between $u$ and $v$ (the inverse of effective resistance). We then use the min-degree assumption to note that all cuts that contain $\ll d$ vertices on one side are of size at least $\Omega(d^2)$ and therefore conclude that any low energy line embedding must map large groups of vertices close together. The latter fact, together with the connectivity assumption implies a lower bound on the conductance of the edge $(u, v)$ (see proof of Lemma~\ref{lem:min-cut-R} in Section~\ref{sec:n15} for the details).

Lemma~\ref{lem:min-cut-R} immediately yields a solution to our \textsc{HeavyEdges} problem with $s\approx \sqrt{n}$: one simply uses a sketch that recovers all edges with connectivity at  most $\approx \sqrt{n}$ and outputs this list. The latter can be one using a result of ~\cite{ahn2012analyzing}, but the decoding time for that procedure is quadratic in the connectivity parameter (due to the need to subtract recovered spanning forests from the sketch), which unfortunately does not yield a runtime improvement for our setting. However,  we give a simple sketch based on the spanning forest algorithm of~\cite{ahn2012analyzing} that achieves linear runtime in Section~\ref{subsec:find_low}. In Section~\ref{sec:n15} we show how this outline leads to an algorithm with $\wt{O}(n^{3/2})$ space and runtime complexity (this part of the analysis follows the ideas developed in ~\cite{kapralov2017single}). Formally, we prove 

\begin{theorem}
	\label{thm:intro_n32}
There exists an algorithm such that for any $\epsilon > 0$, processes a list of edge insertions and deletions for an unweighted graph $G$ in a single pass and maintains a set of linear sketches of this input in $\wt{O}(\e^{-2}n^{1.5})$
space. From these sketches, it is possible to recover, with high probability, a weighted subgraph $H$ with $O(\frac{1}{\epsilon^2}n\log n)$ edges, such that $H$ is a $(1 \pm \epsilon)$-spectral sparsifier of $G$. The algorithm recovers $H$ in $\wt{O}(\epsilon^{-2}n^{1.5})$ time.
\end{theorem}

Unfortunately, the $\wt{O}(n^{3/2})$ space and time complexity provided by Theorem~\ref{thm:intro_n32} seems to be the limit of the idea of removing low degree vertices and then recovering low connectivity edges, because Lemma~\ref{lem:min-cut-R} is optimal (consider a union of $\approx \sqrt{n}$ cliques of size $\approx \sqrt{n}$ connected by matchings, with the first and the last clique further connected by an edge). This motivates the following goal:
\begin{center}
\fbox{
\parbox{0.9\textwidth}{ \begin{center} Design a sketching algorithm for spectral sparsification in dynamic streams that achieves better than $n^{3/2}$ space and decoding time simultaneously. \end{center}}
}
\end{center}
Such a result would need to use $\ell_2$ heavy hitters sketches to go beyond the $n^{3/2}$ space complexity, but at the same time must avoid brute-force decoding used in~\cite{kapralov2017single}. Our main result achieves exactly that: we give an algorithm with $\wt{O}(n^{1.4+o(1)})$ space and decoding time that uses $\ell_2$ heavy hitters sketches to go beyond the relation between effective resistances and connectivity, but at the same time avoids brute force decoding using a novel scheme for bucketing nodes in a graph based on ball carving in the effective resistance metric of the underlying graph.

\paragraph{Main technique: ball-carving in effective resistance metric.}  We start by recalling the high level approach of~\cite{kapralov2017single}, and then outline the main technical ideas involved in implementing the $\ell_2$-heavy hitters decoding to run faster than the $n^2$ brute-force approach. The algorithm of~\cite{kapralov2017single} is 

\begin{center}
\fbox{
\parbox{0.9\textwidth}{
\textsc{HeavyEdgesBruteForce}($S\cdot B, \wt{G}$)\vspace{0.1in}\\
\vspace{0.1in}
\noindent{\bf Input:} ~~~~Sketch $S\cdot B$ of graph $G=(V, E)$, a coarse spectral sparsifier $\wt{G}$ of $G$\\
\vspace{0.1in}
\noindent{\bf Output:} ~~List $Q\subseteq E$ of size $n\cdot s\cdot \log^{O(1)} n$ that contains all edges $e\in E$ with effective resistance $\geq 1/\log n$\\
\vspace{0.01in}

\noindent Initialize $Q\gets \emptyset$\\
{\bf for}~$u\in V$\\
\text{~~~~}{\bf for}~$v\in V\setminus \{u\}$\\
\text{~~~~~~~~}Compute $\wt{\varphi}=\wt{L}^+(\chi_u-\chi_v)$\text{~~~~~~}$\rhd$ $\wt{\varphi}=$ potentials induced by flow from $u$ to $v$ in $\wt{G}$\\
\text{~~~~~~~~}Decode $S\cdot B\wt{\varphi}$, add result to $Q$\\
\text{~~~~}{\bf end for}\\
{\bf end for}\\
}}
\end{center}

The algorithm above recovers a heavy edge $e=(u, v)$ whenever it routes flow from $u$ to $v$ in the coarse sparsifier $\wt{G}$, so the natural question is whether one can group vertices into clusters, or buckets, to avoid testing all pairs.  

\paragraph{Group testing heavy edges by bucketing.} The main idea underlying our analysis is the following bucketing scheme. Suppose that we are able to partition the vertex set $V$ of $G$ into vertex disjoint subsets $P_1,\ldots, P_t$ such that the effective resistance diameter of every set $P_i$ is smaller than a parameter $\beta\ll 1/\log n$.
Now consider an edge $e=(u, v)$ with $R_e\geq 1/\log n$. Since $\beta\ll 1/\log n$, it must be that the endpoints $u$ and $v$ belong to different elements of the partition $P_1\cup \ldots \cup P_t$! This suggests the following approach: instead of sending flow from $u$ to $v$ for every potential pair of vertices, proceed as follows for every element $P_i$ of the partition $V=P_1\cup \ldots \cup P_t$. First contract the subset $P_i$ of $V$ to a supernode $s$ in the coarse sparsifier $\wt{G}$, obtaining (explicitly) a graph $\wt{G}_i:=\wt{G}/P_i$. Then for every node $v\in V\setminus \{s\}$ compute the potential $\wt{\varphi}_i^v:=\wt{L}_i^+(\chi_v-\chi_s)$ induced by unit electrical flow from $s$ to $v$ in $\wt{G}_i$ and decode the sketch $SB\wt{\varphi}_i^v$.   We record this informally in the \textsc{HeavyEdgesFast} algorithm below. We note that the \textsc{HeavyEdgesFast} primitive below serves as an approximation to our actual \textsc{HeavyEdges} algorithm, namely Algorithm~\ref{alg:main-heavy-edge} from Section~\ref{sec:HEanalysis}.

\begin{center}
\fbox{
\parbox{0.99\textwidth}{
\textsc{HeavyEdgesFast}($S\cdot B, \wt{G}$)\vspace{0.1in}\\

\noindent{\bf Input:} ~~Sketch $S\cdot B$ of graph $G=(V, E)$, a coarse spectral sparsifier $\wt{G}$ of $G$,\text{~~~~~~~~~~~~~~~~} \text{~~~~~~~~~~~~~~}partition $V=P_1\cup \ldots\cup P_t$\\
\noindent{\bf Output:} List $Q\subseteq E$ of size $n\cdot s\cdot \log^{O(1)} n$ that contains all edges $e\in E$ with effective \text{~~~~~~~~~~~~~~}resistance $\geq 1/\log n$\\
\vspace{0.01in}

\noindent Initialize $Q\gets \emptyset$\\
{\bf for}~$i=1$ to $t$\\
$\wt{G}_i\gets \wt{G}/P_i$, $s\gets $supernode corresponding to $P_i$\\
\text{~~~~}{\bf for}~all vertices $v$ in $\wt{G}_i$, $v\neq s$\\
\text{~~~~~~~~}Compute $\wt{\varphi}_v=\wt{L}^+(\chi_v-\chi_s)$\text{~~~~}$\rhd$ $\wt{\varphi}_v=$ potentials induced by flow from $s$ to $v$ in $\wt{G}/P_i$\\
\text{~~~~~~~~}Decode $S\cdot B\wt{\varphi}_v$, add result to $Q$\\
\text{~~~~}{\bf end for}\\
{\bf end for}\\
}}
\end{center}

Note that this approach amounts to recovering all edges that could go from $v$ to any node in $P_i$ at the cost of only one flow computation as opposed to $|P_i|$ computations, and hence is promising, as long as we can prove that this approach correctly recovers heavy edges with one endpoint in $P_i$. Our first crucial observation is that this is method of recovering edges actually works, because if for some vertex $u\in P_i$ one has $R_{uv}\geq 1/\log n$, then the effective resistance between $s$ and $v$ in the contracted graph $\wt{G}_i=\wt{G}/P_i$ will be large. Formally this is guaranteed by

\begin{lemma}{\label{lem:additive}}
	In graph $G=(V_G,E_G)$, suppose that vertex $u\in V_G$ belongs to a set of vertices $C$, where $\text{diam}_{\text{eff}}(C)\le \beta$. Also assume that $H=G\slash C$ and let $c$ denote the corresponding super-node
	, i.e., $H=\left(\{c\}\cup V_G\setminus C, E_H, w_H\right)$ is the resulting graph after contracting vertices of $C$ in $G$. Then for any $v\notin C$ such that $R_{uv}^G\ge \beta$ one has $$R_{cv}^H\ge R_{uv}^G \left(1-\frac{\beta}{R_{uv}^G}\right)^2$$
\end{lemma}
The proof of the lemma is given in Appendix~\ref{app:algo}.

Our actual \textsc{HeavyEdges} algorithm (Algorithm~\ref{alg:main-heavy-edge} in Section~\ref{sec:HEanalysis}) crucially uses this lemma for correctness analysis. Now that we know that the idea of contracting subsets of vertices of low effective resistance diameter leads to a correct algorithm, we need to understand the runtime. The runtime depends on the number of elements $t$ in the partition $V=P_1\cup \ldots \cup P_t$\footnote{We note that in the actual \textsc{HeavyEdges} algorithm the sets $P_1\cup\ldots\cup P_t$ in general {\bf may not} form a partition -- see Section~\ref{sec:algo}. They do, however, form a partition if the minimum cut in the input graph is lower bounded by $d$, for example.}: the contraction can be done in $\wt{O}(n)$ time as the coarse sparsifier $\wt{G}$ is given to us explicitly, and computing all the necessary sketches $SB\wt{\varphi}_i^v$ can also be accomplished in $\wt{O}(n)$ time using standard techniques (see Algorithm~\ref{alg:main-heavy-edge}, line~\ref{line:HH-FHE}, and its analysis in Lemma~\ref{lem:corr-main-HE} in Section~\ref{sec:HEanalysis}), so the over all runtime is $\wt{O}(nt)$. This means that in order to make our approach work, we need to answer the following question:
\begin{center} 
	\fbox{
\parbox{0.9\textwidth}{\begin{center}Can a graph $G=(V, E)$ with minimum degree lower bounded by $d$ partitioned into {\bf few ($t\ll n$)} clusters $V=P_1\cup \ldots \cup P_t$ of low (e.g., $\ll 1/\log n$) effective resistance diameter?\end{center}}}
\end{center}
It is not hard to show that one can always partition $G$ into $t=\wt{O}((n/d)\log n)$ such clusters. Unfortunately, however, the bound $t=\wt{O}(n/d)$ is essentially tight, and is not sufficiently good for our purposes. For tightness, is it easy to construct graphs with minimum degree lower bounded by $d$ where the number of clusters $t$ must be $\Omega(n/d)$ -- just consider a $n/d$ cliques of size $d$, joined by a path to ensure connectivity. This means that without further ideas we cannot ensure that the number of elements  $t$ in our partition is smaller than $n/d$, which means that the runtime of the process above is $\approx n^2/d$, which together with $\approx nd$ time for recovering edges  incident on vertices of degree $\leq d$ gives overall space and time $\approx n^2/d+nd$, which is at least $n^{3/2}$ for all choices of $d$.

The next observation that we need is that {\bf if} using a sketch with $\wt{O}(nd)$ rows we could not only ensure that our graph has minimum degree lower bounded by $d$, but also ensure that the minimum-cut in the graph is $\Omega(d)$, a much better result would follow\footnote{In the actual algorithm we are not able to reduce the problem to the setting when the input graph $G$ has min-cut lower bounded by $d$ at the expense of only $\wt{O}(nd)$ space and time complexity. However, we instead ensure a weaker condition that turns out to be sufficient for our purposes -- see the analysis of the \textsc{BallCarving} algorithm (Algorithm~\ref{alg:ballcarving}) in Section~\ref{sec:BC}.}. In the actual algorithm we are not quite able to ensure that the min-cut in our graph is lower bounded by $d$, but rather use a weaker assumption -- see Section~\ref{sec:BC}. Nevertheless, it is useful for this overview to consider the following question:
\begin{center} 
	\fbox{
\parbox{0.9\textwidth}{\begin{center} Can a graph $G=(V, E)$ with {\bf minimum cut} lower bounded by $d$ partitioned into {\bf few ($t\ll n/d$)} clusters $V=P_1\cup \ldots \cup P_t$ of low (e.g., $\ll 1/\log n$) effective resistance diameter?\end{center}}}\end{center}

The answer to this questions turns out to be yes, and the quantitative bounds are sufficient to achieve a better than $n^{3/2}$ space and time -- this is exactly how our algorithm works. We show that every graph $G=(V, E)$ with minimum cut lower bounded by $d$ can be partition into vertex disjoint subsets $P_1\cup \ldots \cup P_t$ with effective resistance diameter $1/\log n$, say, with the number of partitions $k$ being much smaller than $n/d$ (in contrast with our previous version of this question, where $n/d$ was the best possible bound). This fact is exactly what underlies our runtime and space complexity of $n^{1.4+o(1)}\ll n^{3/2}$. The algorithm for doing this is simple: we repeatly pick vertices of the graph (ball centers) and remove balls of a given effective resistance radius, until no vertex remains. The algorithm is given below:

\begin{center}
\fbox{
\parbox{0.9\textwidth}{
\textsc{BallCarving}($S\cdot B, \wt{G}$)\vspace{0.1in}\\
\vspace{0.1in}
\noindent{\bf Input:} ~~~~A coarse spectral sparsifier $\wt{G}$ of $G$, radius $r\in (0, 1)$\\
\vspace{0.1in}
\noindent{\bf Output:} \text{~~}Partition $V=P_1\cup \ldots\cup P_t$ into vertex disjoint subsets of effective resistance diameter $\ll 1/\log n$\\
\vspace{0.01in}

\noindent Initialize $V_{active}\gets V$\\
$t\gets 0$\\
{\bf while}~$V_{active}\neq \emptyset$\\
\text{~~~~~~~~}$t\gets t+1$\\
\text{~~~~~~~~}$v\gets $ a vertex in $V_{active}$\\
\text{~~~~~~~~}$P_i\gets \B_{\wt{G}}(v, r)\cap V_{active}$\\
\text{~~~~~~~~}$V_{active}\gets V_{active}\setminus P_t$\\
{\bf end while}\\
}}
\end{center}

We note that the \textsc{BallCarving} algorithm above serves as an illustration to our actual \textsc{BallCarving} primitive (Algorithm~\ref{alg:ballcarving}) presented and analyzed in Section~\ref{sec:BC}.

We show
\begin{theorem}[Informal]\label{lm:ballcarving-informal}
For every graph $G=(V, E)$ with {\bf minimum cut} lower bounded by $d=n^{0.4}\log^{O(1)} n$, if $\wt{G}$ is a constant factor spectral approximation to $G$, then the procedure \textsc{BallCarving} above outputs a partitioning $V=P_1\cup \ldots P_t$ into $t\leq n^{0.4}\log^{O(1)} n$.
\end{theorem}

The fact that the number of parts in the decomposition is only $\wt{O}(n^{0.4})$ even when $d=\wt{O}(n^{0.4})$ is exactly the reason why our ball-carving based approach is able to go beyond the $n^{3/2}$ space and time barrier. 

The proof of Lemma~\ref{lm:ballcarving-informal} is never used for the actual analysis, but follows formally from the following two crucial bounds that we prove.
Theorem~\ref{thm:numpartitions} bounds the number of elements in the collection $P_1\cup\ldots\cup P_t$ output by our \textsc{BallCarving} algorithm (see Algorithm~\ref{alg:ballcarving} in Section~\ref{sec:BC}) is the core tool behind our analysis. Its proof is given in Section~\ref{sec:BC}.
\begin{theorem}\label{thm:numpartitions}
	For every graph $G=(V, E)$, and a set of edges $E'\subseteq E$ if:
	\begin{enumerate}
		\item The minimum degree in $G$ is lower bounded by $d\geq 1$.
		\item For some $0<\beta< 1/\log n$ and integer $k \ge 1$ the vertex set $V$ admits a partition $V=C_1\cup \ldots\cup C_k$ such that for every $i\in [k]$ the subgraph induced by $C_i$ has effective resistance diameter bounded by $\Delta\in (0, 1)$, i.e., $\text{diam}_{\text{eff}}^{\text{Ind}}(C_i)\le \Delta$ such that $\Delta < \frac{\beta}{400\sqrt{k}}$. 
		\item Edge set $E'$ contains all the edges of connectivity no more than $d$ in graph $G$.
		\item $SB$ is the set of needed sketches.
	\end{enumerate}
	
	Then \textsc{BallCarving}($S\cdot B , E' , \beta, {G}$) returns a set of disjoint subsets of vertices with effective resistance diameter bounded by $2\beta$ in the metric of $G$, and there are no more than $\wt{O}\left(\sqrt{k}+\frac{k}{\beta\cdot d}\right)$ such non-singleton partitions.  
\end{theorem}

We note that the result of Theorem~\ref{thm:numpartitions} depends on the quality of the clustering $V=C_1\cup \ldots \cup C_k$ whose existence is assumed by the theorem. Such a clustering is provided by Theorem~\ref{thm:main-partition} below:

\begin{theorem}\label{thm:main-partition}
	For any unweighted graph $G=(V,E)$ that $|V|=n$ and with min-degree at least $n^{0.4}\log^2 n$, set of vertices, $V$, admits a partitioning into $V=C_1\cup \dots\cup C_k$, that for any 
	$$\forall i\in [k], \text{diam}_{\text{eff}}^{\text{Ind}}(C_i)	\le \frac{10}{n^{0.4}}$$
	and $$k \le  c \cdot n\sqrt{\log n}\sqrt{\frac{1}{n^{0.4}\log^2 n}}=c\cdot n^{0.8} \sqrt{\frac{1}{\log n}}=O(n^{0.8})$$
\end{theorem}
We note that Theorem~\ref{thm:main-partition} provides a decomposition of the vertex set $V$ into vertex-disjoint sets $C_1\cup \ldots \cup C_k$ such that every $C_i$ has {\em very low} effective resistance diameter, about the inverse of the minimum degree, {\em as an induced subgraph}. The proof is a quantitative improvement of the work~\cite{alev2017graph} on graph clustering using effective resistances and is provided in Appendix~\ref{sec:partitioning}. Qualitatively, the fact that the number of clusters of diameter $\approx 1/d$ can be made polynomially smaller than $n$ is one of the main observations that enable our algorithm (Algorithm~\ref{alg:main-sparsify}, presented in Section~\ref{sec:algo}) to go $n^{3/2}$ space and decoding time simultaneously.

We note that Theorem~\ref{lm:ballcarving-informal} follows by combining Theorem~\ref{thm:main-partition} with Theorem~\ref{thm:numpartitions}, and observing that if the input graph $G$ does satisfy the minimum cut assumption, then the set $E'$ passed to \textsc{BallCarving} may be empty, in which case $P_1\cup\ldots\cup P_t$ do form a partition, as required. However, we note that Theorem~\ref{lm:ballcarving-informal} is just a toy application of our techniques, and we therefore do not provide the full proof, instead referring the reader to the formal analysis of our \textsc{HeavyEdges} primitive (Algorithm~\ref{alg:main-heavy-edge}) in Section~\ref{sec:algo} as well as the \textsc{BallCarving} (Algorithm~\ref{alg:ballcarving}) analysis in Section~\ref{sec:BC}

The proof of Theorem~\ref{thm:numpartitions} is the technical core of the paper (see~Section~\ref{sec:BC}). The main idea behind the proof is a very natural ball-growing process that helps us lower bound efficiency of ball carving. The simple main observation is that if \textsc{BallCarving}, when run with parameter $r$ as the radius, outputs many partitions, then balls of radius $r/2$ around the corresponding ball centers do not overlap, which as we show is not possible. The proof is by considering a natural ball-growing process that, starting with any node $u\in V$, keeps growing a ball in effective resistance metric up to radius $r/2$. We show that most such balls will capture many vertices. Since the balls are disjoint, this implies that there cannot be too many of them, and consequently there cannot be too many elements in the partition that \textsc{BallCarving} outputs. The resistance metric in question can be thought of as the effective resistance metric of $G$ itself for the purposes of this outline. In the actual algorithm we use the effective resistance metric of the coarse sparsifier $\wt{G}$ since we do not have access to the effective resistance metric of $G$, and show that these two metrics are equivalent for our purposes up to a small loss in parameters (see proof of Theorem~\ref{thm:ballcarving} in Section~\ref{sec:BC}).  We refer the reader to Section~\ref{sec:BC} for mode details.

\subsection{Reducing the cost of randomness}
\label{sec:reducing_cost_rand}

In the previous section, we discussed the primary challenge in obtaining better space vs. decoding time tradeoffs for spectral sparsification in dynamic graph streams. In particular, faster decoding requires an understanding of how to apply ``bucketing'' methods to what is essentially a sparse recovery problem involving graphs. Our primary technical contribution is the first substantial progress towards this understanding. 
However, beyond this contribution, obtaining faster decoding time also requires solving another mostly orthogonal problem: we need a faster way to generate pseudorandom bits for use in our randomized sketching algorithms.  This is an issue that has largely been unaddressed, as decoding speed has not previous been an objective of prior work linear sketching algorithms for sparsification \cite{kapralov2017single}.

\paragraph{Nisan's pseudorandom number generator.}  Like most streaming algorithms, our methods depend heavily on randomness. We compute sketches of the form $S\cdot B$, where $S$ is a \emph{randomly constructed matrix} with ${n\choose 2}$ columns and $s = o(n)$ rows. Naively, just storing $S$ after random initialization would take $\Omega(n^2)$ space, dominating the space complexity of our algorithms. Accordingly, to obtain truly space efficient methods, we need to find a more compact way of representing the random matrix $S$. This is not a challenge unique to graph sketching --  essentially all linear sketching require efficient ways of representing the sketch matrix $S$.

While there are a number of ways of handling this issue (e.g. many algorithms build $S$ using low-independence hash functions), one of the most generic techniques is to generate $S$ using a \emph{pseudorandom number generator} with a small seed. Indyk first applied this idea to algorithms for estimating vector norms in a streaming setting \cite{indyk2000stable}. He showed that any pseudorandom number generator than can ``fool'' a small space algorithm can also fool any linear sketching algorithm with a small sketch size (i.e., with few rows in $S$).

Instantiating Indyk's result with Nisan's well known pseudorandom number generator \cite{nisan1992pseudorandom} allows for $S$ to be generated from a seed of just $\tilde O(N \log R)$ random bits, as long as $S\cdot x$, or in our case, $S\cdot B$ can be stored in $N$ space and $S$ can be generated from $R$ random bits. Instead of storing $S$, we just need to store this small random seed and columns of all of $S$ can be generated ``on-the-fly'' as needed. This is a powerful result: since Indyk's original application, Nisan's generator has become a central tool in streaming algorithm design.

Unfortunately, when runtime is a concern, Nisan's pseudorandom number generator is a costly option for graph streaming algorithms. If $R$ random bits are required to generate $S$, Nisan's generator requires $O(N \log R)$ time to generate even a single random bit from its $O(N\log R)$ length seed. In our setting $R$ is polynomial in $N$, but upwards of $O(n)$ random bits in $S$ need to be accessed during the cost of our decoding algorithms. Generating these random bits ``on-the-fly'' using Nisan's generator would immediately imply an $\Omega(nN) \geq \Omega(n^2)$ runtime for decoding. 

\paragraph{A faster pseudorandom generator.} 
To deal with the cost of generating $S$ in a pseudorandom way, in Section \ref{sec:prg} we present a pseudorandom generator that is \emph{much faster} than Nisan's. In particular, we show that, at least when $R$ is polynomial in $N$, it is possible to construct a generator that still uses a seed of just $O(N \plog N)$ random bits (i.e., only a $\plog N$ factor more than Nisan's) but can generate any pseudorandom bits in $\polylog N$ time instead of $N\log N$ time.

Our generator can be constructed by carefully combining several results from the literature on pseudorandomness. Ultimately, Nisan's pseudorandom generator requires $O(N\log N)$ to generate a single pseudorandom bits from its $O(N\log N)$ length seed because every pseudorandom bit output depends on \emph{every seed bit}. To avoid this cost, we need a generator that is inherently \emph{local}, with each pseudorandom bit only depending on $\plog N$ seed bits. 

While ``locally computable'' pseudorandom number generators have not been studied directly, there do exist locally computable constructions of \emph{extractors}, a closely related object \cite{Vadhan04,lu2002hyper,deVidickJournal}. The goal of an extractor is to extract a small string of nearly uniform random bits from a long stream of \emph{weakly} random bits. In certain cryptographic settings, it is desirable to do so in a way that only bases each output bit on a relatively small number of input bits.

Furthermore, it is actually possible to construct a pseudorandom number generator using an algorithm for randomness extraction. In particular, by plugging a locally computable extractor from \cite{DeVidick10} into a pseudorandom generator of Nisan and Zuckerman \cite{NisanZuckerman96}, we obtain a generator that can compute each pseudorandom bit using just $O(\plog N)$ pseudorandom bits. Naively, this construction can output up to $N^2$ pseudorandom bits using a seed of $\tilde{O}(N)$. We describe a relatively simple iterative process which further exands the output to $N^c$ pseudorandom bits,, while still maintaining a generation time of $\plog N$ if $c$ is constant.

There are likely many possible improvements to our basic construction. We hope that bringing a broader set of tools from the pseudorandomness literature to the streaming algorithms community, we can initiate an exploration of these improvements, which will lead to faster linear sketching

\section{The algorithm and its analysis}\label{sec:algo}
In this section we present our main sketch-based sparsification algorithm (\textsc{Sparsify}, Algorithm~\ref{alg:main-sparsify}, and the main result of the section is 

\begin{theorem}
\label{thm:main_thm}
There exists an algorithm (Algorithm~\ref{alg:main-sparsify}) such that for any $\epsilon > 0$, processes a list of edge insertions and deletions for an unweighted graph $G$ in a single pass and maintains a set of linear sketches of this input in $\wt{O}\left(\e^{-2}n^{1.4+o(1)}\right)$
space. From these sketches, it is possible to recover, with high probability, a weighted subgraph $H$ with $O(\epsilon^{-2}n\log n)$ edges, such that $H$ is a $(1 \pm \epsilon)$-spectral sparsifier of $G$. The algorithm recovers $H$ in $\wt{O}\left(\e^{-2}n^{1.4+o(1)}\right)$ time.
\end{theorem}

The \textsc{Sparsify} (Algorithm~\ref{alg:main-sparsify})  generally follows the approach of~\cite{kapralov2017single}, and the main technical contribution of this section is the \textsc{HeavyEdges} algorithm (Algorithm~\ref{alg:main-heavy-edge}). We now outline the main ideas involved in both algorithms and analysis.

The \textsc{Sparsify} algorithm have recursive structure: given (a sketch of) a graph $G$ to be sparsified, the algorithm proceeds as follows. First it adds a regularization term to the graph (essentially a multiple of the complete graph) and recursively calls itself, obtaining a coarse (large factor approximation) sparsifier $\wt{G}$ of $G$, whose Laplacian (together with the regularization term) we denote by $\wt{K}$ (see line~\ref{alg:line-coarse} of Algorithm~\ref{alg:main-sparsify}). One then invokes the \textsc{HeavyEdges} procedure (see line~\ref{line:RS-callHE} of Algorithm~\ref{alg:main-sparsify}) to recover all edges of large (about $1/\Gamma$, where $\Gamma>1$ is the quality of approximation of $G$ by $\wt{G}$) effective resistance in the sample. One then uses $\wt{G}$ to estimate the effective resistance of every edge recovered by the invocation of \textsc{HeavyEdges}, and keeps those edges that were recovered from the sample corresponding to their actual approximated effective resistance (see line~\ref{line:set-p'e} of Algorithm~\ref{alg:main-sparsify}). This process is similar to the approach of~\cite{kapralov2017single}, with one nontrivial difference: our \textsc{HeavyEdges} procedure needs a spectral approximation to the sampled graph in order to recover high effective resistance (or, heavy) edges -- indeed, this spectral approximation is needed in order to perform ball carving in the effective resistance metric. As a consequence, the overall procedure has somewhat nontrivial recursive structure that we describe next.

\paragraph{Recursive structure of \textsc{Sparsify} and \textsc{HeavyEdges} (recursion tree $\mathcal T$).} We represent the recursive structure of \textsc{Sparsify} and \textsc{HeavyEdges} by a recursion tree $\mathcal T$ that is described in detail in Section~\ref{sec:index}. Every node $a\in \mathcal T$ of the recursion tree corresponds to subsampling $G_a$ of the input graph $G$ that is formally defined in Section~\ref{sec:index}. When an invocation of \textsc{HeavyEdges} or \textsc{Sparsify} performs a recursive call to itself or the counterpart subroutine, it must pass a collection of sketches of a subsampling of its graph $G_a$ as input to the recursive call. We associate the appropriate set of sketches with the corresponding nodes in the tree $\mathcal T$, so that invocations of our subroutines simply get a node $a\in \mathcal T$ as their first input, and the sketches $S_{\leq a} B$ that correspond to the subtree of $a$. See Section~\ref{sec:index} for more details.

\paragraph{Recursive chain of sparsifiers.} Our recursion tree is a natural generalization of the chain of coarse sparsifiers used in~\cite{kapralov2017single}. In particular, 
	 in recursion tree $\TT$, for any node $a\in U$, that corresponds to a call to $\textsc{Sparsify}$, and node $b\in U$, which corresponds to a call to $\textsc{Sparsify}$, and $b$ is a child of $a$,one has
$$K_a \preceq_r K_b \preceq_r \Gamma K_a$$
where $K_a$ and $K_b$ correspond to Laplacian matrix of $G_a$ and $G_b$. If node $a$ does not have a child that corresponds to a call to $\textsc{Sparsify}$, then
$$K_a \preceq_r \Gamma\cdot \lambda_u \cdot I \preceq_r \Gamma K_a.$$
See Remark~\ref{remark-chain} in Section~\ref{sec:index} for the details.

\paragraph{Novel algorithmic techniques: the \textsc{HeavyEdges} primitive (Algorithm~\ref{alg:main-heavy-edge}).} The main algorithmic novelty presented in this section is our \textsc{HeavyEdges} algorithm. An invocation 
$$
\textsc{HeavyEdges}(a,S_{\le a}B,i, \ell,\e, \beta)
$$
 receives a sketch of a graph $G$ corresponding to node $a$ of the recursion tree $\mathcal T$, with $\textsf{label}(a)=(\textsc{HeavyEdges},i,\ell)$\footnote{Parameter $i$ means we are sampling the graph at rate $\frac{1}{\Gamma^i}$ and $\ell$ stands for regularization factor, which is a complete graph with edge weights $\frac{\lambda_u}{n\cdot \Gamma^\ell}=\frac{2}{\Gamma^\ell}$ added to the graph, see Section~\ref{sec:index} for detailed explanation of this labeling scheme.}, as input and outputs a list of edges of $G$ that is guaranteed (with high probability) to contain all edges of effective resistance higher than $\beta$. Also, Facts \ref{fact:ind-T} and \ref{fact:T-num-nodes} ensures that the recursion terminates. The algorithm starts with preprocessing: we peel off low degree vertices (degree smaller than $d=n^{0.4}\log^2 n$, see lines~\ref{ln:set-dd} and~\ref{line:first-lowdeg} of Algorithm~\ref{alg:main-heavy-edge}) and 
 subtract these edges from all sketches in the corresponding subtree $\TT_{\leq a}$ and add them to the output set of edges. We refer to the result of removing these edges from $G_a$ as $G_a'$ (see Algorithm~\ref{alg:main-heavy-edge}). Finally, before running \textsc{BallCarving} we extract a set $E^*$ that contains all edges of connectivity below $d$ with high probability (line~\ref{line:FLCE-FHE}), again adding them to the output set of edges. Next the the \textsc{BallCarving} primitive is invoked to obtain a collection of subsets $P_1\cup \ldots \cup P_t$ of low effective resistance diameter in $G$. 
 
 The main loop of \textsc{HeavyEdges}  performs the following experiment for every $i=1,\ldots, t$, and every $v\in V\setminus P_i$ send a unit of flow from the supernode corresponding to $P_i$ in the contracted version $\wt{K}/P_i$ of $\wt{K}$ ($\wt{K}$ is the coarse sparsifier obtain from the recursive call) to $v$. The corresponding heavy hitters sketch is then decoded and the result added to the output. As we show below, the number of elements in the collection $P_1\cup \ldots \cup P_t$ is at most $\wt{O}(n^{0.4})$ for appropriate setting of parameters, leading to the claimed runtime and space complexity bounds for \textsc{HeavyEdges}.

In the rest of the section we state the algorithms and present the formal analysis. In Section~\ref{subsec:sparsify} we analyze \textsc{Sparsify}(Algorithm~\ref{alg:main-sparsify}), and in Section~\ref{sec:HEanalysis} we analyze \textsc{HeavyEdges} (Algorithm~\ref{alg:main-heavy-edge}). The proof of Theorem~\ref{thm:main_thm} follows rather directly assuming these lemmas:

\begin{proofof}{Theorem~\ref{thm:main_thm}}
\paragraph{Correctness analysis.}We prove the correctness by induction. 
The {\bf base} of induction follows immediately: one can verify that every leaf in the recursion tree corresponds to a call to \textsc{HeavyEdges}. By invoking Lemma \ref{lem:corr-main-HE}, on any of these leaves, we prove that any leaf in the recursion tree $\TT$, succeeds.  We now give the {\bf inductive step: $\TT_{<a} \rightarrow a$.} If node $a$ corresponds to a call to $\textsc{Sparsify}$, by invoking Lemma \ref{lem:corr-main-SP}, and if node $a$ corresponds to a call to $\textsc{HeavyEdges}$, by invoking Lemma \ref{lem:corr-main-HE}, with high probability, the algorithm corresponding to node $a$, succeeds. 
	Therefore, an invocation of $\textsc{Sparsify}(r, S_{\le r}B, 0, \Lambda+1, \e)$ for node $r$ that is the root of $\TT$ succeeds with high probability. Now that we have a graph with $O(\e^{-2}n^{1+o(1)}\log n)$ edges, we can easily run any almost linear time and space, sparsifier algorithm (e.g. \cite{spielman2011graph}) on it, to get a graph $H$ with  $O(\e^{-2}n \log n)$ edges.
	\paragraph{Run-time and space analysis}
	We set $\Gamma=n^\delta$, and $\delta=\frac{1}{\log \log n}$. Note that we have at most $O\left(\frac{1}{\delta}^{\frac{1}{\delta}}\right)=n^{o(1)}$ nodes in the recursion tree (see section \ref{sec:index}), and by Lemma \ref{lem:corr-main-SP} and \ref{lem:corr-main-HE}, each node in the recursion tree runs in $\wt{O}(n^{1.4+7\delta}\e^{-2})$ time and space in addition to the recursive calls. Therefore, in total our algorithm runs in $\wt{O}\left(\frac{1}{\delta}^{\frac{1}{\delta}} n^{1.4+7\delta}\e^{-2} \right)=\wt{O}(\e^{-2}n^{1.4+o(1)})$ time and space. 
\end{proofof}
\begin{remark}
	One should note that since error probability of algorithms called during the procedure is inverse polynomial in $n$, by union bound our algorithm has error probability inverse polynomial in $n$.
\end{remark}

\subsection{Algorithm \textsc{Sparsify} and its analysis}
\label{subsec:sparsify}

\begin{algorithm}

	\caption{\textsc{Sparsify}: outputs a $(1\pm\e)$-spectral sparsifier of $G_a$}\label{alg:main-sparsify}
	\begin{algorithmic}[1]
	\Procedure{Sparsify($a,S_{\le a}B, i, \ell, \e$)}{}	
			\State $\Lambda \gets \ceil{\log_{\Gamma} \frac{\lambda_u}{\lambda_\ell}}$ \label{line:theta-delta} \Comment{$\Lambda=\ceil{\log_{n^\delta} \frac{2n}{8/n^2}}= \Theta(\frac{1}{\delta})$}
		 	\If {$\ell=0$}
		 	\State $\wt{K} \gets \lambda_u I$ \label{line:main-SP-I}
		 	\Else 
		 	\State $b \gets$ The child of node $a$ in recursion tree, such that $\textsf{label}(b)=(\textsc{Sparsify},i,\ell-1)$
		 	\State $\wt{K} \gets\frac{1}{\Gamma(1+\e)}\textsc{Sparsify}(b,S_{\le b}B,i, \ell-1,\epsilon)$ 	\label{alg:line-coarse}
		 	\EndIf 
		 	\If{$\ell-i=\Lambda+1$} $\gamma=0$\Comment  We start to refine the coarse sparsifier $\wt{K}$ from this line \label{line:set-gamma}
		 	\Else	 \text{ }$\gamma=\lambda_u/\Gamma^\ell$\label{line:set-gamma2} \Comment{$\Gamma$ is the base of a geometric sequence of sampling rates}\label{line:RS-for}
		 	\EndIf
		 	\State $q\gets 400 \log n$ \Comment It suffices to get a $(1\pm \frac{1}{2})$ approximation from JL. 
		 	\State $Q\gets q\times {n \choose 2}$ matrix of i.i.d. $\pm 1$s \label{ln:jl-mat}
		 	\State Compute $M\gets \frac1{\sqrt{q}}Q\wt B\wt K^+$\Comment{$\frac{1}{2} R_e^{\tilde{G}} \le ||M\mathbf{b}_e||_2^2 \le \frac{3}{2} R_e^{\tilde{G}}$}
		 	\For{$j=i$ {\bf to}~$\Lambda$} 
		 			 	\State $b \gets$ The child of node $a$ in recursion tree such that $\textsf{label}(b)=(\textsc{HeavyEdges},j,\ell+j-i)$
		 			 	\State $E_j\gets\textsc{HeavyEdges}\left(b, S_{\le b}B, j, \ell+j-i, \e, \frac{1}{500c'\e^{-2} \cdot\Gamma^3 \log n}\right)$\label{line:RS-callHE}\Comment See Algorithm \ref{alg:main-heavy-edge}.
		 	\For{$e=(u,v) \in E_j$}
		 	
		 	\State $R'_e\gets2||M\mathbf{b}_e||_2^2$
			\State $p'_e\gets\min \{1,c'R'_e \log n \epsilon^{-2} \}$ \Comment{$c'$ is the oversampling constant from Lemma \ref{lem:classic_result}}\label{line:set-p'e}
		 	\If{$j=\Lambda$}

		 	\If{$p'_e \le \Gamma^{i-j}$} $W(e,e)\gets \Gamma^{j-i}$ 
		 	\EndIf
		 	
		 	\Else
		 	\If{$p'_e \in (\Gamma^{i-j-1}, \Gamma^{i-j}]$} $W(e,e)\gets \Gamma^{j-i}$ 
		 	\EndIf
		 	\EndIf
		 	\EndFor
		 	\EndFor
		 	\State \Return $\wt{K}_\e= {B_n}^\T W B_n + \gamma I$\label{ln:output} 
		\EndProcedure
	\end{algorithmic}
\end{algorithm}

	\begin{lemma}\label{lem:corr-main-SP}
	Suppose that $\TT=(U,E_\TT)$ is the recursion tree of algorithm's execution (see section \ref{sec:index}). Suppose that $a\in U$ is such that node  $\textsf{label}(a)=(\textsc{Sparsify},i,\ell)$. Let $\gamma=\frac{\lambda_u}{\Gamma^\ell}$. Let $G_a$ denote the graph corresponding to node $a$ in the recursion tree (see Definition \ref{def:G_a}). Let $B'_a \in \R^{\left({{n}\choose{2}}+n\right)\times n}$, and $K_a \in \R^{n\times n}$ denote the vertex edge incidence matrix and Laplacian matrix of graph $G_a$ respectively. Assume that all the recursive calls in the subtree of node $a$ i.e., $\TT_{<a}$ succeeds. Then an invocation of \textsc{Sparsify}($a,S_{\le a}B, i, \ell, \e)$ with high probability returns $\wt{K}_\e=\wt{B}_\e^\top\wt{B}_\e+\gamma I$, where $\wt{B}_\e$ contains only $O(\epsilon^{-2} n^{1+o(1)} \log n)$ reweighted rows of $B'_a$, and $$(1+\e)K_a \preceq_r \wt{K}_\e \preceq_r (1+\e)K_a \text{.}$$ Moreover, an invocation of \textsc{Sparsify}($a,S_{\le a}B, i, \ell, \e)$, requires $\wt{O}(n^{1.4+7\delta}\e^{-2})$ time and space in addition to the space and time of the recursive calls.
\end{lemma}

\begin{proof}
	Suppose that in tree $\TT$, $a\in U$ is such $\textsf{label}(a)=(\textsc{Sparsify},i,\ell)$. 
	We consider two cases.  
	\paragraph{Case 1: $\ell=0$}
	\hfill \break
	In this case we set $\wt{K}=\lambda_u I$ (see line \ref{line:main-SP-I} of Algorithm \ref{alg:main-sparsify}), and node $a$ does not have a child. By Remark~\ref{remark-chain}, we have  $$\frac{1}{\Gamma}\cdot
	K_a \preceq \wt{K}  \preceq K_a $$ 	
	Let $C=\Gamma$ in this case.
	\paragraph{Case 2: $\ell>0$}
	\hfill \break
	As per line~\ref{alg:line-coarse} of Algorithm~\ref{alg:main-sparsify} we set $\wt{K}=\frac{1}{\Gamma(1+\e)}\textsc{Sparsify}(b,S_{\le b}B, i, \ell-1, \e)$, and suppose that the corresponding node for this call in the recursion tree $\mathcal{T}$ is $b$. By the assumption that the call corresponding to $b$ succeeded, we have 

	\begin{equation}
	\label{eq:mn_ep-ap1}
	(1-\e) \cdot K_b \preceq_r \Gamma(1+\e)\wt{K} \preceq_r (1+\e) \cdot  K_b \text{.}
	\end{equation}
	Moreover, by Remark~\ref{remark-chain}, we have 
	\begin{equation}
	\label{eq:mn_Gamma-ap1}
	\frac{1}{\Gamma} \cdot K_a \preceq \frac{1}{\Gamma} \cdot K_b \preceq K_a \text{.}
	\end{equation}
	Putting \eqref{eq:mn_ep-ap1} and \eqref{eq:mn_Gamma-ap1} together we get
	\begin{equation}
	\frac{1-\e}{\Gamma(1+\e)} \cdot K_a \preceq_r  \wt{K} \preceq_r  K_a \text{.}
	\end{equation}
	Let $C=\frac{\Gamma(1+\e)}{1-\e}$ in this case. Therefore in both cases we have
	\begin{align}
	\frac{1}{C} \cdot K_a\preceq_r \wt{K} \preceq_r  K_a \text{.}\label{eq:adfbua1}
	\end{align} 
	To complete the proof we show that, given a coarse sparsifier $\wt{K}$ where $\frac{1}{C} \cdot K_a\preceq_r \wt{K} \preceq_r  K_a$, we will construct $\wt{K}_\e$ from line \ref{line:set-gamma} to \ref{ln:output} of Algorithm~\ref{alg:main-sparsify} such that
	$$(1-\e) K_a \preceq_r \wt{K}_\e \preceq_r (1+\e) K_a\text{.}$$
	Recall that node $a$, has at most $\Lambda+2$ children, where one of them corresponds to a call to $\textsc{Sparsify}$, and the rest correspond to $\textsc{HeavyEdges}$ calls (see section \ref{sec:index} for more details about the recursion tree). 

	Let $G_a$ denote the graph corresponding to node $a$ in the recursion tree (see Definition \ref{def:G_a}). Let $B'_a \in \R^{\left({{n}\choose{2}}+n\right)\times n}$, and $L_a \in \R^{n\times n}$ denote the vertex edge incidence matrix and Laplacian matrix of graph $G_a$ respectively. Also let $B_a$ be the matrix of first $\binom{n}{2}$ rows of $B'_a$. Thus we have
	\[
	B'_a=B_a\oplus \sqrt{\gamma}I.
	\]
	and, 
	\[L_a=(B'_a)^\top(B'_a)={B_a}^\top B_a+ \gamma I \text{.}\]
	For ease of notation, let $K=K_a$. 
	Observe that by \eqref{eq:adfbua1}, we have $\frac{1}{C} K \preceq_r \wt{K} \preceq_r K$.  Hence for any $y \in [{{n}\choose{2}}+n]$, 
	\begin{equation} \label{inq:r_rtild}
	\mathbf{b}_y^\T K^{+}\mathbf{b}_y \preceq_r \mathbf{b}_y^\T \wt{K}^{+}\mathbf{b}_y \preceq_r C \mathbf{b}_y^\T K^{+}\mathbf{b}_y \text{.}
	\end{equation}
	Let $\tau$ be a vector of leverage score for $B'_a$'s rows. Hence, for any $y \in [{{n}\choose{2}}+n]$ we have $\tau_y =  \mathbf{b}_y^\top K^{+} \mathbf{b}_y$. We define $\wt{\tau}_y =  \mathbf{b}_y^\top \wt{K}^{+} \mathbf{b}_y$. Thus by inequality \eqref{inq:r_rtild} for any $y\in [{{n}\choose{2}}+n]$ we have 
	\begin{equation} \label{eq:tau_bound1}
	\tau_y  \leq  \wt{\tau}_y\leq C \tau_y \text{.}
	\end{equation}
	Let $p_y(G_a)=\min\{1, c'\e^{-2}\log n \cdot \tau_y \}$.
	To complete the proof we need to show that for any $y\in[{{n}\choose{2}}+n]$, the algorithm samples the rows of $B_a$ independently and with probability at least $p_y(G_a)$, hence we can apply Lemma \ref{lem:classic_result} afterwards.
	To see this observe that Algorithm \ref{alg:main-sparsify} returns $\wt{K}_\e={B_a}^\top W B_a+ \gamma I$, where we have
	\[{B_a}^\top W B_a+ \gamma I = (W^{\frac{1}{2}} B_a \oplus \sqrt{\gamma}I)^\top (W^{\frac{1}{2}} B_a \oplus \sqrt{\gamma}I) \text{.}\]
	Thus, for any ${{n}\choose{2}}+1 \leq y \leq {{n}\choose{2}}+n$, row $\mathbf{b}_y$ is sampled with probability $1$. Therefore, $\wt{K}_\e$ includes rows corresponding to $\sqrt{\gamma}I$ with probability $1$. 
	
	Hence, it's sufficient to prove that for any $1\leq e \leq {{n}\choose{2}}$, row $\mathbf{b}_e$ is included in $W^{\frac{1}{2}} B_a$ independently with probability at least $p_e(G_a)$ with the proper weight. We consider two cases.
	\paragraph{Case 1: $p_e(G_a) \leq \Gamma^{i-\ceil{\log_{\Gamma} \frac{\lambda_u}{\lambda_\ell}}}$.} 
	In this case for $j=\ceil{\log_{\Gamma} \frac{\lambda_u}{\lambda_\ell}}$ we have $p_e(G_a) \leq \Gamma^{i-j}$. Suppose that node $b$ is the child of $a$, and $\textsf{label}(b)=(\textsc{HeavyEdges},j=\ceil{\log_{\Gamma} \frac{\lambda_u}{\lambda_\ell}},\ell)$, then for any edge $e \in E_{a}$, one has $e \in E_{G_b}$ with probability $\frac{1}{\Gamma^{j-i}}$. On the other hand, by Chernoff bound we can show that degree of each vertex in $G_b$ is less than $d$. So, all such edges will be returned with high probability (see lines \ref{line:first-lowdeg} - \ref{line:last-lowdeg} in Algorithm \ref{alg:main-heavy-edge}). 
	\paragraph{Case 2: $p_e(G_a) \geq \Gamma^{i-\ceil{\log_{\Gamma} \frac{\lambda_u}{\lambda_\ell}}}$.} 
	In this case there exists a $j \in \{i, \ldots,\ceil{\log_\Gamma \frac{\lambda_u}{\lambda_\ell}}-1 \}$, such that $p_e(G_a) \in [\Gamma^{i-j-1},\Gamma^{i-j}]$. For edge $e\in E_{a}$ suppose that $p_e(G_a) \in [\Gamma^{i-j-1},\Gamma^{i-j}]$. Let $b_0,b_1,b_2 \in U$ be three children of $a$, and we have,
	\begin{align*}
	\textsf{label}(b_0)&=(\textsc{HeavyEdges},j,\ell+j-i)\\
	\textsf{label}(b_1)&=(\textsc{HeavyEdges},j-1,\ell+j-i-1)\\
	\textsf{label}(b_2)&=(\textsc{HeavyEdges},j-2,\ell+j-i-2)
	\end{align*}

	Thus, by Lemma \ref{lem:main_heavy_sample} we have with high probability 
	\begin{align*}
	R_e^{G_{b_0}} &\geq \frac{1}{500c' \e^{-2}\cdot\Gamma \log n }\\ 
	R_e^{G_{b_1}} &\geq \frac{1}{500c' \e^{-2}\cdot\Gamma^2 \log n }\\
	R_e^{G_{b_2}} &\geq \frac{1}{500c' \e^{-2}\cdot\Gamma^3 \log n }\text{.}
	\end{align*}
	
	Thus, by the assumption since $b_0,b_1$ and $b_2$ have succeeded then $E_j$, $E_{j-1}$ and $E_{j-2}$ contains edge $e$ with high probability. \Navid{For the corner cases, I think it suffice to say if $b_0$ and $b_1$ and $b_2$ exist.... Nothing major I think.}
	
	Observe that in line \ref{ln:jl-mat} and \ref{line:set-p'e} Algorithm \ref{alg:main-sparsify} we set $R'_e=2||M\mathbf{b}_e||_2^2$, hence $p'_e=\min \{1,c'R'_e \log n \epsilon^{-2} \}$. Thus by Johnson-Lindenstrauss Lemma  with high probability we have 
	\[\wt{\tau}_e\le R'_e=2||M\mathbf{b}_e||_2^2 \le 3 \mathbf{b}_e^\T \wt{K}^{+} \mathbf{b}_e = 3\wt{\tau}_e \text{.}\] 
	Moreover by \eqref{eq:tau_bound1} we have $\tau_e  \leq  \wt{\tau}_e\leq C \tau_e$, thus we get
	\[\tau_e  \leq  R'_e\leq 3C \tau_e \text{.}\]
	Hence, $p_e \leq p'_e \leq 3C  p_e$, which implies $p'_e \in [\Gamma^{i-j-1},3C\cdot\Gamma^{i-j}]$. Therefore, since $3C\leq \Gamma^2$, for exactly one $k$ in $\{j-2,j-1,j\}$ we have $p'_e \in [\Gamma^{i-k-1},\Gamma^{i-k}]$.
	Recall that $(SB)$ denote the sketches of $B_i$ where the rows are sampled independently at rate $\Gamma^{i-k}$.  Therefore, edge $e$ is included to $\wt{K}_\e$ with probability $$\Gamma^{i-k}\geq \Gamma^{i-j}\geq p_e(G_a)$$
	and with weight $\Gamma^{k-i}$. Therefore we by Lemma \ref{lem:classic_result} we have $\wt{K}_\e \ape K$. Moreover $\wt{B}_\e=W^{\frac{1}{2}} B_a$ contains at most $( \sum_{e\in E_a} R'_e \log n \e^{-2})$ non-zeros with high probability. Note that \[\sum_{e\in E_a} R'_e \leq  3C \sum_{e\in E_a} \tau_e \leq 3Cn \text{.}\] Hence, overall $\wt{B}_\e$ contains $O(C\e^{-2} n \log n)=O(\e^{-2}n^{1+o(1)}\log n)$ non zeros with high probability. \Navid{I should also add the analysis of $j=\Lambda$.}
	\paragraph{Run-time and space analysis:} 
	 For any $E_j$ in line \ref{line:RS-callHE} in Algorithm \ref{alg:main-sparsify}, since the corresponding node $b$ in the recursion tree $\TT$ succeeded by the assumption, $|E_j|=\wt{O}(n^{1.4+7\delta})$. So, the run-time and space claim holds.  
	 
	 \paragraph{Maintenance of sketches:} Note that Algorithm \ref{alg:main-sparsify} takes sketch $S\cdot B$ as an input where it corresponds to the different sketches that are used in different subroutines. More precisely, $S$ is a randomly constructed matrix with ${n \choose 2}$ columns that corresponds to the concatenation of the following matrices: The sampling matrix i.e., $\Pi \in \mathbb{R}^{{n \choose 2}\times {n \choose 2}}$ (Section~\ref{sec:index}), the sketch to find the edges with connectivity at most $\lambda$, i.e., $S^f\Sigma \in\mathbb{R} ^{\lambda\cdot\poly(\log n)\times {n \choose 2}}$ (Section~\ref{subsec:find_low}), the \textsc{SparseRecovery} sketch to recover $k$-sparse vectors, i.e., $S^r\in\mathbb{R} ^{k\cdot\poly(\log n)\times {n \choose 2}}$ (Section~\ref{subsec:sp-rec}), and the \textsc{HeavyHitter} sketch to find the edges that are heavy with respect to parameter $\eta$, i.e., $S^h\in\mathbb{R} ^{\eta^{-2}\cdot\poly(\log n)\times {n \choose 2}}$ (Section~\ref{subsec:hv-hit}). 
	
As per line~\ref{line:first-lowdeg} of Algorithm~\ref{alg:main-heavy-edge} we set $k=d=\wt{O}(n^{0.4})$, in line~\ref{line:FLCE-FHE} of Algorithm~\ref{alg:main-heavy-edge} we set $\lambda=d=\wt{O}(n^{0.4})$, and in line~\ref{line:HH-FHE} we set $\eta=\Theta\left(\sqrt{\frac{\beta}{C}}\right)=\frac{1}{\poly( \Gamma\cdot \log n )}$. Recall that $\Gamma=n^\delta$. Therefore, overall the number of random bits needed for all the matrices, in an invocation of Algorithm~\ref{alg:main-sparsify} is at most $R=\wt{O}(n^2+nk+n\lambda+n\cdot\eta^{-2})=\wt{O}(n^2)$, in addtion to the random bits needed for the recursive calls.

To generate matrix $\Pi$, $S^f\Sigma$, and $S^h$ we use the fast pseudorandom numbers generator that is introduced in Section \ref{sec:prg}. Observe that the space used by Algorithm~\ref{alg:main-sparsify} is $s=\wt{O}(n^{1.4+7\delta})$ in in addition to the space used by the recursive calls. Since $R=O(n^2)$, we have $R=O(s^2)$. Therefore, by Theorem~\ref{high_level_result} we can generate seed of $O(s\cdot \poly(\log s))$ random bits in $O(s\cdot \poly(\log s))$ time that can simulate our randomized algorithm.
	 
To generate the \textsc{SparseRecovery} sketch to recover $k$-sparse vectors, i.e., $S^r\in\mathbb{R} ^{k\cdot\poly(\log n)\times {n \choose 2}}$ (Section~\ref{subsec:sp-rec}), we can not use our fast pseudorandom numbers generator, since these bits need to be accessed again during the decoding time (see line \ref{ln:update-sketch} of Algorithm~\ref{alg:main-heavy-edge}). However, Algorithm \textsc{SparseRecovery} uses low-independence hash functions for those bits instead.
Moreover, the random matrix $Q \in\mathbb{R} ^{\Theta(\log n)\times {n \choose 2}}$ for JL (line~\ref{ln:jl-mat} of Algorithm \ref{alg:main-sparsify}) can be generated using $\log n$-wise independent hash functions.
\end{proof}

\subsection{Algorithm \textsc{HeavyEdges} and its analysis}\label{sec:HEanalysis}

We first state the algorithms (see Algorithm~\ref{alg:main-heavy-edge} below).

\begin{algorithm}
	
	\caption{\textsc{HeavyEdges}($a,S_{\le a}B,i, \ell,\e, \beta$)}  
	\label{alg:main-heavy-edge} 
	\begin{algorithmic}[1]
		
		\Procedure{HeavyEdges($a,S_{\le a}B,i, \ell,\e, \beta$)}{}\Comment{node $a$ is a node in $\TT$, $G_a$ is the corresponding graph, and $S_{\le a}B$ corresponds to the sketches of nodes in subtree $\TT_{\le a}$. See section \ref{sec:index} for details.} 
		\State $E'=\emptyset$	
		\State $d\gets n^{0.4}\log^2 n$	\label{ln:set-dd}
		\While {there exists a vertex $v$ with  $\deg(v) < d$}\label{line:first-lowdeg}
		\State $E_{v} \gets$ \textsc{SparseRecovery}($S_aB,d,v$)	\Comment{See Section \ref{subsec:sp-rec}}
		\State Update sketches $S_{\le a}B$, by removing edges in $E_v$ \label{ln:update-sketch}
		\State Update degrees. 
		\State $E' \gets E' \cup E_{v}$\label{line:last-lowdeg}
		\EndWhile
		\If{$\ell= 0$} \Comment{Let $G_a'$ be the graph after removing low degree vertices in lines \ref{line:first-lowdeg} - \ref{line:last-lowdeg}}
		\State $\wt{K}\gets \lambda_u I$\label{line:leaf-case-K}
		\State $C\gets \Gamma$ \label{line:leaf-case-C}
		\Else
		\State $b\gets \text{The child of node } a \text{ in tree } \TT$
		\State $\wt{K} \gets \frac{1}{\Gamma(1+\e)}\textsc{Sparsify}(b,S_{\le b}B,i, \ell-1,\epsilon)$ \label{line:HE-CallSp} \Comment{See Algorithm \ref{alg:main-sparsify}}
		\State $C\gets \frac{\Gamma(1+\e)}{1-\e}$\label{line:HE-CallSp-C}
		\EndIf		
		\State $E^* \gets  \textsc{FindLowConnectivityEdges}(S_{a}B, d)$\Comment{See Algorithm \ref{spanning-forest-ALG}, Section~\ref{subsec:find_low}}\label{line:FLCE-FHE}
		\State $E' \gets E' \cup E^*$ \label{line:FLCE-add}
		\State $(P_1,P_2,\dots,P_t) \gets \text{\textsc{BallCarving}}(S_{a} B,E^*,\beta/6,\wt{K})$ \Comment{See Algorithm \ref{alg:ballcarving}}
		\For{$i=1$ {\bf to}~$ t$} \label{line:BPstart}
		\State $Y_{P_i} \gets [\mathbf{1}_{P_i},(\mathbf{1}_v)_{v\in V_G \setminus P_i}]$\label{line:Y+p} \label{line:contraction}
		\State $\wt{K}_{P_i}^+ = (Y_{P_i}^\top \wt{K} Y_{P_i})^+$ \label{line:K+p}
		\State $\hat{u} \gets$ The super-node of $P_i$  
		\For{$v\in V_G\setminus P_i$}
		\State $E'\gets E' \cup \text{\textsc{HeavyHitter}}(S_{a}B Y \wt{K}_{P_i}^+(\chi_{\hat{u}}-\chi_{v}), \frac{1}{2}\sqrt{\frac{\beta}{3C}})$ \label{line:HH-FHE}\Comment{See Lemma \ref{lem:corr-main-HE}}
		\EndFor
		\EndFor
		
		\State \Return $E'$
		\EndProcedure
	\end{algorithmic}
\end{algorithm}

\begin{lemma}\label{lem:corr-main-HE}
	Suppose that $\TT=(U,E_\TT)$ is the recursion tree of algorithm's execution (see section \ref{sec:index}). Suppose that node $a\in U$ is such that $\textsf{label}(a)=(\textsc{HeavyEdges},i,\ell)$, and let $G_a$ be an unweighted graph corresponding to node $a$. Assume that all the recursive calls (if there is any) in  the subtree of node $a$ i.e., $\TT_{<a}$ succeed. Then the algorithm corresponding to the node $a$ succeeds, so that \textsc{HeavyEdges}$(a, S_{<a}B, i , \ell , \epsilon, \beta)$ returns a set of edges $E'$, that contains every edge $e$ with $R_{e}^{G_a}\ge\beta$ with high probability. Moreover, an invocation of \textsc{HeavyEdges}$(a, S_{<a}B, i , \ell , \epsilon, \beta)$ requires $\wt{O}(n^{1.4+7\delta}\e^{-2})=\wt{O}(n^{1.4+o(1)}\e^{-2})$ time and space in addition to the space and time of the recursive calls.
\end{lemma}
\begin{proof}
In line \ref{ln:set-dd}, we set $d=n^{0.4}\log^2 n$. Observe that Algorithm \ref{alg:main-heavy-edge} first recover all the edges connected to the low degree vertices, i.e., vertex with degree at most $d$, iteratively, and removes them until no such vertex remains (see lines~\ref{line:first-lowdeg} - \ref{line:last-lowdeg}). The recovering process is done using \textsc{SparseRecovery} algorithm where its correctness is guaranteed by Lemma \ref{lem:sparse-recovery}. Suppose that vertex $v$ with degree at most $d$, is the first vertex going to be removed in the first iteration of the loop. Then by Lemma~\ref{lem:sparse-recovery}, Algorithm \textsc{SparseRecovery} recovers all the neighbors of the vertex $v$ exactly with probability at least $1-n^{-10}$. Therefore, with probability at least $1-n^{-10}$ this is a deterministic process, so that we can use the same sketches to recover the neighbors of vertices in next iterations. Thus, by union bound over all iterations, with probability at least $1-n\cdot n^{-10}$, all the edges connected to the low degree vertices are recovered. Therefore, if edge $e$ is connected to one of the low degree vertices, it is recovered with probability at least $1-n^{-9}$.  Now in the resulting graph every vertex has degree at least $d=n^{0.4}\log^2 n$. Note that we can maintain degrees easily using linear sketches.
	
	For an edge $e$ such that $R_e^{G_a}\ge\beta$, Suppose that it has not been added to $E'$ in the procedure of removing low degree vertices (See lines \ref{line:first-lowdeg} - \ref{line:last-lowdeg} of Algorithm \ref{alg:main-heavy-edge}). By the discussion above, it is guaranteed that from line \ref{line:last-lowdeg} on, there is no edge connected to a vertex with degree less than $d$. We call this new graph $G'_a$. In this case, if an edge $e$ is such that $R_e^{G_a}\ge \beta$, after removing the low degree vertices, by Fact \ref{fact:mon}, one has $R_e^{G'_a}\ge \beta$. In what follows, we show that eventually edge $e$ will be added to $E'$.

	Recall that for any edge $e$ such that $R_e^{G'_a}\ge \beta$, both end points of this edge cannot be inside a partition with effective resistance diameter less than $\beta$, which informally means that in the contraction part of the algorithm we never swallow an edge with effective resistance at least $\beta$. So, we argue that for an edge $e$, such that $R_e^{G'_a}\ge \beta$, if it has not been added to $E'$ in line \ref{line:FLCE-add} of Algorithm \ref{alg:main-heavy-edge}, it will be added by line \ref{line:HH-FHE} of Algorithm \ref{alg:main-heavy-edge}.

	Suppose that $e=(u,v)$ and $u$ (w.l.o.g) belongs to a partition, say $P^*$. In that case, by the argument above, we know that $v \notin P^*$. Now it remains to show that with high probability this edge will appear in line \ref{line:HH-FHE} of Algorithm \ref{alg:main-heavy-edge}. 
	
	We consider two cases.
	
		\paragraph{Case 1: $\ell=0$}
	\hfill \break
	In this case we set $\wt{K}=\lambda_u I$ (see line \ref{line:leaf-case-K} of Algorithm \ref{alg:main-heavy-edge}), and node $a$ is a leaf in tree $\TT$. By Remark~\ref{remark-chain}, we have  $$\frac{1}{\Gamma}\cdot
	K_a \preceq \wt{K}  \preceq K_a $$ 	
	\paragraph{Case 2: $\ell>0$}
	\hfill \break
	As per line~\ref{line:HE-CallSp} of Algorithm~\ref{alg:main-heavy-edge} we set $\wt{K}=\frac{1}{\Gamma(1+\e)}\textsc{Sparsify}(b, S_{\le b}B, i, \ell-1, \e)$,  and suppose that the corresponding node for this call in the recursion tree $\mathcal{T}$ is $b$. By the assumption, we have 
	\begin{equation}
	\label{eq:mn_ep-ap}
	(1-\e) \cdot K_b \preceq_r \Gamma(1+\e)\wt{K} \preceq_r (1+\e) \cdot  K_b\text{.}
	\end{equation}
	Moreover, by Remark~\ref{remark-chain}, we have 
	\begin{equation}
	\label{eq:mn_Gamma-ap}
	\frac{1}{\Gamma} \cdot K_a\preceq \frac{1}{\Gamma} \cdot K_b \preceq K_a \text{.}
	\end{equation}
	Putting \eqref{eq:mn_ep-ap} and \eqref{eq:mn_Gamma-ap} together we get
	\begin{equation}
	\frac{1-\e}{\Gamma(1+\e)} \cdot K_a\preceq_r  \wt{K} \preceq_r  K_a \text{.}
	\end{equation}
	Note that by lines \ref{line:leaf-case-C} and \ref{line:HE-CallSp-C}, we have
	\begin{align}
	\frac{1}{C} \cdot K_a\preceq_r \wt{K} \preceq_r  K_a \text{.}\label{eq:adfbua}
	\end{align} 
	Therefore $R_{uv}^{G'_a}\le R_{uv}^{\wt{G}}$ for any $u,v \in V$, so the partition $P^*\ni u$ of effective resistance diameter $\beta/3$ using metric of $\wt{G}$, has effective resistance diameter at most $\beta/3$ in the metric of $G'_a$.
	 Let $H=G'_a\slash P^*$, and let $\hat{u}$ denote the super node, then by Lemma \ref{lem:additive}, for an edge $e=(u,v)$ for some $v\in V\setminus P^*$, such that $R_{e}^{G'_a}\ge \beta$, we get 
	 \begin{align}\label{eq:reha}
	 R_{\hat{u}v}^{H}\ge \beta/3.
	 \end{align}

	 So, in what remains we show that using \textsc{HeavyHitter} 
	in line \ref{line:HH-FHE} of Algorithm \ref{alg:main-heavy-edge}, we will recover these edges. 
	\paragraph{Correctness of \textsc{HeavyHitter}:}
	Recall that for $P\subseteq V$, we have $Y_{P}= [\mathbf{1}_{P},(\mathbf{1}_v)_{v\in V_G \setminus P}]$ (See line \ref{line:Y+p} of Algorithm~\ref{alg:main-heavy-edge}). Also, note that $K_P=Y^\top K Y$ is the Laplacian matrix of graph $H:=G'_a\slash P$ (See line \ref{line:K+p} of Algorithm~\ref{alg:main-heavy-edge}). By equation \eqref{eq:Eff-Res} from Section \ref{sec:prelim} we have $R^H_{\hat{u}v} = (\chi_{\hat{u}}-\chi_{v})^\T K_P^{+} (\chi_{\hat{u}}-\chi_{v})$. By the discussion above, we have,
	\begin{equation}\label{eq:RH_hat}
	R^H_{\hat{u}v} \geq \frac{\beta}{3}
	\end{equation}
	Note that $\frac{1}{C} K \preceq \wt{K} \preceq K$, thus we have $\frac{1}{C} Y^\top K Y \preceq Y^\top \wt{K} Y\preceq Y^\top K Y$, hence  
	\begin{equation}
	\label{eq:k-ckp}
	\frac{1}{C} K_P \preceq \wt{K}_P \preceq K_P\text{.}
	\end{equation} 
	Therefore we have
	\begin{equation}\label{eq:kp}
	K_P^+ \preceq \wt{K}_P^+ \preceq  CK_P^+
	\end{equation}
	Let $\wt{R}^H_{\hat{u}v} = (\chi_{\hat{u}}-\chi_{v})^\T K_P^{+} (\chi_{\hat{u}}-\chi_{v})$, thus by equation \eqref{eq:kp} we get $R^H_{\hat{u}v} \leq \wt{R}^H_{\hat{u}v} \leq CR^H_{\hat{u}v}$. Hence, by equation \eqref{eq:RH_hat} we have $\wt{R}^H_{\hat{u}v} \geq \frac{\beta}{3}$.
	
	Let $\mathbf{x}_{\hat{u}v} \in \R^{{n}\choose{2}}$ denote a vector such that each nonzero entry in $\mathbf{x}_{\hat{u}v}$ contains the voltage difference across some edge in $G$ when one unit of current is forced from $\hat{u}$ to $v$. Thus we have 
	$$\mathbf{x}_{\hat{u}v}=B Y {K}_{P}^+(\chi_{\hat{u}}-\chi_{v})\text{.}$$
	By equation \eqref{eq:Eff-Res} from Section \ref{sec:prelim} we have
	$$\mathbf{x}_{\hat{u}v}(uv)=\mathbf{x}_{\hat{u}v}(\hat{u}v)=(\chi_{\hat{u}}-\chi_{v})^\T K_P^{+} (\chi_{\hat{u}}-\chi_{v})=R^H_{\hat{u}v}\text{.}$$
	However, note that we do not have access to exact vector $\mathbf{x}_{\hat u v}$. Now let $\wt{\mathbf{x}}_{\hat{u}v}=B Y \wt{K}_{P}^+(\chi_{\hat{u}}-\chi_{v})$. Thus we have
	\begin{align*}
	||\wt{\mathbf{x}}_{\hat{u}v}||_2^2&=(\chi_{\hat{u}}-\chi_{v})^\top\wt{K}_{P}^+ (BY)^\top  B Y \wt{K}_{P}^+(\chi_{\hat{u}}-\chi_{v}) \\
	&\preceq (\chi_{\hat{u}}-\chi_{v})^\top\wt{K}_{P}^+ K_P \wt{K}_{P}^+(\chi_{\hat{u}}-\chi_{v}) && \text{Since }  (BY)^\top  B Y \preceq K_P\\
	&\leq C\cdot (\chi_{\hat{u}}-\chi_{v})^\top\wt{K}_{P}^+ \wt{K}_P \wt{K}_{P}^+(\chi_{\hat{u}}-\chi_{v}) && \text{By equation \eqref{eq:k-ckp} } K_P \preceq C\wt{K}_P \\
	&= C\cdot (\chi_{\hat{u}}-\chi_{v})^\top\wt{K}_{P}^+ (\chi_{\hat{u}}-\chi_{v}) \\
	&= C \wt{R}^H_{\hat{u}v}
	\end{align*}
	Moreover we have
	$$\wt{\mathbf{x}}_{\hat{u}v}(uv)=\wt{\mathbf{x}}_{\hat{u}v}(\hat{u}v)=(\chi_{\hat{u}}-\chi_{v})^\T \wt{K}_P^{+} (\chi_{\hat{u}}-\chi_{v})=\wt{R}^H_{\hat{u}v}\text{.}$$
	Hence we get, $$ \frac{\wt{\mathbf{x}}_{\hat{u}v}(uv)}{||\wt{\mathbf{x}}_{\hat{u}v}||_2}\geq \sqrt{\frac{\wt{R}^H_{\hat{u}v}}{C}}\geq \sqrt{\frac{\beta}{3C}}\text{.}$$
	We set $\eta=\frac{1}{2}\sqrt{\frac{\beta}{3C}}$, thus if $\wt{\mathbf{x}}_{\hat{u}v}(uv) \geq 2\eta ||\wt{\mathbf{x}}_{\hat{u}v}||_2$ our sparse recovery sketch must return $uv$ with high probability, by Lemma~\ref{lem:HH}.
	
	\paragraph{Run-time analysis:} 
	Each call to $\textsc{SparseRecovery}(S_aB,d,v)$, where $v$ is a vertex in graph that has degree less than $d$, needs $\wt{O}(d)$ time and space (see section \ref{sec:Sparse-Rec}). Since \textsc{FindLowConnectivityEdges}$(SB,d)$ calls \textsc{SpanningForest(SB)}, $100d\cdot \log n$ times (see Algorithm \ref{spanning-forest-ALG}), and each $\textsc{SpanningForest}$ runs in $O(n\log^3n)$ time and space. So, in total this line runs in $\wt{O}(n\cdot d)=\wt{O}(n^{1.4})$ time and space. In conclusion, lines \ref{line:first-lowdeg} - \ref{line:last-lowdeg} in Algorithm \ref{alg:main-heavy-edge}, guarantees that $G'_a$ has no vertex with degree less than $d$. Therefore, the first condition in Theorem \ref{thm:ballcarving} is satisfied. 
	
	By Theorem \ref{thm:main-partition}, for the choice of $d=n^{0.4}\log^2 n$, on a graph that has min degree at most $10n^{0.4}\log^2 n$, set of vertices $V$ admits a partitioning $V=C_1\cup\dots\cup C_k$ that one has
	$$\forall i\in [k], \text{diam}_{\text{eff}}^{\text{Ind}}(C_i)	\le \frac{10}{n^{0.4}}$$
	and
	 $$k=\wt{O}(n^{0.8})$$  
	 Thus, one can simply verify that $\frac{10}{n^{0.4}}< \frac{\frac{\beta}{6(2\Gamma)}}{400\sqrt{k}}$, for $\beta=\frac{1}{500c'\e^{-2} \cdot\Gamma^3 \log n}$ and $\Gamma=n^{\delta}$. So the second condition in Theorem \ref{thm:ballcarving} is satisfied. 
	 
	 Also, since we are assuming that every node in $\TT_{\le a}$ (if any) succeeded, then 	\begin{align}
	 \frac{1}{C} \cdot K_a\preceq_r \wt{K} \preceq_r  K_a \text{.}
	 \end{align} 
	 So, the third condition in Theorem \ref{thm:ballcarving} is satisfied, since $C\le2\Gamma$.
	 
	 Also in line \ref{line:FLCE-FHE} of Algorithm \ref{alg:main-heavy-edge}, by Lemma \ref{lem:recover-low-connectivity}, set $E^*$ has all edges with connectivity at least $d$, so the fourth condition of Theorem \ref{thm:ballcarving} is also satisfied. 
	  So, since all the conditions of Theorem \ref{thm:ballcarving} are satisfied,  $\textsc{BallCarving}(S B,E^*,\beta/6,\wt{K})$ runs in $\wt{O}(n^{1.4+4\delta}\e^{-2})$ time and space. 
	  
	  Contraction in line \ref{line:contraction} of Algorithm \ref{alg:main-heavy-edge} is linear time and space. Actually, we do not calculate $\wt{K}_{P_i}^{+}$ in the next line. We use Laplacian solvers in line \ref{line:HH-FHE}.
	Running each heavy hitter needs $O(\frac{C}{\beta}\polylog(n))=O(\e^{-2}n^{3\delta}\polylog(n))$ time and space, by invoking Lemma \ref{lem:HH} with $\eta=\sqrt{\frac{\beta}{3C}}$. Also, we are invoking it $O(n^{1.4+4\delta})$ times, so $\wt{O}(n^{1.4+7\delta}\e^{-2})=\wt{O}(n^{1.4+o(1)}\e^{-2})$ time and space is needed. 
	
	\paragraph{Maintenance of sketches:} Note that Algorithm \ref{alg:main-heavy-edge} takes sketch $S\cdot B$ as an input where it corresponds to the different sketches that are used in different subroutines. More precisely, $S$ is a randomly constructed matrix with ${n \choose 2}$ columns that corresponds to the concatenation of the following matrices: The sampling matrix i.e., $\Pi \in \mathbb{R}^{{n \choose 2}\times {n \choose 2}}$ (Section~\ref{sec:index}), the sketch to find the edges with connectivity at most $\lambda$, i.e., $S^f\Sigma \in\mathbb{R} ^{\lambda\cdot\poly(\log n)\times {n \choose 2}}$ (Section~\ref{subsec:find_low}), the \textsc{SparseRecovery} sketch to recover $k$-sparse vectors, i.e., $S^r\in\mathbb{R} ^{k\cdot\poly(\log n)\times {n \choose 2}}$ (Section~\ref{subsec:sp-rec}), and the \textsc{HeavyHitter} sketch to find the edges that are heavy with respect to parameter $\eta$, i.e., $S^h\in\mathbb{R} ^{\eta^{-2}\cdot\poly(\log n)\times {n \choose 2}}$ (Section~\ref{subsec:hv-hit}). 
	
As per line~\ref{line:first-lowdeg} of Algorithm~\ref{alg:main-heavy-edge} we set $k=d=\wt{O}(n^{0.4})$, in line~\ref{line:FLCE-FHE} of Algorithm~\ref{alg:main-heavy-edge} we set $\lambda=d=\wt{O}(n^{0.4})$, and in line~\ref{line:HH-FHE} we set $\eta=\Theta\left(\sqrt{\frac{\beta}{C}}\right)=\frac{1}{\poly( \Gamma\cdot \log n )}$. Recall that $\Gamma=n^\delta$. Therefore, overall the number of random bits needed for all the matrices, in an invocation of Algorithm~\ref{alg:main-heavy-edge} is at most $R=\wt{O}(n^2+nk+n\lambda+n\cdot\eta^{-2})=\wt{O}(n^2)$, in addition to the random bits needed for the recursive calls.

To generate matrix $\Pi$, $S^f\Sigma$, and $S^h$ we use the fast pseudorandom numbers generator that is introduced in Section \ref{sec:prg}. Observe that the space used by Algorithm~\ref{alg:main-heavy-edge} is $s=\wt{O}(n^{1.4+7\delta})$ in in addition to the rspace used by the recursive calls. Since $R=O(n^2)$, we have $R=O(s^2)$. Therefore, by Theorem~\ref{high_level_result} we can generate seed of $O(s\cdot \poly(\log s))$ random bits in $O(s\cdot \poly(\log s))$ time that can simulate our randomized algorithm.
	 
To generate the \textsc{SparseRecovery} sketch to recover $k$-sparse vectors, i.e., $S^r\in\mathbb{R} ^{k\cdot\poly(\log n)\times {n \choose 2}}$ (Section~\ref{subsec:sp-rec}), we can not use our fast pseudorandom numbers generator, since these bits need to be accessed again during the decoding time (see line \ref{ln:update-sketch} of Algorithm~\ref{alg:main-heavy-edge}). However, Algorithm \textsc{SparseRecovery} uses low-independence hash functions for those bits instead.
Moreover, the random matrix $Q \in\mathbb{R} ^{\Theta(\log n)\times {n \choose 2}}$ for JL (line~\ref{ln:jl-mat} of Algorithm \ref{alg:main-sparsify}) can be generated using $\log n$-wise independent hash functions.
\end{proof}

\subsection{Tree-based indexing scheme for sketches}\label{sec:index}
Our main sparsification algorithm (Algorithm~\ref{alg:main-sparsify}) is a recursive algorithm, such that for any invocation of Algorithm~\ref{alg:main-sparsify} we need fresh random bits and independent sketches. To that end we index our sketches and subroutines using the recursion tree of algorithm's execution.

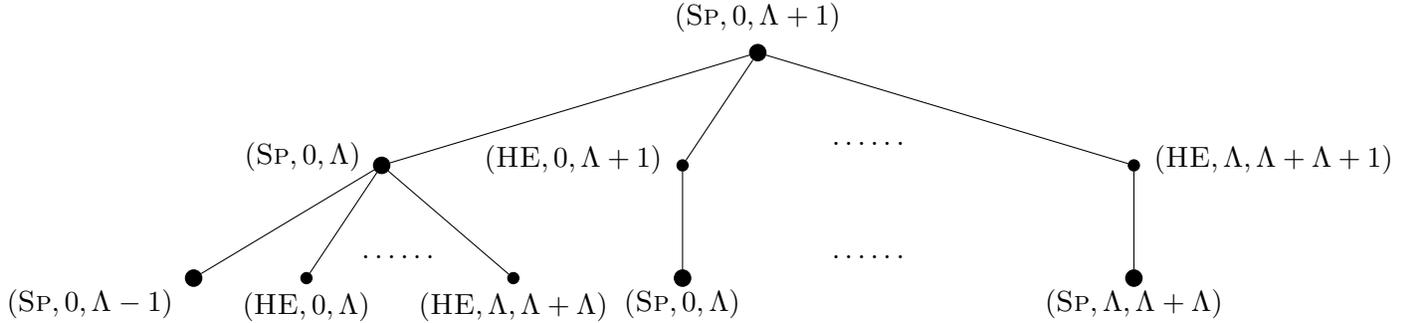
\begin{figure}[h!]

\begin{tikzpicture}[level/.style={sibling distance=50mm/#1}]
\tikzset{
	arn/.style = {circle, white, draw=black, fill=black, inner sep = 1.5},
	arn_i/.style = {circle, white, draw=white, fill=white, inner sep = 0},
	arn_l/.style = {circle, white, draw=black, fill=black, inner sep = 2.2},
	photon/.style={draw=black, very thick, dashed},
	electron/.style={draw=black, very thick, line width=0.08cm},
	tr/.style={buffer gate US,thick,draw,fill=gray!60,rotate=90, anchor=east,minimum width=2.25cm},
	br/.style={buffer gate US,thick,draw,fill=gray!60,rotate=90, anchor=east,minimum width=4.5cm}
}
\node [arn_l] (sp0){}
child { node [arn_l] (sp1){}
	child { node [arn_l] (sp2){}
}
child { node at (-1,0) [arn] (he1){}
}
child { node [arn] at (-0.75,0) (he2){}
}}
child { node [arn] at (-1, 0 ) (he3){}
child { node [arn_l] (sp3){}
}	
}
child { node [arn] (he4){}
child { node [arn_l] (sp4){}
}
}
;

\node [] at (1.5,-1.5) [label=above:{\dots\dots}] {};
\node [] at (1.5,-3) [label=above:{\dots\dots}] {};
\node [] at (-4.75,-3) [label=above:{\dots\dots}] {};

\node [] at (sp0) [label=above:{$(\textsc{Sp},0,\Lambda+1)$}] {};
\node [] at (he3.north) [label=left:{$(\textsc{HE},0,\Lambda+1)$}] {};
\node [] at (he4.north) [label=right:{$(\textsc{HE},\Lambda,\Lambda+\Lambda+1)$}] {};
\node [] at (sp1.north) [label=left:{$(\textsc{Sp},0,\Lambda)$}] {};
\node [] at (sp2.north) [label=below left:{$(\textsc{Sp},0,\Lambda-1)$}] {};
\node [] at (sp3.north) [label=below:{$(\textsc{Sp},0,\Lambda)$}] {};
\node [] at (sp4.north) [label=below:{$(\textsc{Sp},\Lambda,\Lambda+\Lambda)$}] {};
\node [] at (he1.north) [label=below:{$(\textsc{HE},0,\Lambda)$}] {};
\node [] at (he2.north) [label=below:{$(\textsc{HE},\Lambda,\Lambda+\Lambda)$}] {};
\end{tikzpicture}
\caption{Illustration of the first three levels of the recursion tree $\TT$. Nodes corresponding to invocations of \textsc{Sparsify} are marked \textsc{Sp}, nodes corresponding to invocations of \textsc{HeavyEdges} are marked \textsc{HE}.}\label{fig:1}
\end{figure}

Let $\TT=(U, E_\TT)$ be the recursion tree of the algorithm, where $U$ denotes the set of nodes in recursion tree $\TT$ and $E_\TT$ denotes the edges of tree $\TT$. We have two types of nodes in $\TT$, nodes corresponding to calls to \textsc{Sparsify} (see Algorithm~\ref{alg:main-sparsify}) and nodes corresponding to calls to \textsc{HeavyEdges} (see Algorithm~\ref{alg:main-heavy-edge}). For every node $a\in U$ we associate a label $\textsf{label}(a)$:  

\begin{itemize}
	\item{\bf Nodes corresponding to $\textsc{Sparsify}$}  For nodes $a\in U$ that correspond to invocations of \textsc{Sparsify}, we have $\textsf{label}(a)=(\textsc{Sparsify}, i, \ell)$, where $i $ and $\ell$ correspond to the sampling level and the regularization at that point. Recall that $\Lambda=\lceil \log_\Gamma\frac{\lambda_u}{\lambda_\ell}\rceil$ (see line \ref{line:theta-delta} in Algorithm~\ref{alg:main-sparsify}), then any node $u\in U$, that $\textsf{label}(u)=(\textsc{Sparsify}, i, \ell)$, has $(\Lambda - i+1)$ children, $b_q\in U$ for  $q\in [\Lambda-i+1]$, corresponding to each call of \textsc{HeavyEdges} (see line \ref{line:RS-callHE} in Algorithm \ref{alg:main-sparsify}) one has $\textsf{label}(b_q)=(\textsc{HeavyEdges},i+q-1,\ell+q-1)$. if $\ell >0$, then node $a$ also has another child, $b^*\in U$, corresponding to a call to \textsc{Sparsify} (see line \ref{alg:line-coarse} of Algorithm \ref{alg:main-sparsify}) and $\textsf{label}(b^*)=(\textsc{Sparsify},i,\ell-1)$.

	\item{\bf Nodes corresponding to $\textsc{HeavyEdges}$} For nodes $a\in U$ that correspond to invocations of \textsc{HeavyEdges}, we have $\textsf{label}(a)=(\textsc{HeavyEdges}, i, \ell)$, where $i$ and $\ell$ correspond to the sampling level and the regularization at that point. If $\ell > 0$, then this node has one child, $b\in U$, which corresponds to a call of \textsc{Sparsify} (see line \ref{line:HE-CallSp} in Algorithm~\ref{alg:main-heavy-edge}) and $\textsf{label}(b)=(\textsc{Sparsify},i,\ell-1)$.
\end{itemize}
An illustration of the tree $\TT$ is given in Fig.~\ref{fig:1}, and an illustration of the neighborhood of a \textsc{Sparsify} node in $\TT$ is given in Fig.~\ref{fig:2} below.

\begin{figure}[h!]
\begin{center}
\begin{tikzpicture}[level/.style={sibling distance=70mm/#1}]
\tikzset{
	arn/.style = {circle, white, draw=black, fill=black, inner sep = 1.5},
	arn_i/.style = {circle, white, draw=white, fill=white, inner sep = 0},
	arn_l/.style = {circle, white, draw=black, fill=black, inner sep = 2.2},
	photon/.style={draw=black, very thick, dashed},
	electron/.style={draw=black, very thick, line width=0.08cm},
	tr/.style={buffer gate US,thick,draw,fill=gray!60,rotate=90, anchor=east,minimum width=2.25cm},
	br/.style={buffer gate US,thick,draw,fill=gray!60,rotate=90, anchor=east,minimum width=4.5cm}
}
\node [arn] (he1){}
child { node [arn_l] (sp1){}
child { node [arn_l] (sp2){}
}
	child { node [arn]at (-2,0) (he2){}
}
	child { node [arn] (he3){}
}
}
;

\node [] at (he1.north) [label=left:{$(\textsc{HE},i,\ell)$}] {};
\node [] at (sp1.north) [label=left:{$(\textsc{Sp},i,\ell-1)$}] {};
\node [] at (he2.north) [label=below:{$(\textsc{HE},i,\ell-1)$}] {};
\node [] at (he3.north) [label=below:{$(\textsc{HE},\Lambda,\ell+\Lambda-1)$}] {};
\node [] at (sp2.north) [label=below left:{$(\textsc{Sp},i,\ell-2)$}] {};
\node [] at (0.5,-3) [label=above:{\dots\dots}] {};

\end{tikzpicture}
\end{center}
\caption{An illustration of a \textsc{Sparsify} node in $\TT$, its children and the parent in $\TT$.}\label{fig:2}
\end{figure}
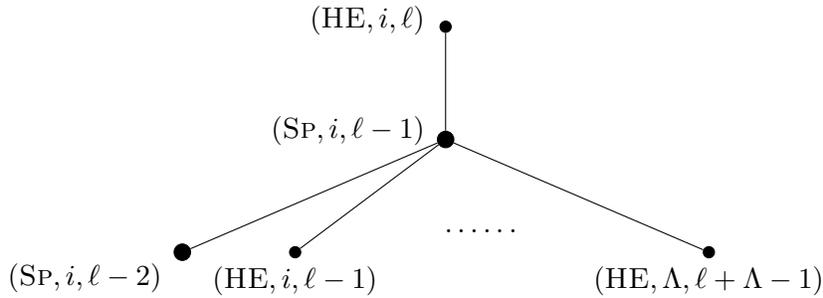

One can easily conclude the following facts. 
\begin{fact}\label{fact:ind-T}
	In tree $\TT$ if node $a$  is ancestor of node $a'$ and if we have $\textsf{label}(a)=(\textsc{Sparsify}, i, \ell)$ and $\textsf{label}(a')=(\textsc{Sparsify}, i', \ell')$ then $i-\ell \le (i'-\ell')-1$ 
\end{fact}
As a corollary, we can state the following fact. 
\begin{fact}\label{fact:T-num-nodes}
	Tree $\TT$ with root $a\in U$, that $\textsf{label}(a)=(\textsc{Sparsify},0,\Lambda+1)$, has depth at most $2(\Lambda+2)$. Also, by the discussion above, every node in tree $\TT$ has at most $\Lambda+2$ children. In conclusion subtree $\TT$ has at most $(\Lambda+2)^{2(\Lambda+2)}$ nodes. Note that by the choice of $\Lambda={\log\log n}$, then $\Lambda^\Lambda=n^{o(1)}$, so we have at most $n^{o(1)}$ nodes in the tree. 
\end{fact}

For any node $a \in U$, that corresponds to $\textsc{HeavyEdges}$ and $\textsf{label}(a)=(\textsc{HeavyEdges},i,\ell)$, for some integers $i$ and $\ell$, let its parent be node $b$, that $\textsf{label}(b)=(\textsc{Sparsify},i',\ell')$, for some integers $i'\le i$ and $\ell'\le \ell$. Then, we define a binary hash function $h_a: \binom{n}{2} \rightarrow \{0,1\}$, that maps any edge $e$ to $1$ with probability $\frac{1}{\Gamma^{i-i'}}$. Then, for any matrix $D \in \R ^{\binom{n}{2}\times n}$, that each row corresponds to an edge, let $D_a$ be $D$
with all rows except those with $h_a (e) = 0$ zeroed out. So $D_a$ is $D$ with rows sampled independently at rate $\frac{1}{\Gamma^{i-i'}}$. This operation also can be done by linear operators. We build a diagonal matrix $\Pi_a\in \R^{\binom{n}{2}\times \binom{n}{2}}$, based on hash functions $h_a$ that serves as a sampling matrix as follows.
$$
\Pi_a(e,e):= h_a(e) 
$$
Then clearly $D_a=\Pi_a  D$. Also, for any node $b\in U$ that corresponds to a call to $\textsc{Sparsify}$, we set $\Pi_b=I$.

\begin{definition}
\label{def:G_a}
Let $a_0$ denote the root of the recursion tree, and for any node $a\in U$, suppose that $(a,a_k,\dots,a_1,a_0)$ denote the unique path from root to node $a$. We define $\wt{\Pi}_a:=\Pi_a\Pi_{a_k}\dots \Pi_{a_0}$.

Let $G$ denote a graph corresponding to the root of the recursion tree. Let $B$ denote the vertex edge incidence matrix of graph $G$. For any node $a$ in the recursion tree, we define $B_a=\wt\Pi_aB$, and $B'_a=B_a\oplus \sqrt{\gamma}$. Let $G_a$ denote a graph corresponding to node $a$ in the recursion tree. Therefore, the vertex edge incidence matrix of $G_a$ is $B'_a$.  Let $K_a$ denote the Laplacian of matrix $G_a$, thus $K_a=({B'_a})^\top (B'_a)$.
\end{definition}

Any invocation of Algorithm~\ref{alg:main-sparsify} and Algorithm~\ref{alg:main-heavy-edge} takes matrix $S\cdot B$ as an input where $S=S_a\wt\Pi_a$, such that $\wt\Pi_a$ is the sampling matrix, and $S_a$ denotes the different sketches used in the subroutines (see Section \ref{sec:util} for more details).

We initialize our algorithm by invocation of $\textsc{Sparsify}(a_0,S_{\le {a_0}}B,0,\Lambda+1)$, where $a_0$ is the root of tree $\TT$. Let $G_{a_0}=G$, and observe that we have no regularization for $G_{a_0}$ (see line \ref{line:set-gamma} in Algorithm~\ref{alg:main-sparsify}). Moreover, by the fact that $\wt{\Pi}_{a_0}=I$, we sample all the edges of $G$, hence, $B_{a_0}=B$. 

As a conclusion of our discussion above, we present the following lemma, which relates the recursion tree $\TT$ to the chain of coarse sparsifiers, defined in Lemma~\ref{lem:chain_coarse}.

\begin{remark}\label{remark-chain}
	 In recursion tree $\TT$, for any node $a\in U$, that corresponds to an invocation of Algorithm $\textsc{Sparsify}$, and node $b\in U$ that is a child of node $a$,  and corresponds to an invocation of Algorithm  $\textsc{Sparsify}$, by Lemma~\ref{lem:chain_coarse} \eqref{itm:last}, \eqref{itm:mid} the following holds:
$$K_a \preceq_r K_b \preceq_r \Gamma K_a$$
where $K_a$ and $K_b$ correspond to laplacian matrix of $G_a$ and $G_b$. Note that if node $a$ does not have a child that corresponds to a call to $\textsc{Sparsify}$, in that case, again by Lemma~\ref{lem:chain_coarse} \eqref{itm:base}
$$K_a \preceq_r \Gamma\cdot \lambda_u \cdot I \preceq_r \Gamma K_a$$
\end{remark}

\begin{fact}
	Suppose that node $a\in U$ corresponds an invocation of \textsc{HeavyEdges} and $\textsf{label}(a)=(\textsc{HeavyEdges},i,l)$, for some $i$ and $l$. Let node $b \in U$ be the parent of node $a$ and $\textsf{label}(b)=(\textsc{Sparsify},i',l')$. Thus, for any edge $e$ that if $\wt{\Pi}_{b}(e,e)=1$, then $\wt{\Pi}_{a}(e,e)=1$ with probability $\frac{1}{\Gamma^{i-i'}}$. On the other hand, if $\wt{\Pi}_{b}(e,e)=0$, then $\wt{\Pi}_{a}(e,e)=0$.
\end{fact}

\newpage
\section{Analysis of \textsc{BallCarving}}\label{sec:BC}
In this section, we introduce and analyze \textsc{BallCarving} (see Algorithm~\ref{alg:ballcarving}) the algorithm that, given a coarse spectral approximation $\wt{G}$ to $G$ explicitly, partitions \Navid{-----} vertices of $G$ into clusters of low effective resistance diameter. More specifically, the algorithm first removes from $G$ a subset $E'$ of edges that is guaranteed to contain all edges with connectivity bounded by $d$ (see line~\ref{line:span-forest} of Algorithm~\ref{alg:ballcarving}), computes connected components in the resulting graph $G'$ and processes these components as follows. We mark all vertices in $G$ active (add them to the set $V_{active}$, line~\ref{line:vactive-ordering}), which means that they are viable candidates for being ball centers, and add all vertices to a set $V'$, which means that they are eligible for being output as part of a partition (the set $V'$ is introduced mostly for technical reasons -- see proof of Theorem~\ref{thm:ballcarving} in Section~\ref{sec:comparing-metrics} below).

We consider vertices in $V_{active}$ in some fixed order and, upon picking a new vertex (line~\ref{pick_node} of Algorithm~\ref{alg:ballcarving}) for consideration distinguish between two cases. If the effective resistance from $u$ to every other vertex in the same connected component in $E\setminus E'$ is small (line~\ref{line:if-BC} of Algorithm~\ref{alg:ballcarving}), then we skip to the next vertex, since this means that every `heavy' edge incident on $v$ must have the other endpoint in a different component and hence is reported in $E'$, at the same time removing the entire component from the active set (these vertices will no longer be considered as candidate ball centers).  Otherwise we (essentially) carve a ball in effective resistance metric of $\wt{G}$ (the coarse sparsifier) around $u$ covering only the vertices on $V'$, i.e. not covered by previously selected balls (line~\ref{line:cal-Ruv-M} of Algorithm~\ref{alg:ballcarving}). We note that computing effective resistances exactly would be too computationally expensive, and we instead use an approximation based on dimensionality reduction: in line~\ref{line:M-BC} we compute an embedding of vertices of $\wt{G}$ into Euclidean space (using the Johnson-Lindenstrauss lemma) such that Euclidean distances squared approximate effective resistance in $\wt{G}$ to a constant factor. Computing $M$ takes time nearly linear in the number of edges of $\wt{G}$ using Laplacian solvers (this is a standard technique, originating from~\cite{spielman2011graph}). Thus, in reality we perform ball-carving in a slightly distorted version of the effective resistance metric in $\wt{G}$.

\paragraph{Analysis of \textsc{BallCarving}: proof outline.} As outlined in Section~\ref{sec:overview}, the crux of our runtime analysis is bounding the number of partitions generated by \textsc{BallCarving}. Deriving this upper bound is the main goal of this section. The main result of the section is 

\begin{theorem}\label{thm:ballcarving}
	For every graph $G=(V, E)$, and a set of edges $E'\subseteq E$ if:
	\begin{enumerate}
		\item The minimum degree in $G$ is lower bounded by $d\geq 1$.
		\item For some $\Gamma\ge 1 $ and $0<r\le 1/\log n$ and integer $k \ge 1$ the vertex set $V$ admits a partition $V=C_1\cup \ldots\cup C_k$ such that for every $i\in [k]$ the subgraph induced by $C_i$ has effective resistance diameter bounded by $\Delta\in (0, 1)$, i.e., $\text{diam}_{\text{eff}}^{\text{Ind}}(C_i)\le \Delta$ such that $\Delta < \frac{r/\Gamma}{800\sqrt{k}}$. 
		\item Graph $\wt{G}$ is such that, for any pair of vertices $(u,v)\in V\times V$, $R_{uv}^{G^\gamma}\le R_{uv}^{\wt{G}} \le 2\Gamma \cdot R_{uv}^{G^\gamma}$, where $G^\gamma$ is $G$ plus regularization $\gamma$, i.e., $L_{G^\gamma}=L_G+\gamma I$.
		\item Edge set $E'$ contains all the edges of connectivity no more than $d$ in graph $G$.
		\item $SB$ is the set of needed sketches.
	\end{enumerate}
	Then \textsc{BallCarving}($S\cdot B , E' , r, \wt{G}$) returns a set of disjoint subsets of vertices with effective resistance diameter bounded by $2r$ in the metric of $\wt{G}$ \Cam{this means the non-induced effective resistance  right? Write  down using an equation so its clear.}, and there are no more than $\wt{O}\left(\sqrt{k}+\frac{k\cdot\Gamma}{r\cdot d}\right)$ such non-singleton partitions, and runs in $\wt{O}(n\sqrt{k}+n\frac{k\Gamma}{r\cdot d}+n/\e^2)$ time and space.
\end{theorem}

Despite the fact that Theorem~\ref{thm:ballcarving} is the main result of the section, the proof of Theorem~\ref{thm:ballcarving} is mostly technical given the following core result:

\noindent{\em {\bf Theorem~\ref{thm:numpartitions}} (Restated from Section~\ref{sec:overview})
	For every graph $G=(V, E)$, and a set of edges $E'\subseteq E$ if:
	\begin{enumerate}
		\item The minimum degree in $G$ is lower bounded by $d\geq 1$.
		\item For some $0<\beta< 1/\log n$ and integer $k \ge 1$ the vertex set $V$ admits a partition $V=C_1\cup \ldots\cup C_k$ such that for every $i\in [k]$ the subgraph induced by $C_i$ has effective resistance diameter bounded by $\Delta\in (0, 1)$, i.e., $\text{diam}_{\text{eff}}^{\text{Ind}}(C_i)\le \Delta$ such that $\Delta < \frac{\beta}{400\sqrt{k}}$. 
		\item Edge set $E'$ contains all the edges of connectivity no more than $d$ in graph $G$.
		\item $SB$ is the set of needed sketches.
	\end{enumerate}

	Then \textsc{BallCarving}($S\cdot B , E' , \beta, {G}$) returns a set of disjoint subsets of vertices with effective resistance diameter bounded by $2\beta$ in the metric of $G$, and there are no more than $\wt{O}\left(\sqrt{k}+\frac{k}{\beta\cdot d}\right)$ such non-singleton partitions. 
}

We note that the difference between Theorem~\ref{thm:ballcarving} and Theorem~\ref{thm:numpartitions} lies in the effective resistance metric used for ball carving. Indeed, Theorem~\ref{thm:numpartitions} deals with the somewhat unrealistic scenario that \textsc{BallCarving} is invoked on a graph $G$ and is given the effective resistance metric of $G$ as input. This is unrealistic, since our \textsc{Sparsify} procedure calls \textsc{BallCarving} in order to prepare partitions of the vertex set of a graph $G$ (which is not known at the time) into low effective resistance diameter pieces, with the help of an explicitly provided {\bf coarse} sparsifier $\wt{G}$. At the same time, in order to improve exposition, we first prove Theorem~\ref{thm:numpartitions} and that Theorem~\ref{thm:ballcarving} is implied by essentially observing an inclusion relation by effective resistance balls in $\wt{G}$ and $G$. This (mostly technical) proof is given in Section~\ref{sec:comparing-metrics}.

\paragraph{Significance of Theorem~\ref{thm:numpartitions} and proof techniques.} 

In order to put the result of Theorem~\ref{thm:numpartitions} in perspective, we note that the theorem relies on a partitioning of the vertex set of the graph into vertex disjoint clusters $C_i$ each of which has low effective resistance diameter as an induced subgraph. Such a decomposition is provided by Theorem~\ref{thm:main-partition} (proved in Section~\ref{sec:partitioning}), which we restate below for convenience of the reader:

\noindent {\em {\bf Theorem~\ref{thm:main-partition}} (Restated)
	For any unweighted graph $G=(V,E)$ that $|V|=n$ and with min-degree at least $n^{0.4}\log^2 n$, set of vertices, $V$, admits a partitioning into $V=C_1\cup \dots\cup C_k$, that for any 
	$$\forall i\in [k], \text{diam}_{\text{eff}}^{\text{Ind}}(C_i)	\le \frac{10}{n^{0.4}}$$
	and $$k \le  c \cdot n\sqrt{\log n}\sqrt{\frac{1}{n^{0.4}\log^2 n}}=c\cdot n^{0.8} \sqrt{\frac{1}{\log n}}=O(n^{0.8})$$
}

We use Theorem~\ref{thm:main-partition} together with Theorem~\ref{thm:ballcarving} in Section~\ref{sec:HEanalysis} to establish correctness of our main primitive \textsc{HeavyEdges}.

We now provide the intuition behind the proof of Theorem~\ref{thm:numpartitions}. The upper bound of $\wt{O}\left(\sqrt{k}+\frac{k}{\beta\cdot d}\right)$ on the number of clusters produced by \textsc{BallCarving} should essentially be thought of as a bound of $\wt{O}\left(\frac{k}{\beta\cdot d}\right)$ (this term is ultimately balanced against the $\sqrt{k}$ term in our parameter settings, see Lemma~\ref{lem:corr-main-HE}). The intuition behind the bound of $\wt{O}(\frac{k}{\beta \cdot d})$ is very simple: we essentially show that a graph with min-cut lower bounded by $d$ and partitioned into $k$ clusters of effective resistance diameter $\approx 1/d$ can be basically viewed as a graph on $k$ nodes with min-degree $d$ (think of contracting all clusters; see graph $H$ in the proof of Lemma~\ref{thm:light-clusters}). One can show (this is roughly what Section~\ref{sec:BG} does) that a ball of effective resistance radius $\beta$ around a `typical' node in such a graph contains $\wt{\Omega}(\beta d)$ vertices, which implies that one cannot have more than $\wt{O}(\frac{k}{\beta \cdot d})$ elements in the partition, since ball centers in any such partition are well-separated in the effective resistance metric. The explanation above reflects the essence of the analysis in Section~\ref{sec:BG} modulo several important technical points such as 
\begin{description}
\item[{\bf (A)}] We never actually have a min-cut assumption, but are able to carry the analysis through for connected components that remain in $G$ after removing the edge set $E'$ that contains low connectivity edges (see line~\ref{line:span-forest} of Algorithm~\ref{alg:ballcarving});
 \item[{\bf (B)}] The ball-growing from Section~\ref{sec:BG} (which is only an analysis construct, and is never algorithmically executed) is not performed in the graph obtained by collapsing the clusters to supernodes, but in the original graph, while carefully keeping track of the supernodes that are covered by our effective resistance balls. 
 \end{description}
 See Section~\ref{sec:BG} for more details. Section~\ref{sec:PG}, which is presented before the ball-growing analysis in Section~\ref{sec:BG}, make the intuition alluded to above that a graph partitioned into $k$ clusters of very low ($\approx 1/d$) effective resistance diameter can essentially be viewed as equivalent to the graph obtained by collapsing clusters into supernodes. The technical conditions that are proved in Section~\ref{sec:PG} are specifically tailored to the application in Section~\ref{sec:BG}.

\begin{algorithm}[h!]
	\caption{\textsc{BallCarving($S B,E',r,\wt{G}$)}\Comment{$SB$ is the sketch of graph $G=(V,E)$.}}\label{alg:ballcarving}
	\begin{algorithmic}[1]
		\State Update $SB$ by removing edges in $E'$ 
\Comment{So, sketch $SB$ becomes a sketch of $G':=(V,E\setminus E')$}
		\State Run \textsc{SpanningForest}$(SB)$ and get connected components $C_1,\dots,C_\ell$ of $G'$		 \label{line:span-forest}
		\State Ordered set $V_{\text{active}} \leftarrow V_G$ ordered by $\pi$  \Comment $V_{\text{active}}$ gets $V_G$ with an arbitrary ordering, $\pi$. However the ordering is arbitrary, we use this ordering in proof of Theorem~\ref{thm:ballcarving}. \label{line:vactive-ordering}
		\State $i \gets 0$
		\State $V' \gets V_G$
		\State $q\gets 10^6 \log n$ \Comment It suffices to get a $(1\pm \frac{1}{5})$ approximation from JL
		\State $Q\gets q\times {n \choose 2}$ is a random $\pm$1 matrix 
		\State  $M\gets \frac1{\sqrt{q}}Q\widetilde B\widetilde L^+$ \Comment $M$ is such that  $R_{uv}^{G^\gamma}\le\frac{5}{4}||M(\chi_u-\chi_v)||_2^2\le 3 \Gamma R_{uv}^{G^\gamma}$\label{line:M-BC}
		\While{$V_{\text{active}}\ne \emptyset$}
		\State Let $u$ be the next vertex of $V_{\text{active}}$. \Comment{We use the ordering $\pi$ of line \ref{line:vactive-ordering}}{\label{pick_node}}
		\State Let $C^*$ be the connected component in $H$ that contains $u$.
		\State $R^{M}_{u,C^*} \gets\frac{5}{4} \max_{v\in C^*} ||M(\chi_u-\chi_v)||_2^2$ \label{line:max-eff-res}  
		\If{$R^M_{u,C^*}\le r/2$} \label{line:if-BC} 
		\State {$V_{\text{active}} \leftarrow V_{\text{active}} \setminus C^*$}	 \Comment Nodes in $C^*$ will not be ball centers in next iterations	\label{line:remove-line13}	 		
		\Else
		\State $i \gets i+1$ \label{line:1st-BC}
		\State $P_i \gets \{u\}$ {\label{when_ballcarve}}
		\For {$v \in V' \setminus\{u\}$}		\Comment Basically, $P_i = \B_M(u,r) \cap V'$
		\If {$||M(\chi_u-\chi_v)||_2^2\le r$} \label{line:cal-Ruv-M} 
		\State Add $v$ to $P_i$
		\EndIf 
		\EndFor
		\State $V_{\text{active}} \leftarrow V_{\text{active}} \setminus P_i$\label{line:last-BC}\Comment Nodes in one partition will not be ball centers in next iterations
		\State $V' \gets V'\setminus P_i$ \Comment $V'$ is the set of remaining vertices
		\EndIf
		\EndWhile
		\Return $(P_1,P_2,\dots,P_i)$   \Comment Note that we do not return the remaining vertices, $V'$
	\end{algorithmic}
\end{algorithm}

\subsection{Analysis of \textsc{BallCarving} with effective resistance metric of $\wt{G}$}\label{sec:comparing-metrics}

In this section we give

\begin{proofof}{Theorem~\ref{thm:ballcarving}}
	We prove the {\bf correctness} of the algorithm using two different invocations of \textsc{BallCarving}. 
	\paragraph{Comparing \textsc{BallCarving}($SB , E' , r, \wt{G}$) and \textsc{BallCarving}($SB , E' , \frac{r}{3\Gamma}, G$):} 
	 Consider an invocation of \textsc{BallCarving} (see Algorithm \ref{alg:ballcarving}) with {\bf radius $r$ and effective resistance metric $\wt{G}$},  let $(P_1^{(1)},\dots,P_{t}^{(1)})$ denote the output sequence of partitions, and let $u_i$ be the center of $P^{(1)}_i$ for every $i\in [t]$ (see line~\ref{when_ballcarve} of Algorithm~\ref{alg:ballcarving}). Define the ordering $\pi$ by letting $u_1,\ldots, u_t$ be the first $t$ vertices in $\pi$, and letting the remaining $n-t$ vertices $V\setminus \{u_1,u_2,\dots,u_t\}$ be in an arbitrary order. Also, for ease of notation, just for this invocation of algorithm, we define $R_{uv}^M=\frac{5}{4}||M(\chi_u-\chi_v)||_2^2$, for $M$ in line \ref{line:M-BC} of Algorithm \ref{alg:ballcarving}. By assumption, we have $$R_{uv}^{G^\gamma}\le R_{uv}^{\wt{G}} \le 2\Gamma \cdot R_{uv}^{G^\gamma}.$$
	 Since we are using JL with $(1\pm \frac{1}{5})$ multiplicative error, we have 
	 $$R_{uv}^{\wt{G}}\le R_{uv}^M\le \frac{3}{2} R_{uv}^{\wt{G}},$$
	 and since $R_{uv}^{G^\gamma}\le R_{uv}^G$, we have,
	 \begin{align}\label{eq:M-G}
	 R_{uv}^M \le 3 \Gamma R_{uv}^{G}
	 \end{align}
	 
	 Now consider an invocation of \textsc{BallCarving} with {\bf radius $\frac{r}{3\Gamma}$ and effective resistance metric $G$}. Note that in this case, since we are just using this invocation for analysis, we can say that we do not use JL in this case, and we are calculating the effective resistances exactly. One should realize that this is w.l.o.g. Further, assume that this invocation uses $\pi$ as the ordering of vertices in line \ref{line:vactive-ordering} to create $V_{\text{active}}$ (recall that the results of Theorem~\ref{thm:ballcarving} and Theorem~\ref{thm:numpartitions} hold regardless of the ordering). Let $(P_1^{(2)},\dots,P_{t'}^{(2)})$ be the output. 
	 
	 Now, we prove by strong induction that for every $\ell\in [t]$, vertex $u_\ell$ is a ball center in invocation of \textsc{BallCarving} with radius $\frac{r}{3\Gamma}$ and effective resistance metric $G$. 
	 \begin{description}
	 	\item[Base: $\ell=1$] The base of the induction is provided by $\ell=1$. Note that at this point, $V_{\text{active}}$ contains all the vertices, and since we are using $\pi$ as the ordering, $u_1$ is the first element in $V_{\text{active}}$. On the other hand, recall that $u_1$ is the ball center of $P_1^{(1)}$, meaning that in the execution of \textsc{BallCarving}($S\cdot B , E' , r, \wt{G}$) line \ref{when_ballcarve} must be called, which ensures that $R_{u_1,C^*}^{M}> r/2$ (see line~\ref{line:if-BC} of Algorithm~\ref{alg:ballcarving}). Also, since by \eqref{eq:M-G} for any $u,v\in V$, one has $ R^{M}_{uv}  \le 3\Gamma \cdot R_{uv}^{G}$,
	 	\begin{equation*}
	 	R_{u_1,C^*}^{G}\ge \frac{r}{6\Gamma} 
	 	\end{equation*}
	 	i.e. the condition in line \ref{line:if-BC} is not satisfied for $u_1$, in the invocation of \textsc{BallCarving} with radius $r/\Gamma$ and effective resistance metric $G$ and ordering $\pi$, thus, line \ref{when_ballcarve} is called in this invocation and $u_1$ is a ball center.

	 	\item[Inductive step: $1,\dots, \ell\to \ell+1$ for any $\ell=1,\dots, t-1$ \ ] We do this part in two parts as follows.

	 	 {\bf Part 1: showing that $u_{\ell+1}$ is active when considered.} We now show that, in the invocation of \textsc{BallCarving}  on $G$, $u_{\ell+1}$ is not covered by any of the balls around $u_i$, for $i \le \ell$, which would prevent $u_\ell$ from being added as a cluster center, i.e. we prove: 
	 	$$
	 	u_{\ell+1} \notin\bigcup_{j\in[\ell]} P_j^{(2)}.
	 	$$
	 	First, since by \eqref{eq:M-G} for any pair of vertices $(u,v)\in V\times V$, $ R_{uv}^{M} \le 3\Gamma \cdot R_{uv}^G$, by Lemma \ref{ball_approx}, for any $u\in V$, we have, $\B_{G}\left(u,\frac{r}{3\Gamma}\right) \subseteq \B_{M}(u,r)$, where $\B_M(u,r)=\{v \in V : R_{uv}^M\le r\}$. Second, we note that
	 	\begin{equation}\label{eq:1hewifhg}
	 	\bigcup_{i\in[\ell]}\B_{M}(u_i,r)=\bigcup_{i\in [\ell]}P_{i}^{(1)}\text{~~~~~and~~~~~~}
	 	\bigcup_{i\in[\ell]}\B_{G}(u_i,\frac{r}{3\Gamma})=\bigcup_{i\in [\ell]}P_{i}^{(2)}
	 	\end{equation}
	 	
	 	Equation ~\eqref{eq:1hewifhg} in turn implies that 
	 	$$
	 	\bigcup_{i\in [\ell]}P_{i}^{(2)} \subseteq \bigcup_{i\in [\ell]}P_{i}^{(1)}.
	 	$$ 
	 	Finally, since $u_{\ell+1}$ is added as a cluster center in the invocation of \textsc{BallCarving} on $G$, one has $u_{\ell+1} \notin \bigcup_{j\in[\ell]} P_j^{(1)}$ and therefore $u_{\ell+1} \notin \bigcup_{j\in [\ell]} P_{j}^{(2)}$. 
	 	We also note that,  in the invocation on $\tilde G$, for any $i \in [\ell]$, $u_i$ is a ball center, so we have not removed any vertex due to execution of line \ref{line:remove-line13}, before  the $\ell$'th iteration. 
	 	
	 	{\bf Part 2: showing that $u_{\ell+1}$ is chosen as a center of a partition.} Note that since $u_{\ell+1}$ is the ball center for $P_{\ell+1}^{(1)}$, then, if $C^*$ is the connected component that $u_{\ell+1}$ belongs to (this component is independent of the effective resistance metric used, and hence is the same in both invocations), one has $R_{u_{\ell+1},C^*}^{M}\ge \frac{r}{2}$ (see line~\ref{line:if-BC} of Algorithm~\ref{alg:ballcarving}). Also, since for any $u,v\in V$, one has $ R_{uv}^{M} \le 3\Gamma \cdot R_{uv}^{G}$, we get
	 	\begin{equation*}
	 	R_{u_{\ell+1},C^*}^{G}\ge \frac{r}{6\Gamma}
	 	\end{equation*} 
	 	i.e. the condition in line \ref{line:if-BC} is not satisfied for $u_{\ell+1}$, in the invocation of \textsc{BallCarving} with radius $\frac{r}{3\Gamma}$ and effective resistance metric $G$ and ordering $\pi$, thus, $u_{\ell+1}$ is a ball center in this invocation (i.e., line \ref{when_ballcarve} is called). 
	 \end{description}
	 
	Thus, if the \textsc{BallCarving} invocation on $\wt{G}$ with radius $r$ returns $t$ partitions, the invocation on $G$ with radius $\frac{r}{3\Gamma}$ and ordering $\pi$ returns $t'\ge t$ partitions. 
	 In turn, by invoking Theorem \ref{thm:numpartitions} with $\beta=\frac{r}{3\Gamma}$, the number of partitions created by \textsc{BallCarving}($S\cdot B , E' , \frac{r}{3\Gamma}, G$) using any ordering, is upper-bounded by $\wt{O}\left(\sqrt{k}+\frac{k\cdot\Gamma}{r\cdot d}\right)$. So, the claim holds.

	 \paragraph{Run-time analysis:}
	 By Lemma \ref{lem:Low-Conn-E}, running \textsc{SpanningForest} procedure on the sketch, in order to find the connected components, needs $\wt{O}(n)$ time and space (see line~\ref{line:span-forest}). 
	
	 Each calculation of $R_{u,C^*}^{M}$ in line \ref{line:max-eff-res} takes $O(|C^*|\polylog(n))$. We show an upper-bound on the number of times we need to compute this quantity. Two possible cases can happen when we reach line~\ref{line:if-BC}:
	 
	 {\bf Case 1: $R_{u,C^*}^{M}\le r/2$.} In this case, since we are removing all vertices $v\in C^*$ from $V_{\text{active}}$, we will never calculate $R_{w,C^*}^{\wt{G}}$ for any $w\in C^*$, after this point. Therefore, when $R_{u,C^*}^{M}\le r/2$, we compute $R_{u,C^*}^{M}$ at most once for each $i \in [\ell]$ such that $C^*=C_i$ (see line \ref{line:span-forest}).

	 {\bf Case 2: $R_{u,C^*}^{M}> r/2$.} In this case, vertex $u$ will become a ball center, and since $|V_{\text{active}}|\le n$, in total this takes $\wt{O}(n)$ time to compute $R_{uv}^{M}$ in line~\ref{line:cal-Ruv-M} for all $v \in V_{\text{active}} \setminus \{u\}$. Note that by the correctness argument above, we get at most $\wt{O}\left(\sqrt{k}+\frac{k\Gamma}{r\cdot d}\right)$ ball centers. 
	 
	  So, in total we spend $\wt{O}\left(\sum_{i\in [\ell]}|C_i|+n\sqrt{k}+n\frac{k\Gamma}{r\cdot d}\right)=\wt{O}\left(n\sqrt{k}+n\frac{k\Gamma}{r\cdot d}\right)$ time in line \ref{line:max-eff-res}.
	 Also, in total it takes $\wt{O}\left(n\sqrt{k}+n\frac{k\Gamma}{r\cdot d}\right)$ time to execute lines \ref{line:1st-BC} - \ref{line:last-BC}. 
	 So, all in all, it takes $\wt{O}\left(n\sqrt{k}+n\frac{k\Gamma}{r\cdot d}\right)$ to run \textsc{BallCarving}($S\cdot B , E' , r, \wt{G}$).
\end{proofof}

\subsection{Partitioning graphs into clusters with light boundaries}
\label{sec:PG}

The main result of this section is the following
\begin{lemma}\label{thm:light-clusters}
For every graph $G=(V, E)$, every $\Delta, \beta>0$ and integer $k\geq 1$, if:
\begin{enumerate}
\item $\Delta < \beta/(400\sqrt{k})$ and $\beta<1/\log n$.
\item $V$ admits a partition, $V=C_1\cup \ldots \cup C_k$ 
such that for any $i\in [k]$,  $\text{diam}_{\text{eff}}^{\text{Ind}}(C_i)\leq \Delta$.
\end{enumerate}
Then there exists a subset $A\subseteq [k]$, such that $|A|\leq 8\sqrt{k}\log n$ and the number of edges with effective resistance at least $\beta$ connecting two distinct clusters with indices in $[k]\setminus A$  is upper-bounded by $(k\log n)/\beta$.
\end{lemma}

Intuitively, if we have our graph partitioned into $k$ balls with very  small effective resistance radius $O(\beta/\sqrt{k})$, we want to argue that the effective resistances of the edges between these balls should look approximately the same as if we simply  collapsed these balls and looked at the effective resistances on the collapsed $k$-node graph. On a $k$ node graph, the sum of effective resistances is always $k-1$, and thus, the number of edges with effective resistance $\ge \beta$  is less than $k/\beta$. Here we show that this fact holds approximately for our clustered graph: as long as we remove a small subset of $\tilde O(\sqrt{k})$ clusters, the number of edges with effective resistance $\ge \beta$ between the remaining clusters is bounded by $\tilde O(k/\beta)$. Formally:

We will prove Lemma \ref{thm:light-clusters} by showing that any flow in the collapsed graph (with $k$ nodes, corresponding  to the clusters $C_1,\ldots, C_k$) can be routed with approximately  the same energy  in the original graph, witnessing that the effective resistances in the original graph can be upper bounded as a function of those in the collapsed graph. We start with a technical lemma: on any graph with low maximum effective resistance (eventually, our low effective resistance diameter clusters) a set of vertex demands can be satisfied with a low energy flow:
\begin{lemma}\label{lm:resistance-diameter-routing}
For every graph $G=(V, E)$, every vector $\sigma \in \R^{|V|}$ of demands such that $\sigma^\top  \mathbf{1}=0$, if $\Delta=\max_{u, v\in V} R_{uv}$, there exists a $\sigma$-flow $f$ with $\sum_{e\in E} f_e^2\leq \Delta\cdot \frac{||\sigma||^2_1}{4}$.
\end{lemma}
In order to prove Lemma \ref{lm:resistance-diameter-routing}, we need the following simple lemma. 
\begin{lemma}\label{lm:decomp}
	For every graph $G=(V, E)$, for every vector $\sigma\in \R^{|V|}$ such that $\sigma^\top  \mathbf{1}=0$ there exists a positive integer $r$ and for $i \in [r]$, $(s_i, t_i)\in V\times V$ and $\alpha_i\in \R^+ $ such that 
	$$\sum_{i=1}^r \alpha_i \mathbf{b}_{s_i t_i}=\sigma \text{ and }||\alpha||_1=||\sigma||_1/2.$$
\end{lemma}
The proof of this lemma is deferred to the Appendix \ref{app:algo}. We now give the proof of Lemma \ref{lm:resistance-diameter-routing}.

\begin{proofof}{Lemma \ref{lm:resistance-diameter-routing}}
	The potential difference induced by the demand vector $\sigma$ on any pair of vertices $(u,v)\in V\times V$, is $\mathbf{b}_{uv}^\top L^+\sigma$. By Lemma~\ref{lm:decomp} the vector $\sigma$ can be written as follows
	\begin{align}
		\sigma&=\sum_i \alpha_i \mathbf{b}_{s_it_i}\label{sigma_linear}\\
		\text{such that } (s_i,t_i)\in V\times V &,\, \alpha_i\in \R^+ \text{ and }\sum_i\alpha_i=\frac{||\sigma||_1}{2}\label{sigma_such}.
	\end{align} 
	Then, we have
	\begin{align*}
		\left|\mathbf{b}_{uv}^\top L^+\sigma\right| &= \left|\mathbf{b}_{uv}L^+\left(\sum_i \alpha_i \mathbf{b}_{s_it_i}\right)\right|&&\text{By equation \eqref{sigma_linear}}\\
		&\le \sum_i\alpha_i\left|\mathbf{b}_{uv}L^+\mathbf{b}_{s_it_i}\right|&&\text{By triangle inequality}\\
		&\le R_{uv} \sum_i \alpha_i&&\text{By Fact \ref{fact:maxphi}}\\
		&= R_{uv} \cdot \frac{||\sigma||_1}{2}&&\text{By conditions in \eqref{sigma_such}}\\
		&\le \Delta \cdot \frac{||\sigma||_1}{2}&& \text{By effective resistance diameter of the graph}
	\end{align*}
			
On the other hand, if vector $f$ is the electrical flow vector induced by demand vector $\sigma$ we can write $f=BL^+\sigma$. Hence,	
	\begin{align*}
	\sum_{e\in E} f(e)^2 = ||f||_2^2 &= \sigma^\top  L^+ B^\top  BL^+ \sigma&&\text{Since $f=BL^+\sigma$}\\
	&= \sigma^\top  L^+ \sigma&&\text{Since $B^\top B=L$}\\
	&= \sum_i \alpha_i \mathbf{b}_{s_it_i}^\top  L^+ \sigma	&&\text{By equation \eqref{sigma_linear}}\\
	&\le \sum_i \alpha_i \Delta \cdot \frac{||\sigma||_1}{2}&& \text{By the discussion above}\\
	&= \Delta\cdot \frac{||\sigma||^2_1}{4} && \text{By the conditions in \eqref{sigma_such}}
	\end{align*}
\end{proofof}

\begin{proofof}{Lemma~\ref{thm:light-clusters}}

We now prove Lemma \ref{thm:light-clusters} by exhibiting an algorithm that removes any cluster with a large number of outgoing high effective resistance edges to nodes in other clusters. We prove that  the algorithm removes just $\tilde O(\sqrt{k})$ clusters, which gives us the lemma.

\begin{algorithm}[H]
	\caption{\textsc{LabelClusters}$(G,C_1, \ldots, C_k, \beta)$\Comment{Note that we do not run this algorithm, we just need it for the analysis.}} {\label{alg:cluster-labeling}} 
	\begin{algorithmic}[1]
		\State Remove all edges of resistance $\leq \beta$ between different clusters, let $E_0\gets$ set of remaining inter-cluster edges.\label{firstStep}
		\State $j\gets 0$, $A\gets \emptyset$
		\While{$|E_j| > \frac{k\log n}{\beta}$} 
		\State $\mathcal I \gets \left\{i\in [k]\setminus A: |\delta_{E_j}(C_i)| \geq \frac{|E_j|}{\sqrt{k}}\right\}$
		\State $A\gets A\cup \mathcal{I}$
		\State $E_{j+1}\gets E_j\setminus \bigcup_{i\in \mathcal{I}} (C_i\times V)$ \Comment{Remove heavy clusters from the graph, obtaining graph $G_{j+1}$}
		\State $j\gets j+1$
		\State {\bf If}~$|E_j|\geq |E_{j-1}|/2$ {\bf then break}
		\EndWhile
		\Return $(A, E_j)$\Comment{Heavy cluster indices and set of heavy edges between light clusters}
	\end{algorithmic}
\end{algorithm}

\begin{definition}
	Given a clustering of vertices $V$ of the graph $G$ into $k$ clusters $C_1,\ldots, C_k$, suppose that $(A,E_j)$ is the output of \textsc{LabelClusters}$(G,C_1, \ldots, C_k, \beta)$. We call cluster $C_i$ a ``heavy cluster", if $i\in A$, and a``light cluster" otherwise.  \Navid{I think it's not a good idea to have a definition in the middle of a proof}
\end{definition}
Note that in Algorithm \ref{alg:cluster-labeling} if a cluster is added to set $A$ , the set of heavy clusters, it never gets removed from it during the execution of the algorithm.  

In line \ref{firstStep}, Algorithm \ref{alg:cluster-labeling}
first removes edges that connect different clusters and have effective resistance smaller than $\beta$. $E_0$ is the set of remaining inter-cluster edges. Note that by Rayleigh's monotonicity law (Fact~\ref{fact:mon}) the effective resistance of edges in $E_0$ in the resulting graph is at least $\beta$.

In lines~3 to~8 the algorithm then repeatedly performs the following operation for $j\geq 0$: remove all clusters $C_i$ from the current graph $G_j$ that have dense boundaries. Formally, remove a cluster if  the number of edges going from it to nodes in other clusters is at least $|E_j|/\sqrt{k}$. If this operation does not decrease the number of inter-cluster edges in the graph by more than a factor of two, stop. Note that since the number of inter-cluster edges in the remaining graph decreases by at least a factor of two in every round, the number of rounds is no larger than $4\log_2 n$. The number of clusters remaining in every round is trivially at most $k$, and by Markov's inequality the number of clusters removed as heavy clusters at every step is no larger than $2\sqrt{k}$. Thus, overall at most $8\sqrt{k}\log_2 n$ clusters are removed (and hence returned as heavy clusters). It remains to show the number of edges connecting distinct light clusters is appropriately small.

Let $J$ denote the last value $j$ before Algorithm~\ref{alg:cluster-labeling} terminates, and let $(A, E_J)$ be its output.  If the algorithm terminated because $|E_J|\leq (k\log n)/\beta$ (see line~3 of Algorithm~\ref{alg:cluster-labeling}), then we are done: the number of high effective resistance edges connecting nodes across clusters outside $A$ is equal to $|E_j|$ and thus bounded by $(k\log n)/\beta$.

We now show that the other termination condition: that $|E_J|\geq |E_{J-1}|/2$ (see line~8 of Algorithm~\ref{alg:cluster-labeling}) can never happen. We prove that, if the algorithm terminates in this way, at least one of the remaining edges has effective resistance in $G$ strictly less than $\beta$, a contradiction since we removed all low effective resistance in line \ref{firstStep} of the algorithm and since by Rayleigh's monotonicity law, the effective resistances only increase as we remove more nodes.

 Intuitively, we show that one can think of the clusters as supernodes in an auxiliary graph $H$ on at most $k$ nodes. A graph $H$ on $k$ nodes cannot have more than $(k-1)/\beta\leq k/\beta$ edges of effective resistance at least $\beta$, so if we could show that effective resistances of edges in $E_J$  are not much larger than effective resistances of the same edges in $H$, we are done. Instead of that we will show it on a sampled version of $E_J$ which we call $E_J'$ (See {\bf Step 1} below for more details). Formally, the proof proceeds over three steps. 
 \Cam{Give intuition for *why* we are doing sampling.}

\Cam{$\delta_{E_J}(C_i)$  is used in Algorithm \ref{alg:cluster-labeling} before it is defined.}

{\bf Step 1.} 
Let $\delta_{E_J}(C_i)$ denote the set of edges of $E_J$ with one end point in cluster $C_i$. Note that for every cluster $C_i$ remaining we have $|\delta_{E_J}(C_i)|\leq |E_{J-1}|/\sqrt{k}\leq 2|E_J|/\sqrt{k}$. 
Let $E^*$ be the set of inter-cluster edges obtained by sampling each $e\in E_J$ independently with probability $\min\{1, (10k/\beta)/|E_J|\}$, and let $\delta_{E^*}(C_i)$ denote the set of edges in $E^*$ with one end point in cluster $C_i$. 

Since $\beta<1/\log n$ by assumption, one has by an application of Chernoff bounds
$|E^*|\geq 5k/\beta$ with high probability. At the same time by an application of Chernoff bounds, using the assumption that $\beta<1/\log n$, we have that for every cluster $i\in [k]\setminus A$ 
$$
{\bf Pr}\left[|\delta_{E^*}(C_i)\cap E^*|>40\sqrt{k}/\beta\right]<n^{-2},
$$
and hence by a union bound  one has $|\delta_{E^*}(C_i)\cap E^*|<40\sqrt{k}/\beta$ simultaneously for all $i\in [k]\setminus A$. We condition on these high probability events in what follows.

{\bf Step 2: defining an auxiliary graph $H$.} Now define a graph $H=(V_H, E_H)$ as follows. Nodes in $V_H=[k]\setminus A$ correspond to clusters $C_i$, $i\in [k]\setminus A$, and the multiplicity of an edge $(a, b)\in H\times H,a\neq b,$ equals the number of edges $(u, v)\in E^*$ such that $u\in C_a$ and $v\in C_b$. In other words, consider the graph induced by the clusters $[k]\setminus A$ in $G$, and contract every cluster to a supernode.  Since $H$ has no more than $k$ nodes, and $|E^*|\geq 5k/\beta$, there must exist an edge $e^*=(a^*, b^*)\in E_H$ with effective resistance at most $\beta/5$ (since the sum of effective resistances in any $k$-node graph is $k-1$. Let $e=(u, v)$ denote one of the corresponding edges in $G$, i.e. $u\in C_a$, $v\in C_b$ and $(u, v)\in E^*$. In the next step we show that the resistance of this edge in $G$ is strictly smaller than $\beta$, providing the required contradiction.

{\bf Step 3: relating resistances in $H$ to resistances in $G$.} Since $e^*=(a^*, b^*)$ has effective resistance at most $\beta/5$ in $H$, by  Fact \ref{fact:minEnergyFlow} there exists a flow $f^*:E_H\to \mathbb{R}_+$ that ships one unit of current from $a^*$ to $b^*$ such that $\sum_{g\in E_H} \left(f^*(g)\right)^2\leq \beta/5$, i.e. the energy of $f^*$ is no larger than $\beta/5$. We now extend this flow to a flow $f$ that ships one unit of current from $u$ to $v$ in $G$ and has comparable energy. This witnesses that $(u,v)$ has low effective resistance, giving our contradiction.

For every cluster $C_i, i\in [k]\setminus A$ let $\sigma_i:C_i\to \mathbb{R}$ denote the demands imposed by $f^*$. Specifically, for every $u\in C_i$ let 
$$
\sigma_i(u):=\sum_{\substack{e\in \delta_{E^*}(u)\\ e\in (V\times\{u\} )\cap E^*}} f^*(e) -\sum_{\substack{e\in \delta_{E^*}(u)\\ e\in (\{u\}\times V )\cap E^*}} f^*(e) 
$$
be the total flow going into node $u$ in $f^*$ (this can be either positive or negative). 

\Cam{I understand the split summation now. But and explanation of the fact that we view edges as having an arbitrary direction is needed in the intro. Otherwise the assumption is that on an undirected graph, $(V \times \{u\}) = (\{u\}\times V)$.}

\Cam{ I think it is confusing is that $f^*(e)$ is a flow over a multigraph $H$, which we don't really have any definitions for. E.g. we haven't defined how the Laplacian works, how effective resistances work etc. I think it makes more sense for f to be a flow over a weighted graph $H$ (where weight of an edge is equal to number of edges in the original graph between the two clusters. Then to define a corresponding flow $f^*_G$, in which you take the flows on the edges in $H$ and split them evenly across each corresponding edge. Then we can talk about the demand $u$ induced by $f^*_G$, can formally write $f = f^*_G + \sum f_i$ etc.}

Since $C_i$ has effective resistance diameter at most $\Delta$ as an induced subgraph by assumption,
 we get by Lemma~\ref{lm:resistance-diameter-routing} there exists a $\sigma_i$-flow $f_i$ in $C_i$ with $||f_i||_2^2\leq \Delta ||\sigma_i||_1^2/4$.  Let 
$$
f:=f^*+\sum_{i\in [k]\setminus A} f_i,
$$
and note that it ships $1$ unit of current from $u$ to $v$ in $G$.

We have 
\begin{equation}\label{eq:904hg9hg4g}
||f||_2^2=||f^*||_2^2+\sum_{i\in [k]\setminus A} ||f_i||_2^2\leq \beta/5+\Delta\sum_{i\in [k]\setminus A} ||\sigma_i||_1^2/4.
\end{equation}
At the same time one has, for every $i\in [k]\setminus A$, 

\begin{equation}\label{eq:gh42g9gh}
\begin{split}
\sum_{u\in C_i}  \sum_{e\in \delta_{E^*}(u)} (f^*(e))^2&\ge \frac{\left(\sum_{e\in \delta_{E^*}(C_i)} \left|f^*(e)\right| \right)^2}{\left|\delta_{E^*}(C_i)\right|}\\
&\ge \frac{||\sigma_i||_1^2}{40\sqrt{k}/\beta}
\end{split}
\end{equation}
where the last transition is due to conditioning on the event that $|\delta_{E^*}(C_i)|\le40\sqrt{k}/\beta$ for all $i\in [k]\setminus A$.

Summing~\eqref{eq:gh42g9gh} over all $i\in [k]\setminus A$, we get

\begin{equation}\label{eq:294hgg}
\begin{split}
\sum_{i\in V_H} ||\sigma_i||_1^2&\leq (40\sqrt{k}/\beta) \sum_{i\in V_H}\sum_{u\in C_i}  \sum_{e\in \delta_{E^*}(u)} (f^*_e)^2\\
&\leq (80\sqrt{k}/\beta) \sum_{e\in E^*} (f^*_e)^2\\
&= (80\sqrt{k}/\beta) ||f^*||_2^2\\
\end{split}
\end{equation}
Where the factor of two in the second line is due to each edge being counted twice when summing over $\delta_{E^*}(u)$.
Putting~\eqref{eq:904hg9hg4g} together with~\eqref{eq:294hgg}, we obtain 
\begin{align*}
||f||_2^2&\leq \beta/5+\Delta\sum_{i\in [k]\setminus A} ||\sigma_i||_1^2/4\\
&\leq \beta/5+20\Delta \frac{\sqrt{k}}{\beta}||f^*||_2^2\\
&\leq \beta/5+4\Delta\sqrt{k}\\
&\leq \beta/5 + \beta/100\\
&<\beta,
\end{align*}
where we used the assumption $\Delta < \beta/(400\sqrt{k})$. This provides the required contradiction, and completes the proof of the theorem.
\end{proofof}

\subsection{Ball growing in effective resistance metric}\label{sec:BG}

Before proving Theorem~\ref{thm:numpartitions}, we prove the following auxiliary lemma:

\begin{lemma}\label{lm:large-cut}\Navid{What about changing the notation of connected components?}
	For every graph $G=(V, E)$, $|V|=n$, every integer $d\geq 1$, if $E'\subseteq E$ is a set of edges that contains all edges in $G$ with connectivity at most $d$, and $C_1,\ldots, C_r$ are the connected components in the graph $G'=(V, E\setminus E')$ (i.e., the graph $G$ with the edges in $E'$ removed), then:
	
	For every $u,v\in V$, if $u$ and $v$ belong to the same connected component $C^*$ in $G'$, for every cut $S$ that $u\in S$ and $v\notin S$ one has 
	$$
	|E\cap (S\times V\setminus S)|> d.
	$$
\end{lemma}
\Cam{We've generally been using the notation $(S \times A)$ to denote the set of edges from $S$ to $A$ in a bunch of places. This should be  defined in the prelims.}
\begin{proof}

Since $u$ and $v$ are in the same connected component in $G'$, there exists a path $p=(u,w_1,w_2,\dots,w_l,v)$ in graph $G'$ from $u$ to $v$, where for any $i \in [l]$, $w_i\in C^*$. Clearly, any cut $(S,V\setminus S)$ that separates $u$ and $v$, cuts at least one of the edges of path $p$.  On the other hand, by definition of $E'$, if an edge $e \notin E'$, then its connectivity is strictly larger than $d$. Therefore, all the edges in $p$ have connectivity strictly larger than $d$. 
\end{proof}

\Cam{Ball notation shuold generally  be subscripted by the graph.}

\Cam{I reworked the structure of this lemma since I was having trouble understanding it. Probably  worth checking if there is time.}

\Cam{Add intuition that this is hard because one of the balls is not induced.}
\begin{lemma}\label{lm:triangle-sets}
For every $G=(V, E)$, $u\in V$, and radius $r\in \mathbb{R}_+$ if $C\subseteq V\setminus \B_G(u, r)$, $q=|E\cap (\B_G(u, r)\times C)|$ and $\text{diam}_{\text{eff}}^{\text{Ind}}(C)\le \Delta$ for some $\Delta\in \R_+$, then for any $v\in C$, 
$$R_{uv}\leq r+1/q+\Delta.$$
\end{lemma}
\begin{proof}
Suppose that $\varphi \in \R^{|V|}$ is the vector of potential induced on vertices when one injects one unit of flow to vertex $u$ and removes it from vertex $v$. In that case we have
$\varphi(u)-\varphi(v)=\mathbf{b}_{uv}^\top L^+ \mathbf{b}_{uv}=R_{uv}$.
 Simply, by Fact \ref{fact:maxphi} for any vertex $w\in \mathbf{B}_G(u,r)$, we have $|\mathbf{b}_{uw}^\top L^+\mathbf{b}_{uv}|\le R_{uw}\le r$. Again, by Fact \ref{fact:maxphi} for any vertex $\wt{w} \in C$, we have $|\mathbf{b}_{v\wt{w}}^\top L^+\mathbf{b}_{uv}|\le R_{v\wt{w}}\le \Delta$. Let $E^*= E\cap (\B_G(u, r)\times C)$ (by assumption $|E^*| = q$). Let $${B}^*:=\{w: w\in V \text{ and } |\mathbf{b}_{uw}^\top L^+\mathbf{b}_{uv}|\le r\}.$$ 
 Note that by Fact \ref{fact:maxphi}, $\B_G(u,r)\subseteq B^*$. Let $\partial B^*  = E \cap (B^*\times  V\setminus B^*)$ be the set of edges connecting $B^*$ to the rest of the graph. 
 Our proof proceeds by considering  two cases:
 
 \medskip
 \textbf{Case 1: $E^* \not\subseteq \partial B^*$.}
 \medskip
 
 In this case, there must be an edge $e = (v_1,v_2)\in E^*$ such that $e\notin \partial B^*$. Since $v_1  \in \B_G(u,r) \subseteq B^*$ it  must be that $v_2 \in B^*$ or else we would have $e \in \partial B^*$.
 
 \Cam{Later $\partial$ it is defined to denote a set of clusters rather than a set of edges. Maybe should use a separate notation?}\Navid{Fixed, please check if you have time}
 
 Thus,
 \begin{align*}
 R_{uv} &= \mathbf{b}_{uv}^\top L^+ \mathbf{b}_{uv}\\
 &=\varphi(u)-\varphi(v)\\
 &=(\varphi(u)-\varphi(v_2))+(\varphi(v_2)-\varphi(v))\\
 &\le |\varphi(u)-\varphi(v_2)|+|\varphi(v_2)-\varphi(v)|\\
 &= |\mathbf{b}_{uv_2}^\top L^+\mathbf{b}_{uv}|+|\mathbf{b}_{v_2v}^\top L^+\mathbf{b}_{uv}|\\
 &\le r+\Delta&& \text{Since } v_2\in B^* \text{ and } v_2 \in C \\
 &\le r+1/q+\Delta,
 \end{align*}
 which completes the lemma in this case.
 
  \medskip
 \textbf{Case 2: $E^*\subseteq \partial B^*$.}
 \medskip
 
 To prove the lemma in this case we show that $E^* \subseteq \partial B^*$  implies that there must be  a pair of vertices $(v_1 \in \mathbf{B}_G(u,r),v_2\in C)$ with $(v_1,v_2) \in E^*$ and $|\mathbf{b}_{v_1v_2}^\top L^+\mathbf{b}_{uv}|\le 1/q$. Given this, the lemma holds since: 
\begin{align*}
R_{uv} &= \mathbf{b}_{uv}^\top L^+ \mathbf{b}_{uv}\\
&=\varphi(u)-\varphi(v)\\
&=(\varphi(u)-\varphi(v_1))+(\varphi(v_1)-\varphi(v_2))+(\varphi(v_2)-\varphi(v))\\
&\le |\varphi(u)-\varphi(v_1)|+|\varphi(v_1)-\varphi(v_2)|+|\varphi(v_2)-\varphi(v)|\\
&= |\mathbf{b}_{uv_1}^\top L^+\mathbf{b}_{uv}|+|\mathbf{b}_{v_1v_2}^\top L^+\mathbf{b}_{uv}|+|\mathbf{b}_{v_2v}^\top L^+\mathbf{b}_{uv}|\\
&\le r+1/q+\Delta 
\end{align*}
where the last inequality is by the fact that $v_1 \in \B_G(u,r)$ so $ |\mathbf{b}_{uv_1}^\top L^+\mathbf{b}_{uv}| \le R_{uv_1} \le r$ , $|\mathbf{b}_{v_1v_2}^\top L^+\mathbf{b}_{uv}|\le 1/q$, and $v_2\in C$ so $|\mathbf{b}_{v_2v}^\top L^+\mathbf{b}_{uv}| \le R_{v_2v}  \le \Delta$.

Thus is just remains to prove that  $e = (v_1 \in \mathbf{B}_G(u,r),v_2\in C) \in E^*$ with $|\mathbf{b}_{v_1v_2}^\top L^+\mathbf{b}_{uv}|\le 1/q$ must exist. Assume for the sake of contradiction that:
\begin{align}
\text{for all pairs } (v_1,v_2)\in E^*, |\mathbf{b}_{v_1v_2}^\top L^+\mathbf{b}_{uv}|> 1/q \label{assumption_flow}
\end{align} 
 
 Note that since $u$ is the only vertex in the graph to which the current is injected, its potential is the maximum potential in the graph. Thus, for any $w\in V$, one has
 \begin{align*}
 \varphi(w)&=\varphi(u)-\left(\varphi(u)-\varphi(w)\right)\\
 &=\varphi(u)-|\mathbf{b}_{uw}^\top L^+\mathbf{b}_{uv}|
 \end{align*} 
 Hence, by the definition of ${B}^*$, we will have the following facts:
 \begin{enumerate}
 	\item{For any $w \in B^*$, $\varphi(w)\ge \varphi(u)-r$}
 	\item{For any $w \in V\setminus B^*$, $\varphi(w)< \varphi(u)-r$}
 \end{enumerate}

  So, for any edge $e= (u_1,u_2) \in \partial B^*$, the direction of the flow is from $u_1$ to $u_2$.
 Also, since by assumption for all pairs $(v_1,v_2)\in E^*$, $|\mathbf{b}_{v_1v_2}^\top L^+\mathbf{b}_{uv}|> 1/q$, 
if we assume $\mathbf{f}$ is the flow vector of this setting, we have the following inequalities for the amount of outgoing flow from $B^*$:
\begin{align*}
\sum_{\substack{e=(u,v)\in \partial B^*\\ u \in B^*,v \in V\setminus B^*}}\mathbf{f}(e)-\sum_{\substack{e=(u,v)\in \partial B^*\\ u \in V\setminus B^*,v \in B^*}}\mathbf{f}(e)&=\sum_{e\in \partial B^*}|\mathbf{f}(e)|&&\text{Since as argued, flow moves strictly out of $B^*$.}\\
&\ge \sum_{e\in  E^*}|\mathbf{f}(e)| &&\text{Since, by  assumption of Case 2, $E^*\subseteq \partial B^*$}\\
&= \sum_{e\in  E^*} |\mathbf{b}_{e}^\top L^+\mathbf{b}_{uv}|&&\text{By the fact that the graph is not weighted}
\\
&> q \cdot \frac{1}{q}&&\text{Since $|E^*|=q$ and assumption \eqref{assumption_flow}} \\
&=1
\end{align*}
 which means more than one unit of current is passing through cut $(B^*,V\setminus B^*)$, which is a contradiction.
 
 \Cam{We can add later  but  the above statement needs a proof I think. That the electrical flow, since it minimizes energy should have no cycles, and so the total flow across any  cut  between $u$ and $v$ is $1$.}
\end{proof}

\Cam{What set of clusters? Ball growing? Ball carving? Or the partition we have been assuming exists in all the theorems?}
\begin{definition}
	Let $\mathcal C:=\{C_1,\ldots, C_k\}$ to be the set of clusters that $G$ is partitioned into (recall that some of the clusters may be singletons). Also, define 
	$$
	{\mathcal C}(u,r):=\left\{C\in \mathcal C: \text{s.t. } \forall v \in C , v \in \B(u, r)\right\}
	$$ 
	to be the subset of clusters that are completely contained in the ball of radius $r$ in effective resistance metric centered at $u$.

\end{definition}
\begin{definition}
	{\em cluster boundary} of an effective resistance ball $\B(u, r)$ around $u$ of radius $r$ is defined as follows:
	\begin{equation}\label{eq:cluster-boundary}
	\partial \B(u, r,\mathcal C):=\left\{C\in \mathcal{C} \setminus {\mathcal C}(u,r): \exists v_1 \in C \text{ such that } (v_1, v_2)\in E\text{~for some~}v_2\in \B(u, r)\right\}.
	\end{equation}
	
\end{definition}

\begin{algorithm}[H]
	\caption{\textsc{BallGrowing}$(G,\mathcal{C},\Delta,u,r)$\Comment{Note that we do not run this algorithm, we just need it for the analysis.}} {\label{alg:ball-growing}}

	\begin{algorithmic}[1]
		\State $t \gets 0$
		\State $r_t \gets 0$
		\While {$r_t < r$}
		\State $\mathcal{C}_0,\mathcal{C}_1,\dots \gets \emptyset$
		\For {$C\in \partial \B(u, r_t)$}
		\State $\wt d_C\gets \left|\{(u, w)\in E: u\in C , w\in \B(u, r_t)\}\right|$
		\State $j\gets \lfloor\log_2\wt d_C\rfloor$
		\State $\mathcal{C}_j\gets \mathcal{C}_j \cup \{C\}$
		\EndFor
		\State Find $j^*$ such that $|{\mathcal C}_{j^*}| \geq \frac{d}{2^{j^*}(2+2\log_2 n) }$
		\State $r_{t+1}\gets \min\{r_t+2^{-j^*}+\Delta \ , r \}$
		\State $t \gets t+1$
		\EndWhile
	\end{algorithmic}
\end{algorithm}	
\begin{lemma}\label{lem:BGrowing}
	For every graph $G=(V, E)$ if:
	\begin{enumerate}
		\item The minimum degree in $G$ is lower bounded by $d\geq 1$.
		\item For some $0<r\le 1/\log n$ and integer $k \ge 1$ the vertex set $V$ admits a partition $V=C_1\cup \ldots\cup C_k$ such that for every $i\in [k]$ the subgraph induced by $C_i$ has effective resistance diameter bounded by $\Delta\in (0, 1)$, i.e., $\text{diam}_{\text{eff}}^{\text{Ind}}(C_i)\le \Delta$. \Navid{Don't we need the condition on $\Delta$?}

		\item Edge set $E'$ contains all the edges of connectivity no more than $d$ in graph $G$.	
	\end{enumerate} and $\wt{C}_1,\ldots, \wt{C}_r$ are the connected components in the graph $G'=(V, E\setminus E')$ (i.e., the graph $G$ with the edges in $E'$ removed), the following holds:
	
		 For every $u\in V$, if $C^*$ is the connected component that $u$ belongs to in $G'$ and there exists $v\in C^*$ such that $R_{uv}> \beta$, then for any $t$ with $r_t< \beta$ in algorithm \textsc{BallGrowing}$(G,\mathcal{C},\Delta,u,r)$ we have:
	\begin{equation}
	|\mathcal{C}(u, r_t)|\geq (r_t-t\cdot \Delta)\cdot \frac{d}{2+2\log_2 n}.
	\end{equation}
\end{lemma}
\begin{proof}
We prove the claim by induction on $t$.
\begin{description}
		\item[Base: $t=0$] The base of the induction is provided by $t=0$. We have $|\mathcal{C}(u, r_t)|\geq 0= r_t\cdot d/(2+2\log_2 n)=0$, as required.
	
	\item[Inductive step: $t\to t+1$] 
	
	For every $C\in \partial \B(u, r_t, \mathcal C)$ (recall the definition of cluster boundary of a ball in~\eqref{eq:cluster-boundary}) let $\wt d_C$ denote the degree of $C$ in $\B(u, r_t)$, i.e. 
	$$
	\wt d_C:=\left|\{(v, w)\in E: v\in C , w\in \B(u, r_t)\}\right|,
	$$
	and note that, as long as Algorithm~\ref{alg:ballcarving} grows a ball from $u$, one has by Lemma~\ref{lm:large-cut}:
	\begin{equation}{\label{cut_assumption}}
	\sum_{C\in \partial \B(v, r, \mathcal C)} \wt d_C\geq d.
	\end{equation}

	We now classify clusters $C\in \mathcal C$ according to their degree $\wt d_C$ in $\B(u, r_t)$, defining for each $j=0, 1, \ldots, \log_2 n$ 
	$$
	{\mathcal C}_j:=\{C\in {\mathcal C}: \wt d_C\in [2^j, 2^{j+1})\}.
	$$
	Since $\sum_{j=0}^{\log_2 n} |{\mathcal C}_j| 2^j\geq (1/2) \sum_{C\in {\mathcal C}} \wt d_C$, there exists at least one $j^*\in \{0,1, 2,\ldots, \log_2 n\}$ such that 
	\begin{equation}\label{eq:c-star-large}
	|{\mathcal C}_{j^*}| 2^{j^*}\geq \frac{d}{2+2\log_2 n}.
	\end{equation}
	We let 
	$$
	r_{t+1}=\min\{r_t+2^{-j^*}+\Delta \ , r \} 
	$$
	If $r_{t+1}=r$, we stop the construction. If the construction is not stopped, we note that by Lemma~\ref{lm:triangle-sets} for every $v \in C\in {\mathcal C}_j$ one has $R_{uv}\leq r_t+1/\wt d_C+\Delta$. Indeed, this is exactly the conclusion of Lemma~\ref{lm:triangle-sets} with $r=r_t$, where we note that $q=\wt d_C$.

	Since for every $C\in \partial \B(u, r_t, \mathcal C)$ one has $C\cap \mathcal{C}(u, r_t)=\emptyset$, we get
	\begin{align}\label{eq:i4ghb2bg23g}
	|{\mathcal C}(u, r_{t+1})|&\geq |{\mathcal C}(u, r_t)|+|\mathcal C_{j^*}|&&\\
	&\geq |{\mathcal C}(u, r_t)|+ 2^{-j^*}\frac{d}{2+2\log_2 n}&&\text{By~\eqref{eq:c-star-large}}\\
	&\geq (r_t-t\cdot \Delta)\cdot \frac{d}{2+2\log_2 n}+2^{-j^*}\frac{d}{2+2\log_2 n}&&\text{By the inductive hypothesis}
	\end{align}
	Since $r_{t+1}=r_t+2^{-j^*}+\Delta$ , one has $r_t=r_{t+1}-2^{-j^*}-\Delta$, and hence, substituting into~\eqref{eq:i4ghb2bg23g}, we get
	\begin{equation*}
	\begin{split}
	|{\mathcal C}(u, r_{t+1})|&\geq (r_t-t\cdot \Delta)\cdot \frac{d}{2+2\log_2 n}+2^{-j^*}\frac{d}{2+2\log_2 n}\\
	&\geq ((r_{t+1}-2^{-j^*}-\Delta)-t\cdot \Delta)\cdot \frac{d}{2+2\log_2 n}+2^{-j^*}\frac{d}{2+2\log_2 n}\\
	&= ((r_{t+1}-(t+1)\cdot \Delta)\cdot \frac{d}{2+2\log_2 n},\\
	\end{split}
	\end{equation*}
	
\end{description}
	establishing the inductive claim.

\end{proof}
	\begin{proofof}{Theorem~\ref{thm:numpartitions}}
		We consider the following experiment. Fix a vertex $u\in V$. We will iteratively grow a ball around $u$ as described in algorithm \ref{alg:ball-growing} (\textsc{BallGrowing}$(G,\mathcal{C},\Delta,u,\frac{\beta}{2})$) in effective resistance metric and show that at least one of the following statements about $\mathbf{B}(u,\beta/2)$ holds
		\begin{description}
			\item[(A)] the number of clusters contained in the ball becomes at least $\Omega(d\cdot \beta/\log n)$,\\ or 
			\item[(B)] at least $\Omega(d/\log n)$ heavy edges connecting pairs of "low degree" clusters are cut by the boundary of the ball, \\ or
			\item[(C)] There is at least one "high degree" cluster in the ball.
		\end{description}
		Conditions {\bf (A)} , {\bf (B)} and {\bf (C)} independently imply the required bound.  Indeed, if $v_1,\ldots, v_k$ are such that $R_{v_i, v_j}\geq \beta$ for $i\neq j$, then balls of radius $\beta/2$ around each $v_i$ are disjoint, and each ball contains either $\Omega(d\cdot \beta/\log n)$ clusters, or cuts at least $\Omega(d/\log n)$ heavy edges connecting "low degree" clusters, which we have at most $\frac{k\log n}{\beta}$ of them, or it contains at least one "high degree" cluster, which we have at most $\wt{O}\left(\sqrt{k}\right)$ of them. Thus, the total number of such balls cannot be larger than $O\left(\sqrt{k}\log n+\frac{k\log n}{\beta \cdot d}+\frac{k\log^2 n}{\beta\cdot d}\right)=\wt{O}\left(\sqrt{k}+\frac{k}{\beta \cdot d}\right)$, as required. 
		
		Suppose that at the time that \textsc{BallGrowing}$(G,\mathcal{C},\Delta,u,\frac{\beta}{2})$ terminated we had $T=t$, i.e., $r_T=\beta/2$. We consider two cases.
		\begin{enumerate}
			\item Case 1: $r_{T-1}\geq \beta/4$. Then the ball $\B(v, r_{T-1})$ already contains at least $(\beta/4-\frac{T-1}{d})\cdot \frac{d}{2+2\log_2 n}=\Omega(d\cdot \beta/\log n)$ clusters by the inductive hypothesis, thus establishing {\bf (A)}. To see this, observe that if $\frac{T-1}{d}\leq \frac{\beta}{100}$ we are clearly done since $r_{T-1}\geq \frac{\beta}{16}$ so that the ball contains at least $\Omega(d\cdot \beta/\log n)$ clusters. Otherwise if $T-1\geq \Omega(d\cdot\beta)$, then because at each step we captured at least one cluster, the ball contains at least $T-1\geq \Omega(d\cdot\beta)$ clusters.)

			\item Case 2: $r_{T-1}\leq \beta/4$. In this case $j^*$ satisfies 
			\begin{align}
			\frac{1}{2^{j^*}} + \frac{1}{d}\geq \frac{\beta}{4} \label{deg_radius_relation}
			\end{align} 
			as otherwise the construction would not have been stopped (since we would have had $r_T=r_{T-1}+2^{-j^*}+1/d<\beta/4$ in that case). Now recall that   
			$$
			\sum_{c\in U_{j^*}} \wt d_c\geq \frac{d}{2+2\log_2 n}
			$$
			by choice of $j^*$. Thus we have
			\begin{equation*}
				\begin{split}
					\sum_{c\in U_{j^*}} \wt d_c&=\sum_{c\in U_{j^*}\cap \mathcal{C}(u, \beta/2)} \wt d_u+\sum_{c\in U_{j^*}\setminus \mathcal{C}(u, \beta/2)} \wt d_c\\
					&\geq \frac{d}{2+2\log_2 n},
				\end{split}
			\end{equation*}
			and thus at least one of the two summands is at least $\frac{d}{4+4\log_2 n}$. If it is the former, then by definition of $U_{j^*}$ and by \eqref{deg_radius_relation}, we have 
			$$
			|\mathcal{C}(u, \beta/2)|\geq 2^{-j^*-1}\sum_{c\in U_{j^*}\cap \mathcal{C}(v, \beta/2)} \wt d_c\geq (\frac{\beta}{8}-\frac{1}{2d})\cdot \frac{d}{4+4\log_2 n},
			$$
			i.e. at least $\Omega(d\cdot \beta/\log n)$ clusters are contained in the ball of radius $\beta/2$ around $u$, establishing {\bf (A)}.
			
			If it is the latter, we have 
			$$
			\sum_{c\in U_{j^*}\setminus \mathcal{C}(u, \beta/2)} \wt d_c\geq \frac{d}{4+4\log_2 n}, 
			$$
			i.e. at least $\frac{d}{4+4\log_2 n}$ edges go from $B(v, r_t)$ to the clusters outside of $B(v, \beta/2)$. Since $r_t\leq \beta/4$ by assumption, all these edges have length at least $\beta/4$ in effective resistance metric. Thus, we have at least $\frac{d}{4+4\log_2 n}$ edges of length at least $\beta/4$ crossing the boundary of the ball $B(v, \beta/2)$. Two cases: Whether there is a "high degree" cluster inside the ball, establishing {\bf (C)}, or at least half of them cut the heavy edges between "low degree" clusters, establishing {\bf (B)}, or at least half of them cut the edges connecting "low degree" clusters contained in the ball to the "high degree" clusters outside the ball. However, we argue that this cannot happen. Since we have $\Omega(d)$ of these edges, then at least one of the "high degree" clusters received $\Omega(\frac{d}{\sqrt{k}\log m_0})$ edges. Also, remember that $d \ge \sqrt{k}/2$ so this "high degree" cluster is actually inside the ball, establishing {\bf (C)}. 
		\end{enumerate}

\end{proofof}

\section{A sketch with $\wt{O}(n^{3/2})$ space and decoding time}\label{sec:n15}
In this section we propose an algorithm that attains $\wt{O}(n^{3/2})$ space and time. The main result of this section is Theorem~\ref{thm:n1.5}

\begin{theorem}
\label{thm:n1.5}
There exists an algorithm such that for any $0<\epsilon <\frac{1}{2}$, processes a list of edge insertions and deletions for an unweighted graph $G$ in a single pass and maintains a set of linear sketches of this input in $\wt{O}(\e^{-2}n^{1.5})$
space. From these sketches, it is possible to recover, with high probability, a weighted subgraph $H$ with $O(\epsilon^{-2}n\log n)$ edges, such that $H$ is a $(1 \pm \epsilon)$-spectral sparsifier of $G$. The algorithm recovers $H$ in $\wt{O}(\epsilon^{-2}n^{1.5})$ time.
\end{theorem}

\subsection{The algorithm}
\label{sec:n1.5alg}
Let $G=(V,E)$ be an unweighted graph. Recall that $B\in \R^{{n \choose 2}\times n}$ is the vertex edge incidence matrix of graph $G$. Suppose that $\Gamma$ is the global parameter of Algorithm~\ref{alg:sparsify}.
Let $B_j$ denote $B$ where its rows are sampled inedependently at rate $\Gamma^{-j}$. 

We define a binary hash function $h_j: \binom{n}{2} \rightarrow \{0,1\}$, that maps any edge $e$ to $1$ with probability $\frac{1}{\Gamma^j}$. Then, for any matrix $D \in \R ^{\binom{n}{2}\times n}$, that each row corresponds to an edge, let $D_j$ be $D$ with all rows except those with $h_j (e) = 0$ zeroed out. So $D_j$ is $D$ with rows sampled independently at rate $\frac{1}{\Gamma^{j}}$. This operation also can be done by linear operators. We build a diagonal matrix $\Pi_j \in \R^{\binom{n}{2}\times \binom{n}{2}}$, for all $j\in{1,2,\ldots, \ceil{\log{\frac{\lambda_u}{\lambda_l}}}}$ based on hash functions $h_j$ that serves as a sampling matrix as follows.
$$
\Pi_j(e,e):= h_j(e) 
$$
Then clearly $B_j=\Pi_j  B$. 

Let $G^\gamma$ denote a graph obtained by adding a complete graph $\frac{\gamma}{n} K_n$ to it.  
Let $B^\gamma, L^\gamma$ denote the vertex edge incidence matrix and Laplacian matrix of graph $G^\gamma$ respectively. Let $\oplus$ denote appending the rows of two matrices together.  Thus $B^\gamma=B\oplus \sqrt{\gamma}I$. Hence, $L^\gamma={B}^\top B+ \gamma I$.

\begin{algorithm}
	\caption{Main Sparsification Algorithm: outputs a $(1\pm\e)$-spectral sparsifier of $G^{\gamma(\ell)}$}\label{alg:sparsify}
	\begin{algorithmic}[1]
	\Procedure{Sparsify($SB, \ell, \e$)}{}	
				
			\If{$\ell= 0$} 
				\State $\wt{K}\gets\lambda_u I$ \label{ln:set-k0}	
			\Else
		 		\State $\wt{K}\gets \frac{1}{\Gamma(1+\e)}\cdot\textsc{Sparsify}(S B,\ell-1,\epsilon)$ 	\label{ln:setk-l1}
			\EndIf
			\If{$\ell= d+1$} $\gamma=0$		\label{ln:set-gama-0}
			\Else \text{ } $\gamma=\frac{\lambda_u}{\Gamma^\ell}$ 	\label{ln:set-gama}
			\EndIf		
			\State $q\gets 400 \log n$ \Comment it suffices for a $(1\pm \frac{1}{2})$-approximation
			\State $Q\gets q\times {n \choose 2}$ matrix of i.i.d. $\pm1$s
		
			\State Compute $M\gets \frac1{\sqrt{q}}Q\wt B\wt K^+$	\label{ln:jln32}	
			\For{$j=0$ {\bf to}~$ \ceil{\log_{\Gamma} \frac{\lambda_u}{\lambda_\ell}}$} \Comment{$\ceil{\log_{\Gamma} \frac{\lambda_u}{\lambda_\ell}}=\Theta(\log n)$}
				\State $E'_j\gets\textsc{HeavyEdges}(SB_j, \frac{\e^2}{500\cdot\Gamma^3 c'})$ \Comment{see Algorithm~\ref{alg:heavy-edge}} 
				\For{$e=(u,v) \in E'_j$} 
					\State $R'_e\gets2||M\mathbf{b}_e||_2^2$\Comment{$ R_e^{\tilde{G}} \le 2||M\mathbf{b}_e||_2^2 \le 3 R_e^{\tilde{G}}$}		
					\State $p'_e\gets\min \{1,c'R'_e \log n \epsilon^{-2} \}$ 				\Comment{$c'$ is the oversampling constant from Lemma \ref{lem:classic_result}}
					\If{$j=\ceil{\log_{\Gamma} \frac{\lambda_u}{\lambda_\ell}}$}

		 				\If{$p'_e \le \Gamma^{-j}$} $W(e,e)\gets \Gamma^{j}$ \label{ln:last_row}
		 				\EndIf
		 			\Else
		 			\If{$p'_e \in (\Gamma^{-j-1}, \Gamma^{-j}]$} $W(e,e)\gets \Gamma^{j}$ \label{ln:mid-row}
		 			\EndIf	
		 			\EndIf			
			\EndFor
		\EndFor
		\State \Return $ B^\T W B + \gamma I$ 
		\EndProcedure
	\end{algorithmic}
\end{algorithm}

\begin{algorithm}
	\caption{Heavy Edges: returns all the edges with effective resistance $\geq \frac{\beta}{\log n}$}\label{alg:heavy-edge}
	\begin{algorithmic}[1]
	\Procedure{HeavyEdges($SB,\beta$)}{}
		\State $E' \gets \emptyset$		
		\State $d \gets \frac{\sqrt{n}\log^2 n}{\beta}$ \label{ln:set-d}
		\State $\lambda \gets 200\sqrt{n}$ \label{ln:set-lamb}
		\While {there exists a vertex $v$ with  $\deg(v) < d$} 
				\State $E_v \gets$ \textsc{SparseRecovery}($SB,d,v$)	
				\Comment \text{recover neghborhood of $v$, See Section~\ref{subsec:sp-rec}} \label{ln:up-sk}
				\State Update sketches and degrees by removing edges in $E_v$
				\State $E'\gets E' \cup E_v$ 
		\EndWhile

		\State $E'\gets E' \cup  \textsc{FindLowConnectivityEdges}(S B,\lambda)$ \Comment \text{See Section~\ref{subsec:find_low}} \label{ln:fnd-low}
	\State \Return $E'$ 
	\EndProcedure
	\end{algorithmic}
\end{algorithm}

\subsection{The analysis}
\begin{proofof}{Theorem~\ref{thm:n1.5}}

Let $\Gamma=2$, $\lambda_u=2n$, $\lambda_\ell=\frac{8}{n^2}$, $d=\ceil{\log_{\Gamma}{\frac{\lambda_u}{\lambda_l}}}$, and for any $\ell \in \{1,2,\ldots d\}$, $\gamma(\ell)=\frac{\lambda_u}{\Gamma^\ell}$. 

Suppose that $$\wt{K}_\e=\textsc{Sparsify}(SB, d+1, \e)\text{.}$$ We will prove that $\wt{K}$ is a $(1 \pm \epsilon)$-spectral sparsifier of $G$.

Define $K=L$, and $K(\ell)=K+\gamma(\ell)I$. Observe that $K(\ell)=L^{\gamma(\ell)}$. 
Let $\wt{K}(d)=\textsc{Sparsify}(SB, d, \e)$. Recall that $K(d)=L^{\gamma(d)}$. Thus, by Lemma~\ref{lem:coarse} we have 
\begin{equation}
\label{eq:ep-apx}
(1-\e) \cdot K(d) \preceq_r \wt{K}(d) \preceq_r (1+\e) \cdot  K(d)\text{.}
\end{equation}
As per line~\ref{ln:set-gama-0} of Algorithm~\ref{alg:sparsify} we set $\gamma=0$. Thus, by Lemma~\ref{lem:chain_coarse}, ~\eqref{itm:last}, we have 
\begin{equation}
\label{eq:Gamma-apx}
\frac{1}{\Gamma} \cdot K\preceq_r \frac{1}{\Gamma} \cdot K(d) \preceq_r K \text{.}
\end{equation}
Putting~\eqref{eq:ep-apx} and~\eqref{eq:Gamma-apx} together we get
\begin{equation}
\frac{1-\e}{\Gamma(1+\e)} \cdot K\preceq_r \frac{1}{\Gamma(1+\e)} \cdot \wt{K}(d) \preceq_r  K \text{.}
\end{equation}
As per line~\ref{ln:setk-l1} of Algorithm~\ref{alg:sparsify} we set $\wt{K}=\frac{1}{\Gamma(1+\e)}\cdot \wt{K}(d)$, also let $C=\frac{\Gamma(1+\e)}{(1-\e)}$.
Hence, by Lemma \ref{lem:rfn-sp} with probability at least $1-\frac{1}{n^2}$, we have that $\wt{K}_\e$ containes $O\left(\frac{\Gamma(1+\e)}{1-\e} \e^{-2} n \log n\right)=O(\e^{-2} n \log n)$ non-zero entries, and
$$(1-\e) \cdot K \preceq_r \wt{K}_\e \preceq_r (1+\e)\cdot K\text{.}$$
Recall that $K=L$, therefore
$$(1-\e) \cdot L \preceq_r \wt{K}_\e \preceq_r (1+\e) \cdot L\text{.}$$
 \paragraph{Run-time and space:} Note that by Lemma~\ref{lem:coarse}, we have that the runtime and space of the algorithm is $\wt{O}(\Gamma\cdot d\cdot\e^{-2}\cdot n^{1.5})$, where $\Gamma=2$, and $d=\Theta(\log n)$. Therefore, the algorithm runs in $\wt{O}(\e^{-2}\cdot n^{1.5})$ space and time.

 \paragraph{Maintenance of sketches:} Note that Algorithm \ref{alg:sparsify} takes sketch $S\cdot B$ as an input where it corresponds to the different sketches that are used in different subroutines. More precisely, $S$ is a randomly constructed matrix with ${n \choose 2}$ columns that corresponds to the concatenation of the following matrices: The sampling matrix i.e., $\Pi \in \mathbb{R}^{{n \choose 2}\times {n \choose 2}}$ (Section~\ref{sec:n1.5alg}), the sketch to find the edges with connectivity at most $\lambda$, i.e., $S^f\Sigma \in\mathbb{R} ^{\lambda\cdot\poly(\log n)\times {n \choose 2}}$ (Section~\ref{subsec:find_low}), the \textsc{SparseRecovery} sketch to recover $k$-sparse vectors, i.e., $S^r\in\mathbb{R} ^{k\cdot\poly(\log n)\times {n \choose 2}}$ (Section~\ref{subsec:sp-rec}).
	
As per line~\ref{ln:set-d} of Algorithm~\ref{alg:heavy-edge} we set $k=d=\wt{O}(\sqrt{n}\cdot\poly{\Gamma})$ where $\Gamma=\Theta(1)$, and in line~\ref{ln:set-lamb} of Algorithm~\ref{alg:heavy-edge} we set $\lambda=\wt{O}(\sqrt{n})$. Therefore, overall the number of random bits needed for all the matrices, in an invocation of Algorithm~\ref{alg:main-sparsify} is at most $R=\wt{O}(n^2+nk+n\lambda)=\wt{O}(n^2)$, in addition to the random bits needed for the recursive calls.

To generate matrix $\Pi$, $S^f\Sigma$, and $S^h$ we use the fast pseudorandom numbers generator that is introduced in Section \ref{sec:prg}. Observe that the space used by Algorithm~\ref{alg:main-sparsify} is $s=O(n^{1.5}\cdot\poly(\log n)\e^{-2})$ in in addition to the space used by the recursive calls. Since $R=O(n^2)$, we have $R=O(s^2)$. Therefore, by Theorem~\ref{high_level_result} we can generate seed of $O(s\cdot \poly(\log s))$ random bits in $O(s\cdot \poly(\log s))$ time that can simulate our randomized algorithm.
	 
To generate the \textsc{SparseRecovery} sketch to recover $k$-sparse vectors, i.e., $S^r\in\mathbb{R} ^{k\cdot\poly(\log n)\times {n \choose 2}}$ (Section~\ref{subsec:sp-rec}), we can not use our fast pseudorandom numbers generator, since these bits need to be accessed again during the decoding time (see line \ref{ln:up-sk} of Algorithm~\ref{alg:sparsify}). However, Algorithm \textsc{SparseRecovery} uses low-independence hash functions for those bits instead. Moreover, the random matrix $Q \in\mathbb{R} ^{\Theta(\log n)\times {n \choose 2}}$ for JL (line~\ref{ln:jln32} of Algorithm \ref{alg:sparsify}) can be generated using $\log n$-wise independent hash functions.
\end{proofof}

\begin{lemma}
\label{lem:coarse}
For every $G=(V, E)$, every $\Gamma>1$, $0<\lambda_l\leq \lambda_u$, $d = \ceil{\log_\Gamma \frac{\lambda_u}{\lambda_\ell}}$, if $L$ is the Laplacian of $G$, the maximum eigenvalue of $L$ bounded from above by $\lambda_u$ and minimum nonzero eigenvalue bounded from
below by $\lambda_\ell$, and for every $\ell \in \{0, 1, 2, \ldots, d\}$ one has $\gamma(\ell)=\frac{\lambda_u}{\Gamma^\ell}$, the following conditions hold.

For every $\ell\in \{0, 1, 2, \ldots, d\}$ an invocation of Algorithm \ref{alg:sparsify} i.e., $\textsc{Sparsify}(SB, \ell, 
 \e)$ outputs a $(1\pm\e)$-spectral sparsifier to $L^{\gamma(\ell)}$ with probability at least $1-n^{-2}$. More precisely, if $\wt{L}^{\gamma(\ell)}=\textsc{Sparsify}(SB, \ell, 
 \e)$, then with probability at least $1-n^{-2}$, $\wt{L}^{\gamma(\ell)}$ contains at most  $O\left(\frac{\Gamma(1+\e)}{1-\e} \e^{-2} n \log n\right)$ non-zeros and
 $$(1-\e) \cdot L^{\gamma(\ell)} \preceq_r \wt{L}^{\gamma(\ell)} \preceq_r (1+\e) \cdot L^{\gamma(\ell)}\text{.}$$
Moreover, $\textsc{Sparsify}(SB, \ell, 
 \e)$ runs in $\wt{O}(\ell\cdot\Gamma\cdot  n^{1.5}\cdot \e^{-2})$ space and time.
\end{lemma}
\begin{proof} 
Let $K=L$, and $K(\ell)=K+\gamma(\ell)I$. Observe that $K(\ell)=L^{\gamma(\ell)}$. 
We prove the lemma by induction on $\ell$.
\paragraph{Base case: $\ell=0$}
\hfill \break
For $\ell=0$, as per line~\ref{ln:set-k0} of Algorithm~\ref{alg:sparsify} we set $\wt{K}(0)=\lambda_u I$. 

As per line~\ref{ln:set-gama} of Algorithm~\ref{alg:sparsify} we set $\gamma=\lambda_u$. Thus by Lemma~\ref{lem:chain_coarse},~\eqref{itm:base}, we have $$\frac{1}{\Gamma}\cdot K(0) \preceq \lambda_u I \preceq K(0) \text{.}$$
Therefore, we can apply Lemma \ref{lem:rfn-sp}. Hence, with probability at least $1-\frac{1}{n^2}$, $\wt{K}(0)$ contains $O(\e^{-2}\Gamma n \log n)$ non-zero entries and we have
\[ (1-\e) \cdot  K(0) \preceq_r \wt{K}(0) \preceq_r (1+\e)\cdot  K(0).\]

\paragraph{Inductive step: $\ell-1\to \ell$}
\hfill \break
Let $\wt{K}(\ell-1)=\textsc{Sparsify}(SB, \ell-1, \e)$. Recall that $K(\ell-1)=L^{\gamma(\ell-1)}$. Thus, by induction hypothesis we have 
\begin{equation}
\label{eq:ep-ap}
(1-\e) \cdot K(\ell-1) \preceq_r \wt{K}(\ell-1) \preceq_r (1+\e) \cdot  K(\ell-1)\text{.}
\end{equation}
As per line~\ref{ln:set-gama} we set $\gamma=\frac{\lambda_u}{\Gamma^\ell}$. Thus, by Lemma~\ref{lem:chain_coarse}, ~\eqref{itm:mid}, we have 
\begin{equation}
\label{eq:Gamma-ap}
\frac{1}{\Gamma} \cdot K(\ell)\preceq \frac{1}{\Gamma} \cdot K(\ell-1) \preceq K(\ell) \text{.}
\end{equation}
Putting~\eqref{eq:ep-ap} and~\eqref{eq:Gamma-ap} together we get
\begin{equation}
\frac{1-\e}{\Gamma(1+\e)} \cdot K(\ell)\preceq_r \frac{1}{\Gamma(1+\e)} \cdot \wt{K}(\ell-1) \preceq_r  K(\ell) \text{.}
\end{equation}
As per line~\ref{ln:setk-l1} 
of Algorithm~\ref{alg:sparsify} we set $\wt{K}=\frac{1}{\Gamma(1+\e)}\cdot \wt{K}(\ell-1)$, and let $C=\frac{\Gamma(1+\e)}{(1-\e)}$.
Hence, by Lemma \ref{lem:rfn-sp} with probability at least $1-\frac{1}{n^2}$, we have that $\wt{K}(\ell)$ containes $O\left(\frac{\Gamma(1+\e)}{1-\e} \e^{-2} n \log n\right)$ non-zero entries, and
$$(1-\e) \cdot K(\ell) \preceq_r \wt{K}(\ell) \preceq_r (1+\e)\cdot K(\ell)\text{.}$$
Recall that $K(\ell)=L^{\gamma(\ell)}$, therefore
$$(1-\e) \cdot L^{\gamma(\ell)} \preceq_r \wt{L}^{\gamma(\ell)} \preceq_r (1+\e) \cdot L^{\gamma(\ell)}\text{.}$$

Note that Algorithm $\textsc{Sparsify}(SB, \ell, \e)$ is a recursive algorithm where its space and runtime is given by the space and runtime of invocation of $\textsc{Sparsify}(SB, \ell-1, \e)$ and $\textsc{HeavyEdges}((SB)_j, \frac{\e^2}{500\cdot\Gamma^2 c'})$ for all $j\in \{1,\ldots,\ceil{\log{\frac{\lambda_u}{\lambda_l}}}\}$. By Lemma~\ref{lem:E_j-contains}, we have that the space and runtime of Algorithm $\textsc{HeavyEdges}(SB, \beta)$ is $\wt{O}(n^{1.5}\cdot\beta^{-1})$, moreover $\ceil{\log{\frac{\lambda_u}{\lambda_l}}} =\Theta(\log n)$. Therefore,  the space and run time of Algorithm $\textsc{Sparsify}(SB, \ell, \e)$ is $\wt{O}(\ell \cdot n^{1.5}\cdot \beta^{-1})$. Note that $\beta=\frac{\e^2}{500\cdot\Gamma^2 c'}$, where  $c'$ are constant. Hence, Algorithm $\textsc{Sparsify}(SB, \ell, \e)$ runs in $\wt{O}(\ell\cdot  \Gamma\cdot n^{1.5}\cdot \e^{-2})$ space and time.
\end{proof}
We will need Lemma~\ref{lem:classic_result} that we use in the proof of Lemma~\ref{lem:rfn-sp}. It is well known that by sampling the edges  of $B$ according to their effective resistance, it is possible to obtain a matrix $\wt{B}$ such that $(1-\e)B^\top B \preceq \wt{B}^\top\wt{B} \preceq (1+\e)B^\top B$ with high probability (see Lemma~\ref{lem:classic_result}).
\begin{lemma} \label{lem:rfn-sp}
Let $B \in \R^{{{n}\choose{2}}\times n}$ be the vertex edge incidence matrix of an unweighted graph $G$. Let $\gamma \geq 0$, and consider
$K = B^\top B + \gamma I$. 
Let $C>1$, and $\wt{K}$ be a spectral approximation to $K$ with $O(n\log n)$ non-zeros such that $\frac{1}{C} K \preceq_r \wt{K} \preceq_r K$.
An invocation of Algorithm~\ref{alg:sparsify} such that the parameters $C$, $\wt{K}$ and $\gamma$ of the algorithm satisfy above conditions, returns $\wt{K_\e}=\wt{B}_\e^\top\wt{B}_\e+\gamma I$, where $\wt{B}_\e$ contains only $O(C\epsilon^{-2} n \log n)$ reweighted rows of $B$, and $(1+\e)K \preceq_r \wt{K}_\e \preceq_r (1+\e)K$ with probability at least $1-\frac{1}{n^2}$.
\end{lemma}

\begin{proof}
Let $G^\gamma$ denote a graph obtained by  adding a complete graph $\frac{\gamma}{n} K_n$ to $G$. 
Let $B^\gamma\in \R^{\left({{n}\choose{2}}+n\right)\times n}$, and $L^\gamma \in \R^{n\times n}$ denote the vertex edge incidence matrix and Laplacian matrix of the graph $G^\gamma$ respectively. Thus we have 
\[
B^\gamma=B\oplus \sqrt{\gamma}I.
\]
and, 
\[L^\gamma=(B^\gamma)^\top(B^\gamma)={B}^\top B+ \gamma I =L+\gamma I\text{.}\]
For any $y \in [{{n}\choose{2}}+n]$,  let $\mathbf{b}_y$ denote the row $y$ of matrix $B^\gamma$.
Note that $K=L^\gamma$. 
Observe that $\frac{1}{C} K \preceq_r \wt{K} \preceq_r K$, hence for any $y \in [{{n}\choose{2}}+n]$,  
\begin{equation} \label{inq:zr_rtild}
\mathbf{b}_y^\T K^{+}\mathbf{b}_y \preceq_r \mathbf{b}_y^\T \wt{K}^{+}\mathbf{b}_y \preceq_r C \mathbf{b}_y^\T K^{+}\mathbf{b}_y \text{.}
\end{equation}

Let $\tau$ be a vector of leverage score for $B^\gamma$'s rows. Hence, for any $y \in [{{n}\choose{2}}+n]$ we have $\tau_y =  \mathbf{b}_y^\top K^{+} \mathbf{b}_y$. We define $\wt{\tau}_y =  \mathbf{b}_y^\top \wt{K}^{+} \mathbf{b}_y$. Thus by inequality~\eqref{inq:zr_rtild} for any $y\in [{{n}\choose{2}}+n]$ we have 
\begin{equation} \label{eq:tau_bound}
\tau_y  \leq  \wt{\tau}_y\leq C \tau_y \text{.}
\end{equation}
Let $p_y(G^\gamma)=\min\{1, c'\e^{-2}\log n \cdot \tau_y \}$.
To complete the proof we need to show that for any $y\in[{{n}\choose{2}}+n]$, Algorithm \ref{alg:sparsify}
samples the rows of $B^\gamma$ independently and with probability at least $p_y(G^\gamma)$, hence, we can apply Lemma \ref{lem:classic_result} afterwards.
To see this observe that Algorithm \ref{alg:sparsify} returns 
$\wt{K}_\e=B^\top W B+ \gamma I$, where we have
\[B^\top W B+ \gamma I = (W^{\frac{1}{2}} B \oplus \sqrt{\gamma}I)^\top (W^{\frac{1}{2}} B \oplus \sqrt{\gamma}I) \text{.}\]

Thus, for any ${{n}\choose{2}}+1 \leq y \leq {{n}\choose{2}}+n$, row $\mathbf{b}_y$ is sampled with probability $1$. Therefore, $\wt{K}_\e$ includes rows corresponding to $\sqrt{\gamma}I$ with probability $1$. 

Hence, it's sufficient to prove that for any $1\leq e \leq {{n}\choose{2}}$, row $\mathbf{b}_e$ is included in $W^{\frac{1}{2}} B$ independently with probability at least $p_e(G^\gamma)$ with the proper weight.

Observe that since $\Gamma^{-\ceil{\log_{\Gamma} \frac{\lambda_u}{\lambda_\ell}}} \leq \frac{1}{n^2}$, therefore for any edge $e$ there exists a $j \in \{0, \ldots,\ceil{\log_\Gamma \frac{\lambda_u}{\lambda_\ell}} \}$, such that $p_e(G^\gamma) \in [\Gamma^{-j-1},\Gamma^{-j}]$. For edge $e\in E$, suppose that $p_e(G^\gamma) \in [\Gamma^{-j-1},\Gamma^{-j}]$. 

Let $R_e^G=\mathbf{b}_e^\top L^{+}\mathbf{b}_e$ denote the effective resistance of edge $e$ in graph $G$. We define $p_e(G)=\min\{1, c'R_e^G\log n \e^{-2}\}$.
Recall that $\tau_e=R_e^{G^\gamma}$, and $p_e(G^\gamma)=\min\{1, c'R_e^{G^\gamma}\log n \e^{-2}\}$. Since $G^\gamma$ is obtained by adding a complete graph $\frac{\gamma}{n}K_n$ to $G$, hence by Fact \ref{fact:mon} we have $R_e^{G}\geq R_e^{G^\gamma}$. Therefore we have 
$$p_e(G) \geq p_e(G^\gamma) \geq \Gamma^{-j-1}\text{.}$$
Let $G_j$, $G_{j-1}$ and $G_{j-2}$ denote the graph obtained by sampleing edges of graph $G$ at rate $\Gamma^{-j}$, $\Gamma^{-{j-1}}$ and $\Gamma^{-{j-2}}$respectively. 
Therefore, since $p_e(G)\geq \Gamma^{-j-1}$, by Lemma \ref{lem:heavy_sample} we have 
$$R_e^{G_j} \geq \frac{\e^2}{500\cdot\Gamma c' \log n}$$ 
$$R_e^{G_{j-1}} \geq \frac{\e^2}{500\cdot\Gamma^2 c'\log n }$$
$$R_e^{G_{j-2}} \geq \frac{\e^2}{500\cdot\Gamma^3 c'\log n }$$
with probability at least $1-n^{-100}$.  
Let $\beta=\frac{\e^2}{500\cdot\Gamma^3 c'}$. Therefore, since $\beta \in (0,1)$, and $R_e^{G_j} \geq \frac{\beta}{\log n}$, we can apply Lemma~\ref{lem:E_j-contains}. Thus, $\textsc{HeavyEdges}((SB)_j,\beta)$ contains edge $e$ with probability at least $1-n^{-8}$. Moreover, since $R_e^{G_{j-1}} \geq \frac{\beta}{\log n}$, and $R_e^{G_{j-2}} \geq \frac{\beta}{\log n}$ then by Lemma~\ref{lem:E_j-contains} $\textsc{HeavyEdges}((SB)_{j-1},\beta)$ and $\textsc{HeavyEdges}((SB)_{j-2},\beta)$ contains edge $e$ with probability at least $1-n^{-8}$.

Observe that in Algorithm \ref{alg:sparsify} we set $R'_e=||M\mathbf{b}_e||_2^2$, and $p'_e=\min \{1,c'R'_e \log n \epsilon^{-2} \}$. Thus by Johnson-Lindenstrauss Lemma with high probability we have 
\[\wt{\tau}_e\le R'_e=2||M\mathbf{b}_e||_2^2 \le 3 \mathbf{b}_e^\T \wt{K}^{+} \mathbf{b}_e = 3\wt{\tau}_e \text{.}\] 
Moreover by~\eqref{eq:tau_bound} we have $\tau_e  \leq  \wt{\tau}_e\leq C \tau_e$, thus we get
\[\tau_e  \leq  R'_e\leq 3C \tau_e \text{.}\]
Hence, $p_e \leq p'_e \leq 3Cp_e$, which implies $p'_e \in [\Gamma^{-j-1},3C\cdot\Gamma^{-j}]$. Therefore, since $3C\leq \Gamma^2$, for $k=j-2$ or $k=j-1$, or $k=j$ we have $p'_e \in [\Gamma^{-k-1},\Gamma^{-k}]$.
Reacll that $(SB)_k$ denote the sketches of $B$ where the rows are sampled independently at rate $\Gamma^{-k}$.  Therefore, edge $e$ is included to $\wt{K}_\e$ with probability at least $$\Gamma^{-k}\geq \Gamma^{-j}\geq p_e(G^\gamma)$$
and with weight $\Gamma^{k}$. Therefore we by Lemma \ref{lem:classic_result} we have $(1+\e)\cdot \preceq_r \wt{K}_\e \preceq_r (1+\e)\cdot K$. Moreover $\wt{K}_\e$ contains at most $( \sum_{e\in E} R'_e \log n \e^{-2})$ non-zeros with high probability. Note that \[\sum_{e\in E} R'_e \leq  3C \sum_{e\in E} \tau_e \leq3Cn \text{.}\] Hence, overall $\wt{B}_\e$ contains $O(C\e^{-2} n \log n)$ non zeros with high probability.
\end{proof} 
\begin{lemma}
\label{lem:heavy_sample}
Let graph $G=(V,E)$ be a graph with $n$ vertices. For any integer $k,j$ such that $k\leq j$, suppose that graph $G_{j-k}=(V,E_{j-k})$ is obtained from $G$ by sampling edges of graph $G$ independently at rate $\Gamma^{-{(j-k)}}$. Suppose that edge $e=(u,v)$ is indcluded in $E_{j-k}$. Let $R_e^{G}$, $R_e^{G_{j-k}}$ denote the effective resistance of edge $e$ in graph $G$ and $G_{j-k}$ respectively. Let $p_e=\min \{1,c'R_e^{G} \log n \epsilon^{-2} \}$. If $p_e\geq \Gamma^{-j-1}$, then $R_e^{G_{j-k}} \geq \frac{\e^2}{500\cdot\Gamma^{k+1} c' \log n}$ with probability at least $1-n^{-100}$. 
\end{lemma}
\begin{proof}
Suppose that in graph $G$, we inject $\frac{1}{R_e^{G}}$ unit of flow to node $u$ and extract it from $v$. Let vector $\varphi=\frac{1}{R_e^{G}}L^{+}\mathbf{b}_e$ denote the potentials induced at the vertices in graph $G$. Therefore we have 
\begin{equation} \label{eq:delta-phii}
\varphi(u)-\varphi(v)=1
\end{equation}
By Lemma \ref{lem:reff-energy} we get
\begin{equation}
\label{eq:R-phi}
\sum_{(a,b) \in E} \left(\varphi(a)-\varphi(b)\right) ^2 = \frac{1}{R_e^{G}}
\end{equation}
Putting~\eqref{eq:R-phi}, and ~\eqref{eq:delta-phii} together we get
\begin{equation}
\frac{1}{R_e^{G}} = 1+ \sum_{(a,b) \in E\setminus \{e\}} \left(\varphi(a)-\varphi(b)\right) ^2
\end{equation}
Note that $E_{j-k}$ is obtained by sampling edges in $E$ with rate $\Gamma^{-(j-k)}$, therefore since $e\in E_{j-k}$ we have 
\begin{align}
\label{eq:exp}
\mathbb{E}\left[\sum_{(a,b)\in E_{j-k}\setminus \{e\}} \left(\varphi(a)-\varphi(b)\right)^2  \right]&=\Gamma^{-(j-k)}\cdot\sum_{(a,b) \in E\setminus \{e\}} \left(\varphi(a)-\varphi(b)\right) ^2 \nonumber \\
&= \Gamma^{-(j-k)}\cdot \left(\frac{1}{R_e^{G}} -1 \right) 
\end{align}
Note that $p_e=\min \{1,c'R_e^{G} \log n \epsilon^{-2} \}$, thus $p_e\leq c'R_e^{G} \log n \epsilon^{-2}$. Moreover by the assumption of the lemma we have $p_e(G)\geq \Gamma^{-j-1}$. Therefore, 
\begin{equation}
\label{eq:R_lwbnd}
R_e^{G} \geq \frac{p_e}{c'\log n \e^{-2}} \geq \frac{\Gamma^{-j-1}}{c'\log n \e^{-2}}
\end{equation}
Therefore we have
\begin{align}
\mu&=\mathbb{E}\left[\sum_{(a,b)\in E_j \setminus \{e\}} \left(\varphi(a)-\varphi(b)\right)^2\right] \nonumber \\
&=\Gamma^{-(j-k)}\cdot \left(\frac{1}{R_e^{G}} -1 \right) && \text{By~\eqref{eq:exp}}\nonumber \\
&\leq \frac{\Gamma^{-(j-k)}}{R_e^{G}} \nonumber \\
&\leq \frac{\Gamma^{-(j-k)}}{\left( \frac{\Gamma^{-j-1}}{c'\log n \e^{-2}} \right)} && \text{By~\eqref{eq:R_lwbnd}} \nonumber \\
&\leq \Gamma^{k+1} c'\log n \epsilon^{-2}\text{.} \label{eq:exp_up}
\end{align}
Observe that for any $(a,b)\in E$, $\left(\varphi(a)-\varphi(v)\right)^2 \in [0,1]$, we can apply standard multiplicative Chernoff bound to show the the concentration \cite{hoeffding1963probability}. Let $\delta=\frac{300\log n}{\mu}$. Then we have
\begin{align*}
\Pr{\left[\sum_{(a,b)\in E_{j-k}\setminus \{e\}} \left(\varphi(a)-\varphi(b)\right)^2 > (1+\delta) \mu \right]}  &\leq \text{exp}{\left(\frac{-\delta\mu}{3}\right)} \leq n^{-100}
\end{align*}
Therefore with probability at least $1-n^{-100}$, we have
\begin{align}
\label{eq:cons_up}
\sum_{(a,b)\in E_{j-k}\setminus \{e\}} \left(\varphi(a)-\varphi(b)\right)^2 
&\leq  (1+\delta)\mu \nonumber\\
&=\left(1+\frac{300\log n}{\mu}\right)\mu && \text{Since }\delta=\frac{300\log n}{\mu}\nonumber\\
&=\mu+300\log n \nonumber \\
&\leq 300\log n + \Gamma^{k+1} c'\log n \epsilon^{-2} &&\text{By~\eqref{eq:exp_up}}
\end{align}
Therefore since $e\in E_{j-k}$ with probability at least $1-n^{-100}$, we have
\begin{equation}
\label{eq:max-phi}
\sum_{(a,b)\in E_{j-k}} \left(\varphi(a)-\varphi(b)\right)^2=1+\sum_{(a,b)\in E_{j-k}\setminus \{e\}} \left(\varphi(a)-\varphi(b)\right)^2 \leq 500\cdot\Gamma^{k+1} c'\e^{-2} \log n
\end{equation}
Moreover by Fact \ref{fact:eff-pot} we have,
\begin{equation}
\label{eq:R_j-R}
\sum_{(a,b) \in E_{j-k}} \left(\varphi(a)-\varphi(b)\right) ^2 \geq  \frac{1}{R_e^{G_{j-k}}}
\end{equation}
Putting~\eqref{eq:R_j-R} and~\eqref{eq:max-phi} we get
\[
R_e^{G_{j-k}} \geq \frac{\e^2}{500\cdot\Gamma^{k+1} c' \log n}
\text{.}\]
\end{proof}
\begin{lemma}
\label{lem:E_j-contains}
Let $G=(V,E)$ be an unweighted graph. For edge $e\in E$, let $R_e$ denote the effective resistance of edge in graph $G$. For any $\beta\in (0,1)$, if $R_e \geq\frac{\beta}{\log n}$ then an invocation of Algorithm~\ref{alg:heavy-edge} i.e., $\textsc{HeavyEdges}(SB, \beta)$ contains edge $e$ with probability at least $1-n^{-8}$. Moreover, Algorithm~\ref{alg:heavy-edge} runs in $\wt{O}(n^{1.5}\cdot\beta^{-1})$-space and time.
\end{lemma}
\begin{proof}
As per line~\ref{ln:set-d} of Algorithm~\ref{alg:heavy-edge} we set $d=\frac{\sqrt{n}(\log n)^2}{\beta}$. Observe that Algorithm \ref{alg:heavy-edge} first recover all the edges connected to the low degree vertices, i.e., vertex with degree at most $d$, iteratively, and removes them until no such vertex remains. The recovering process is done using \textsc{SparseRecovery} algorithm where its correctness is guaranteed by Lemma \ref{lem:sparse-recovery}. Suppose that vertex $v$ with degree at most $d$, is the first vertex going to be removed in the first iteration of the loop. Then by Lemma~\ref{lem:sparse-recovery}, Algorithm \textsc{SparseRecovery} recovers all the neighbors of the vertex $v$ exactly with probability at least $1-n^{-10}$. Therefore, with probability at least $1-n^{-10}$ this is a deterministic process, so that we can use the same sketches to recover the neighbors of vertices in next iterations. Thus, by union bound over all iterations, with probability at least $1-n\cdot n^{-10}$, all the edges connected to the low degree vertices are recovered. Note that we can maintain degrees easily using linear skecthes.

Therefore, if edge $e$ is connected to the one of the low degree vertices it is recovered with probability at least $1-n^{-9}$.  Otherwise, let $G'=(V',E')$ be a graph obtained by removing the low degree vertices from graph $G$. Therefore, $e\in E'$. Observe that 
\begin{equation} \label{eq:min-deg}
\min_{v \in V'} \{\text{deg}(v)\} \geq d=\frac{\sqrt{n}(\log n)^2}{\beta}\text{.}
\end{equation}
Therefore, since $\min_{v\in V'}\{\text{deg}(v)\}\geq \frac{\sqrt{n}(\log n)^2}{\beta}$, and $R_e \geq \frac{\beta}{\log n}$, by Lemma~\ref{lem:min-cut-R} we have that 
\begin{equation}\label{eq:lambda-bound}
\lambda_e\leq 200\sqrt{n}\text{.}
\end{equation}
As per line~\ref{ln:set-lamb} of Algorithm~\ref{alg:heavy-edge} we set $\lambda=200\sqrt{n}$. Let
$E''\subseteq E'$ denote the set of edges of edge-connectivity at most $\lambda$ in $G'$ (see Definition~\ref{def:edg-conn}). Therefore Therefore, by~\eqref{eq:lambda-bound} we have 
\begin{equation}
\lambda_e\leq \lambda \text{.}
\end{equation}
As per line~\ref{ln:fnd-low} of Algorithm~\ref{alg:heavy-edge}, we invoke  $\textsc{FindLowConnectivityEdges}(S B,\lambda)$. By Lemma~\ref{alg:low-connectivity-edges} we have that with probability at least $1-n^{-100}$, Algorithm~\ref{lem:recover-low-connectivity} recovers all the edge with edge-connectivity at most $\lambda$. Therefore since $\lambda_e \leq \lambda$, algorithm~\ref{alg:heavy-edge} outputs edge $e$ with probability at least $1-n^{-8}$.

Note that Algorithm~\ref{alg:heavy-edge}, invoke Algorithm \textsc{SparseRecovery} for any vertex that is removed i.e., at most $n$ times. By Lemma~\ref{lem:sparse-recovery} we that the space and runtime of Algorithm \textsc{SparseRecovery} to recover a $d$-sparse vector is at  most $O(d\cdot\poly(\log n))$. Moreover, by Lemma~\ref{lem:recover-low-connectivity}, the space and run time of Algorithm \textsc{FindLowConnectivityEdges} is bounded by $O(\lambda n\cdot \poly(\log n))$. 
Recall that $d=\frac{\sqrt{n}(\log n)^2}{\beta}$, and $\lambda=200\sqrt{n}$.
Therefore, overall the space and runtime of Algorithm~\ref{alg:heavy-edge} is at most $\wt{O}(nd+\lambda n)=\wt{O}(n^{1.5}\cdot \beta^{-1})$.
\end{proof}

In the following lemma we show that if the minimum degree of graph $G=(V,E)$ is lower bounded by $d\approx \sqrt{n}$, then every edge $e\in E$ with effective resistance $R_e^{G} \geq \frac{1}{\log n}$, say, necessarily has connectivity at most $\wt{O}(\sqrt{n})$. More formally,
\begin{lemma}
\label{lem:min-cut-R}
For any $\beta \in(0,1)$, let $G=(V,E)$ be a graph with $n$ vertices such that for any vertex $v \in V$, we have $\deg(v) \geq \frac{\sqrt{n}(\log n)^2}{\beta}$. Then for any edge $e\in E$, with edge-connectivity $\lambda_{e} \geq 200\sqrt{n}$ (see Definition~\ref{def:edg-conn}), we have $R_e^{G} \leq \frac{\beta}{\log n}$. 
\end{lemma}
\begin{proof}
Suppose that in graph $G$, we inject $\frac{1}{R_e^{G}}$ unit of flow to node $u$ and extract it from $v$. Let vector $\varphi=\frac{1}{R_e^{G}}L^{+}\mathbf{b}_e$ denote the potentials induced at the vertices in graph $G$. By Lemma \ref{lem:reff-energy} we have
\begin{equation}
\label{eq:R-phi-G}
\sum_{(a,b) \in E} \left(\varphi(a)-\varphi(b)\right) ^2 = \frac{1}{R_e^{G}}
\end{equation}

Let $d=\min_{v\in V}\{\text{deg}(v)\}$, therefore by the assumption of the lemma $d\geq \frac{\sqrt{n}\cdot(\log n)^2}{\beta}$.

We say that vertex $a$ is \textit{isolated} if there are fewer than $\frac{d}{2}$ vertices such at  distance less than $\frac{d}{n\log^2 n}$ from $a$ in the potentials embedding, i.e., $\big|\{b: |\varphi(a)-\varphi(b)|< \frac{d}{n\log^2 n}\}\big|\leq\frac{d}{2}$.

Thus, for any isolated vertex such $a$ we have
\[\sum_{(a,b) \in E} \left(\varphi(a)-\varphi(b)\right) ^2 \geq \frac{d}{2} \cdot\left(\frac{d}{n\log^2 n}\right)^2\]

Let $I$ denote the set of isolated vertices. Thus if $|I|\geq \frac{2n\log n}{d}$ we have

\[\sum_{(a,b) \in E} \left(\varphi(a)-\varphi(b)\right) ^2 \geq |I|\cdot \frac{d}{2} \cdot\left(\frac{d}{n\log^2 n}\right)^2 \geq \frac{d^2}{n\log^3 n} > \frac{\log n}{\beta^2} \text{.}\]

where the last inequality follows from the assumption $d\geq \frac{\sqrt{n}\log^2 n}{\beta}$.

Now we consider the case that $|I|< \frac{2n\log n}{d}$. 

Let $U$ be a uniformly random sample of $ \frac{100n\log n}{d}$ points. The probability that every non-isolated point has a neighbor at distance at most $\frac{d}{n\log^2 n}$ from it, in set $U$ is 
$$1- n\left(1-\frac{d}{2n}\right)^{\frac{100n\log n}{d}} = 1-\frac{1}{n^{\Omega(1)}}$$.

Thus with high probability every point has at least one neighbor point in $I$ or $U$ at distance $\frac{d}{n\log^2 n}$ from it. Let $r=|U|+|I|$, and let $I_1,\ldots, I_r$ denote intervals of width $\frac{2d}{n\log^2 n}$ around points in $I \cup U$. Note that, $|U|+|I|\leq \frac{200 \cdot n\log n}{d}$, thus we have
\begin{equation}\label{eq:1}
\left|\bigcup_{j=1}^r I_j\right|\leq r\cdot \frac{2d}{n\log^2 n} \leq \frac{400}{\log n},
\end{equation}

Now let $ J_1,\ldots, J_q, q\leq r$ denote the set of {\bf disjoint} intervals such that 
$$
\bigcup_{j=1}^q J_j=\bigcup_{j=1}^r I_j,
$$
ordered such that $J_{j+1}$ is to the right of $J_j$ for every $j$, and let $\Delta_0,\Delta_1,\ldots, \Delta_q$ denote the distances between the intervals. Specifically, let $J_0=[0, 0], J_{q+1}=[1, 1]$ (intervals of length $0$ around $0$ and $1$ respectively) for convenience, and let $\Delta_j$ denote the distance between the right endpoint of $J_j$ and the left endpoint of $J_{j+1}$. Note that by~\eqref{eq:1} 
$$
\sum_{j=0}^q \Delta_j\geq 1- \frac{400}{\log n} \geq \frac{1}{2}
$$
for sufficiently large $n$. Now since the sweep cut separating $J_j$ from $J_{j+1}$ has at least $\lambda$ edges and the fact that there is no point outside these intervals with high probability, we get that the conductance is at least 
$$
\sum_{(a,b) \in E_G} \left(\varphi(a)-\varphi(b)\right) ^2 \geq \sum_{j=0}^q \Delta_j^2\cdot \lambda\geq \lambda\cdot \frac1{4(q+1)},
$$
since $\sum_{j=0}^q \Delta_j^2$ is maximized subject to $\sum \Delta_j\geq 1/2$ when $\Delta_j=\frac{1}{2(q+1)}$. Since $q\leq r\leq |I|+|U|\leq \frac{200n\log n}{d}$, we get that the conductance is at least 
$$
\sum_{(a,b) \in E} \left(\varphi(a)-\varphi(b)\right) ^2 \geq \lambda\cdot \frac1{4(q+1)}\geq \frac{\lambda d}{200 n\log n}\geq \frac{\log n}{\beta}
\text{.}$$
where in the last inequality we used the assumptions $\lambda \geq 200 \sqrt{n}$, and $d\geq \frac{\sqrt{n}\log^2 n}{\beta}$. Therefore by~\eqref{eq:R-phi-G} we have $R_e^{G} \leq \frac{\beta}{\log n}$.
\end{proof}

\section{Faster pseudorandom numbers for sketching algorithms}
\label{sec:prg}

Like many sketching and streaming algorithms, our algorithm results rely crucially on randomness. In particular, they use many more random bits than they have space to store. Moreover, most of these random bits are not used in a ``read once'' way. For example, many are used to initialize persistent random hash functions, which must access the \emph{same set of random bits} every time a particular edge is updated in our graph sketch.

Naively, after random initialization, we need to store each of these persistent hash functions. Doing so, however, would require $\tilde{O}(n^2)$ space for a graph with $n$ nodes, which would dominate the space complexity of our methods.  To cope with this issue, we need a more compact way of representing persistent random hash functions, a challenge arising in the design of most randomized streaming algorithms, both for graph problems and other applications \cite{Muthukrishnan2005}.

There are several techniques to deal with the issue. One approach is to prove that an algorithm can be implemented with \emph{limited independence} hash functions, which can take exponentially fewer bits to represent than fully random hash functions \cite{carterWegman79}. However, proving that limited independence hashing still gives a correct algorithm can be a significant burden. For example, Indyk's well known streaming algorithm for $\ell_p$ norm estimation \cite{indyk2000stable, Indyk06} was only shown to work with limited independence a decade after its introduction \cite{KaneNelsonWoodruff10}. Moreover, many streaming algorithms for graph problems are not known to work with limited independence (see e.g. \cite{ahn2012graph,kapralov2017single}) and this same challenge carry's over to a variety of other problems \cite{rudraUurtamo2010,belazzouguiZhang16,chakraborty2016,blasiok2017}.

In these cases, a more powerful `black box' technique is needed to reduce the costly requirement of ``storing randomness''.
For algorithms based on linear sketching (like those presented in this paper) one such technique is the application of \emph{pseudorandom number generators} \cite{nisan1992pseudorandom, NisanZuckerman96}, which have been widely used in streaming algorithms since Indyk's original application to streaming norm estimation \cite{Indyk06}. A pseudorandom number generator can obviate the need to persistently store random hash functions altogether.

\subsection{Simulating small space randomized algorithms} 

The goal of a pseudorandom number generator (PRG) is to deterministically generate a large string of \emph{pseudorandom} bits from a much smaller seed of \emph{truly random} bits. For certain algorithms, including those that use bounded memory, it is possible to show that using these pseudorandom bits instead of a full set of truly random bits leads to very little degradation in performance.

While we are not interested in reducing the total number of random bits used by our algorithms, PRGs offer an additional advantage: they can reduce the space required to store randomness when random bits need to be accessed repeatedly. In particular, we only need to store the PRG's small random seed and can then generate pseudorandom bits ``on-the-fly'', as they are needed. 

Towards this goal, a pseudorandom number generator designed by Nisan has been especially popular in streaming applications \cite{nisan1992pseudorandom}.
For any algorithm that use no more than $S$ bits of space, Nisan's PRG generates $R$ pseudorandom bits from a seed of $O(S\log R)$ truly random bits. If these pseudorandom bits are used to simulate truly random bits, the algorithm's failure probability increases by at most $2^{-s}$. Furthermore, space to store the pseudorandom bits only increases the algorithm's space complexity from $S$ to $O(S\log R)$.

This PRG can be used to reduce the randomness requirements of any \emph{linear sketching} algorithm\footnote{This is a broad class: all known turnstile streaming algorithms (i.e. those handling insertions and deletions) are based on linear sketching \cite{LiYiWoodruff14}.} that 1) does not access more than $S$ random bits on every sketch update and 2) does not use more than $S$ space beyond what is required to store randomness. This claim is not immediate: naively, streaming algorithms that use a large number of persistent random bits do not run in small space. However, it can be proven via a reordering argument from \cite{Indyk06}, which we discuss further in Section \ref{sec:prg_linear_sketch}. 


\subsection{The computation cost of PRGs} 

Since our results use linear sketching, we can apply Nisan's PRG to eliminate the assumption that hash functions and other random bits are chosen truly at random. In particular, for graphs on $n$ nodes, our algorithms use $R = \poly(n)$ random bits and $S = \poly(n)$ space, beyond what is required to store randomness. So Nisan's PRG allows for  implementations in \emph{total space} $O(S\log S)$. 

However, in contrast to prior work on streaming spectral sparsification \cite{kapralov2017single}, we are also interested in the \emph{time complexity} of our sketching methods, both in terms of update and recovery time. When runtime is a concern, Nisan's PRG provides an unsatisfying solution: for our application it is prohibitively slow.

In particular, using a seed of $O(S\log R)$ random bits, Nisan's PRG requires $O(S\log R)$ time to generate any specific pseudorandom bit. This runtime is good when $S$ is logarithmic in the natural problem parameters. For example, for many streaming problems involving length $n$ vectors, $S = \polylog(n)$ and $R = \poly(n)$. In this setting, 
the total generation time for Nisan's PRG is just $\polylog(n)$ per bit. This is very good considering that, if $\poly(n)$ random bits are stored persistently, it takes $O(\log n)$ time just to specify which bit we would like the PRG to generate.

However, our algorithms and other graph streaming algorithms, use significantly more than $\polylog(n)$ space \cite{Feigenbaum05,McGregor14}. Specifically, in our setting, both $S$ and $R$ are polynomial in $n$ for graphs on $n$ nodes, so Nisan's PRG requires $O(S\log S)$ time per random bit. Since $\polylog(S)$ random bits are accessed on every edge update, this leads to an update time of $\tilde{O}(S)$. We would like to reduce the PRG's cost to $\polylog(S)$ time per bit, which would improve our update time to $\polylog(n)$.

\subsection{Main Result}
\label{sub:main_result}
The goal of this section is to demonstrate that this significant runtime improvement can be achieved using a different pseudorandom generator than Nisan's. In particular, we prove:

\begin{theorem}
	\label{high_level_result}
	For any constants $q,c > 0$, there is an explicit PRG that draws on a seed of $O(S\plog(S))$ random bits and can simulate any randomized algorithm running in space $S$ and using $R = O(S^q)$ random bits. This PRG can output any pseudorandom bit in $O(\log^{O(q)} S)$ time and the simulated algorithm fails with probability at most $S^{-c}$ higher than the original.
\end{theorem}
This result is a direct corollary of the more detailed Theorem \ref{thm:prg_general}, which we prove in Section \ref{sec:multilevel}.

Theorem \ref{high_level_result} implies a method for reducing the randomness required by a large class of linear sketching algorithms (including those presented in this paper). A formal statement appears as Theorem \ref{thm:prg_for_sketching} of Section \ref{sec:prg_linear_sketch}. In short, as long as the sketching algorithm uses $S$ space and, for any update to a particular entry (i.e. an edge in our case, or a vector entry in a vector streaming algorithm) the algorithm only accesses at most $S$ persistent random bits, than it can be simulated using a PRG with seed $O(S\plog S)$. As in Theorem \ref{high_level_result}, this PRG can produce a single pseudorandom bit in just $O(\plog S)$ time.

To prove Theorem \ref{high_level_result}, we will use the Nisan-Zuckerman PRG \cite{NisanZuckerman96}.
A well known alternative to Nisan's construction, this PRG uses as a component any \emph{randomness extractor}, which is a tool for converting a string of weakly random bits to a string of nearly uniformly random bits by using a small amount of additional randomness.
The Nisan-Zuckerman PRG can be made efficient by using a $\emph{locally computable extractor}$.

 Notice that Nisan's PRG required a seed of $O(S\log R)$ truly random bits and could generate a pseudorandom bit in $O(S\log R)$ time. Ultimately, this is because every pseudorandom bit generated depends on every seed bit. If we hope to beat this generation time, it is essential that our PRG accesses few seed bits for each pseudorandom bit generated -- specifically we can only access at most $O(\plog S)$ bits from our $O(S\plog S)$ length seed.

Fortunately, locally computable extractors with exactly this property have been studied for cryptographic applications. Several constructions are sufficient for Theorem \ref{high_level_result}. We will analyze a specific construction of De and Vidick \cite{deVidickJournal}, but note that a slight alteration of Lu's modification of Trevisan's extractor \cite{lu2002hyper,Trevisan01} and of Vadhan's extractor from \cite{Vadhan04} can also be used.

\subsection{Notation and Preliminaries}
Towards Theorem \ref{high_level_result}, we first introduce definitions and preliminaries necessary for the Nisan-Zuckerman PRG construction. As mentioned, this construction relies black-box on any algorithm for ``randomness extraction'', which is the goal of taking an input set of weakly random bits and outputting a (smaller) set of nearly uniform random bits. Ideally, the number of uniform random bits extracted reflects the amount of ``hidden randomness'' in the weakly random stream. Formally, we quantify randomness via the \emph{minimum entropy}:

\begin{definition}[Minimum Entropy -- ``min-entropy'']	\label{def:minentropy}
	A distribution $\mathcal{D}$ over $\{0,1\}^N$ has min-entropy $k$ if, for $X$ drawn from $\mathcal{D}$, and for  all $x\in \{0,1\}$,
	\begin{align*}
	\Pr[X = x] \leq 2^{-k}.
	\end{align*}
\end{definition}
When $\mathcal{D}$  is uniform over $\{0,1\}^N$, it has min-entropy $N$. An example of a distribution with min-entropy $k < N$ is any distribution that's uniform over $Z \subset \{0,1\}^N$ where $|Z| \geq 2^{k}$.  Intuitively, the min-entropy is the number ``hidden bits'' of randomness in a draw from $\mathcal{D}$.

To define randomness extraction, we also use the standard \emph{total variation distance} to measure closeness of distributions:
\begin{definition}[Total Variation Distance]
	The total variation distance between two distributions $\mathcal{D} $ and $\mathcal{D} '$ over $\{0,1\}^N$ is denoted $\|\mathcal{D}  - \mathcal{D} '\|_{TV}$ and defined as:
	\begin{align}
		\|\mathcal{D} - \mathcal{D} '\|_{TV} = \frac{1}{2}\sum_{x \in \{0,1\}^N} | \Pr_{\mathcal{D} }(x) - \Pr_{\mathcal{D} '}(x)|. 
	\end{align} 
	Equivalently, 
	\begin{align}\label{eq:useful_tv}
		\|\mathcal{D}  - \mathcal{D} '\|_{TV} = \sup_{A \subseteq \{0,1\}^N} | \Pr_{x\sim \mathcal{D}} (x\in A) - \Pr_{x\sim \mathcal{D} '}(x\in A) |. 
	\end{align} 
	We say that two distributions are ``$\epsilon$-close'' if $\|\mathcal{D}  - \mathcal{D} '\|_{TV}  \leq \epsilon$.
\end{definition}

\begin{definition}[Randomness Extractor]
	\label{def:extractor}
	A $(k,\epsilon)$-extractor (with additional parameters $N,t,m$) is a function $Ext(X,Y): \{0,1\}^N \times  \{0,1\}^t \rightarrow  \{0,1\}^m$ such that, for any random variable $X\in \{0,1\}^N$ with minimum entropy $k$, if $Y$ is chosen from the uniform distribution over $\{0,1\}^t$, $Ext(X,Y)$ is $\epsilon$-close to the uniform distribution over $\{0,1\}^m$. 
\end{definition}
In words, an extractor converts a weak random source $X$ to a set of $m$ bits that is close to uniformly random. To do so, it requires an additional source of uniformly random bits, $Y \in  \{0,1\}^t$. Typically, $t$ should be thought of as much smaller than $m$ and $N$.

Given $t$ bits of true randomness and ``$k$ bits'' of randomness hidden within $X$, we could imagine an extractor that outputs $m = k+t$ random (or nearly random bits). This is not quite possible since some of the input randomness is hidden: it can be shown that at best $m = k + t - O(\log(1/\epsilon))$ \cite{Radhakrishnan00boundsfor}.
Furthermore, a certain minimum amount of additional randomness is \emph{required} to access the hidden randomness in $X$. For example, even to output $t + 1$ nearly random bits (i.e. extracting just 1 random bit from $X$) we need to set $t \geq \log(n-k) + O(\log(1/\epsilon))$ \cite{Radhakrishnan00boundsfor}.

Beyond measuring distance to uniformity, we also use the total variation distance to measure how closely the output of a randomized algorithm matches the output of the same algorithm instantiated with pseudorandom bits. In particular, we define:
\begin{definition}[Pseudorandom Simulation]
	\label{def:simulation}
	Consider the final memory state $f \in \{0,1\}^{|S|}$ of any algorithm $A$ using $S$ bits of space and $R$ uniformly random bits of randomness. $f$ is a random variable whose distribution depends on this randomness. Denote this distribution, which is over bit strings in  $\{0,1\}^{|S|}$, by $\mathcal{F}$. Let $\mathcal{F}'$ be the distribution over final memory states induced by running $A$ with pseudorandom bits $Z \in \{0,1\}^R$. We say that $Z$ can be used to \emph{simulate} $A$ with error $\epsilon$ if:
	\begin{align*}
	\|\mathcal{F} - \mathcal{F}'\|_{TV} \leq \epsilon.
	\end{align*}
\end{definition}

\begin{claim}
	Suppose a randomized algorithm $A$ succeeds with probability $(1-\delta)$. If $A$ is simulated to error $\epsilon$ with pseudorandom bits, the simulated algorithm succeeds with probability $(1-\delta - \epsilon)$.
\end{claim}
\begin{proof}
	This is an immediate consequence of Definition \ref{def:simulation} and the characterization of TV distance in \eqref{eq:useful_tv}. In particular, if $\mathcal{C}$ is the set of final memory states that produced a successful output, then $\Pr_{x\sim \mathcal{F}}[x \in \mathcal{C}] - \Pr_{x\sim \mathcal{F}'}[x \in \mathcal{C}] \leq \epsilon$.
\end{proof}

\subsection{Nisan-Zuckerman PRG and locally computable extractors}

The Nisan-Zuckerman PRG uses a randomness extractor as a black-box component, although in a somewhat unusual way. It uses a length $N$ input that has \emph{truly random bits}, but applies an extractor that only requires min-entropy $N/2$. However, it applies this extractor multiply times to the same input (using different auxiliary seeds $Y$). In particular, 

\begin{construction}[Basic Nisan-Zuckerman PRG]\label{const:nz1} Given an $(N/2, \epsilon)$-extractor $Ext(X,Y): \{0,1\}^N \times  \{0,1\}^t \rightarrow  \{0,1\}^S$, let $X \in \{0,1\}^N$ and $Y_1,Y_2,...Y_\ell \in \{0,1\}^t$ be chosen uniformly at random. Let $Z \in \{0,1\}^{S \cdot \ell}$ be chosen as:
	\begin{align*}
	Z = Ext(X,Y_1), Ext(X,Y_2), \ldots, Ext(X,Y_\ell).
	\end{align*}
\end{construction}

\begin{theorem}[Lemma 2 of \cite{NisanZuckerman96}]
	\label{thm:nz_onelevel}
	If $Z$ is chosen according to Construction \ref{const:nz1} and is used as the randomness input for a space $S$ algorithm requiring $S\cdot \ell$ random bits, then the distribution over final states of the algorithm is within total variation distance  $\ell \cdot (\epsilon + 2^{-S})$ from the distribution over outputs when uniform random bits are used.
\end{theorem}

There are many possible extractor constructions which can be plugged into Theorem \ref{thm:nz_onelevel}, including one introduced in Nisan and Zuckerman's original paper. However, we want an extractor that can efficiently compute any \emph{particular} bit of $Z$. To do so, we use the following ``locally computable extractor result'':

\begin{theorem}[Theorem 1.1 of \cite{DeVidick10}]\label{thm:local_extract}
	For any $ \beta > 0$, there is an $(N/2, N^{-\beta})$-extractor $Ext: \{0,1\}^N \times \{0,1\}^t \rightarrow \{0,1\}^m$ where $t = O(\beta^2\log^4 N)$, $m \geq \frac{1}{8} N$, and each output bit is computable in $O(\log^z N)$ time for some parameter independent constant $z$. The extractor runs in $N + t$ space.
\end{theorem}

Combining Theorem \ref{thm:nz_onelevel} and Theorem \ref{thm:local_extract} and we have:

\begin{theorem}
	\label{thm:prg_ssquared}
	There is a pseudorandom generator that, given a seed of $O(S + \frac{R}{S}\log^4 S)$ truly random bits, can generate $R$ pseudorandom bits which can be used to simulate any space $S$ algorithm up to error $RS^{-c}$, for any constant $c$. For a fixed constant $z$, the generator can produce any specified pseudorandom bit in $O(log^z S)$ time while only accessing $O(\log ^2 S )$ seed bits.
\end{theorem}

\begin{proof}
	For a fixed constant $\omega\leq 8$, we can set $N = \omega S$ so that $m \geq S$ in Theorem \ref{thm:local_extract}. We then plug this extractor construction into PRG Construction \ref{thm:nz_onelevel}. We have parameters $t = O(\log^4 S)$,  $N = O(S)$, $m \geq S$, and we select $\ell = \frac{R}{S}$. This allows use to produce $R$ pseudorandom bits from a seed of length $O(S + \frac{R}{S}\log^4 S)$. It follows from Theorem \ref{thm:nz_onelevel} that, for any space $S$ algorithm, these bits can be used in a simulation with total variation distance:
	\begin{align*}
		\ell(N^{-\beta} + 2^{-S}) = O(RS^{-(\beta - 1)}).
	\end{align*}
	We can thus achieve simulation error $\leq RS^{-c}$ as long as $t = O((c+1)^2\log^4 N)$.
\end{proof}

\subsection{Multilevel PRG}
\label{sec:multilevel}
Theorem \ref{thm:prg_ssquared} immediately gives a way of quickly generating up to $S^2$ pseudorandom bits for a space $S$ algorithm using a seed of length $\tilde{O}(S)$. This is already sufficient for most of the applications in this paper which, for a graph on $n$ nodes,  use $\tilde{O}(n^2)$ random bits and $O(n^{1 + \gamma})$ space for $\gamma > 0$. 

We also state a more general result which can be applied to algorithms that use more than $S^2$ bits of randomness. In particular, we show how to simulate any space $S$ algorithm using $\tilde{O}(S)$ truly random bits whenever that algorithm requires $S^q$ bits of randomness, for any $q$. As long as $q$ is taken to be constant, each pseudorandom bit can still  be generated in $O(\polylog(S))$ time.

\begin{theorem}[Detailed version of Theorem \ref{high_level_result}]
	\label{thm:prg_general}
	There is a pseudorandom generator that, given a seed of $O(S\log^4 S)$ truly random bits, can generate $S^q$ pseudorandom bits, which can be used to simulate any space $S$ algorithm up to error $S^{-c}$, for any constant $c$. The generator can produce any specified pseudorandom bit in $O(\log^{z + 3q} S)$ time, where $z$ is a parameter independent constant.
\end{theorem}

This result is achieved using the  same construction suggested in \cite{NisanZuckerman96}. 

\begin{construction}[Multilevel Nisan-Zuckerman PRG]\label{const:nz2}
	Suppose we want to construct a PRG that can simulate, with error $S^{-c}$, any algorithm $A$ running in space $S$ and requiring $S^q$ random bits. 
	Let $\omega = \lceil q/.9\rceil$. Construct a chain of pseudorandom number generators,
	\begin{align*}
	P_0, P_1, \ldots, P_\omega, 
	\end{align*}
	each instantiating Construction \ref{const:nz1} as in Theorem \ref{thm:prg_ssquared}. $P_i$'s parameters are set to simulate a space $9^i \cdot S$ algorithm that uses $R = S^{q-.9i}$ random bits, with error $\frac{S^{-c}}{\omega+1}$. To generate any of the $S^q$ pseudorandom bits required to execute $A$, we use run $P_0$, but simulated with bits from $P_1$, which itself is simulated with bits from $P_2$, which is simulated with bits from $P_3$, so on and so forth.
\end{construction}

\begin{proof}[Proof of Theorem \ref{thm:prg_general}]
	We claim that the multilevel PRG of Construction \ref{const:nz2} satisfies Theorem \ref{thm:prg_general}.
	We first note that, if $P_0$ was run with fully random inputs $X, Y_1, \ldots, Y_\ell$, then its output bits would simulate $A$ with error $S^{-c}/t$. Moreover, we note that $P_0$ generates pseudorandom bits in blocks of size $S$ by repeatedly applying the local extractor of Theorem \ref{thm:local_extract} to a string of length $N\leq 8S$. Doing so requires space $8S + O(\log^4 S)$ since, in addition to storing $X \in \{0,1\}^N$, for each block we need to draw a seed of $O(\log^4 S)$ random bits for $Y_i$. 
	
	So, for sufficiently large $S$, $P_0$ uses $9 S$ space. Accordingly, $P_0$ itself can be simulated using a small seed PRG to  further reduce randomness. In particular, $P_0$ uses $O(S + S^{q-1}\log^4(S))$ random bits, which is $\leq S^{q-.9}$ for sufficient large $S$. Furthermore, it runs in space $9S$, so $P_1$ can simulate it with error $\frac{S^{-c}}{\omega+1}$. To so do, $P_1$ uses $O\left(9S + \frac{S^{q-.9-1}}{9}\log^4(9 S)\right) \leq S^{q-2\cdot .9}$ random bits and  runs in space $9^2 S$. Continuing this argument, we can each $P_i$  in Construction \ref{const:nz2} is sufficient to simulate $P_{i-1}$ with error $\frac{S^{-c}}{\omega+1}$. 

	Let $\mathcal{D}$ be the output distribution of $A$ run with truly random bits, let $\mathcal{D}_0$ be the output distribution when $A$ is simulated using $P_0$, which itself uses truly random bits, and in general, let $\mathcal{D}_i$ be the output distribution obtained by using $P_0, \ldots, P_i$ to simulate $A$, with $P_i$ using truly random bits. Since the TV distance obeys triangle inequality, we have:
	\begin{align*}
	\|\mathcal{D} - \mathcal{D}_t\|_{TV} \leq \|\mathcal{D} - \mathcal{D}_0\|_{TV} + \|\mathcal{D}_0 - \mathcal{D}_1\|_{TV} + \ldots + \|\mathcal{D}_{t-1} - \mathcal{D}_t\|_{TV}. 
	\end{align*}
	If each PRG is run with error parameter $\frac{S^{-c}}{\omega+1}$, we can thus bound $\|\mathcal{D} - \mathcal{D}_t\|_{TV} \leq S^{-c}$. 
	
	So Construction \ref{const:nz2} gives a valid pseudorandom generator for proving Theorem \ref{thm:prg_general}. We just need to check its seed length and runtime. To verify seed length, we only need to consider $P_\omega$, which is the only PRG using truly random bits. $P_\omega$'s seed length is upper bounded by $O(9^\omega S + S \log^4 S)$, which is just $O(S \log^4 S )$ as long as $q$ is constant, and thus $\omega = O(q)$ is constant.
	
	To verify runtime, we note that, following Theorem \ref{thm:prg_ssquared} and assuming $q$ is constant, each pseudorandom generator only requires $O(\log^z S)$ computation time and accesses to $O(\log^2 S)$ seed bits to generate a single pseudorandom bit. Accordingly, to produce a single pseudorandom bit for $A$, $P_0$ requires $O(\log^4 S)$ time, $P_1$ requires $O(\log^z S \cdot \log^2 S)$ time, $P_2$ requires $O(\log^z S \cdot \log^2 S \cdot \log^2 S)$ time, etc. $P_\omega$ of course perform the most work, requiring $O(\log^z S \cdot \log^{2\omega} S)$ time, so we can upper bound our total runtime by $O(q \log^z S \cdot \log^{2\omega} S) = O(\log^{z + 3q} S)$ as desired.
\end{proof}

	\subsection{Pseudorandomness for linear sketches}
	\label{sec:prg_linear_sketch}
	Theorem \ref{high_level_result} establishes that any space $S$ algorithms that uses $S^q$ bits, for constant $q$, can be simulated by a pseudorandom number generator with bit generation time $O(\plog S)$ and seed length $O(S\plog S)$. In this section, we restate an argument from \cite{Indyk06} that allows use to apply this result to small space linear sketching algorithms. We begin with a general definition of linear sketching:
	
	\begin{definition}[Linear sketching algorithm]
		\label{def:linear_sketch}
		A linear sketching algorithm $A$ gives a method for processing a vector $x \in \R^N$. The algorithm is characterized by a (typically randomized) \emph{sketch matrix} $\Pi \in \R^{m\times N}$ and by a possibly randomized  \emph{decoding function} $f: \R^m \rightarrow O$ where $O$ is some output domain. Algorithm $A$ executes by first computing $\Pi x$ and then outputting $f(\Pi x)$. Note that $f$ \emph{only} takes $\Pi x$ as input -- $f$ {cannot depend} on $\Pi$ in any other way, e.g. it cannot share randomness with $\Pi$. 
	\end{definition}

	While our methods are framed as sketching edge-vertex incidence matrices instead of vectors, it's not hard to see that they can be reformulated (by reordering rows/columns in $\Pi$) to fit Definition \ref{def:linear_sketch}, with $x\in \{0,1\}^{n\choose 2}$ being an indicator vector for which edges are contained in the input graph $G$.
	
	The advantage of linear sketching algorithms is that they can be immediately adapted to the dynamic streaming setting, where instead of receiving $x$ at one time, we receive updates of the form $(i,z)$, indicating that $x$ should be updated so that $x_i \leftarrow x_i + z$. In out setting $z$ will always be $\pm 1$, indicating that a new edge was added, or an existing edge removed. Upon receiving an update, we can simply update $\Pi x \leftarrow \Pi x + z\Pi_i$, where $\Pi_i$ is the $i^\text{th}$ column of $\Pi$. At the end of our stream of updates, we will have computed $\Pi x$ for the final form of $x$.

	As long as each entry of $\Pi x$ can be stored using a bounded number of bits, a linear sketching algorithm uses just $S = O(m)$ space to store its running accumulation of $\Pi x$. However, naively it also requires $O(mN)$ space to store $\Pi$ : an update to position $x_i$ may come at any time, in which case we need to access column $\Pi_i$. Fortunately, the linearity of the sketch allows for a reduction in this complexity.
	
	\begin{theorem}
		\label{thm:prg_for_sketching}
		Suppose $A$ is a linear sketching algorithm (satisfying Definition \ref{def:linear_sketch}) into dimension $m$. Suppose that for any valid input vector $x$, $\Pi x$ can be stored in $S$ space. Additionally, suppose each column $\Pi_i$ can be stored in $S$ space and generated using 1) independent uniform random bits 2) persistent information stored in at most $S$ space. As long as $A$ requires no more than $S^q$ random bits (i.e. $\Pi$ can be generated using $S^q$ random bits), then it can be simulated using the pseudorandom number generator of Theorem \ref{high_level_result}. This PRG uses a seed of length $O(S \plog S)$, has error $S^{-c}$ for any constant $c$, and can output any particular pseudorandom bit in $O(\plog S)$ time.
	\end{theorem}
	\begin{proof}
		The argument follows \cite{indyk2000stable}.
		Suppose $A$ is run on a stream of updates to indices $i_1, i_2, \ldots, i_t \in \{1,\ldots, N\}$ in $x$. Consider reordering this stream so that all updates to index $1$ come first, followed by all updates to index $2$, etc., until we finish with all updates to index $N$. Now consider $A$ executed on this reordered stream. We claim that in this case, $A$ can be executed entirely in $O(S)$ space. 
		
		The reason is that, instead of storing all of $\Pi$, we can simply generate $\Pi_i$ when it is needed and then delete it from memory as soon as we are done with updates to index $i$. By assumption, each $\Pi_i$ can be stored in $S$ space and generated using information stored in an $S$ sized block of persistent memory, so doing so uses at most $O(S)$ space on top of what it required to store our running accumulation of $\Pi x$. It follows from Theorem \ref{high_level_result} that, at least when our stream is appropriately ordered, $A$ can be simulated using a PRG with a seed of length $O(S\plog S)$.
		
		Finally, we note that, since $\Pi x$ is a linear map, the output of $\Pi x$, and thus $f(\Pi x)$, does not depend \emph{at all} on the ordering of our stream. In particular, if $A$ run on the ordered stream with a pseudorandom generator outputs a particular solution $f(\Pi x)$, then $A$ run on the unordered stream, also with a pseudorandom generator, outputs the exact same solution $f(\Pi x)$. So if $A$ run on the ordered stream with a PRG succeeds with probability $(1-\delta)$, so does $A$ run on the original input stream with a PRG. 
	\end{proof}


\begin{thebibliography}{KLP{\etalchar{+}}16b}

\bibitem[AALG17]{alev2017graph}
Vedat~Levi Alev, Nima Anari, Lap~Chi Lau, and Shayan~Oveis Gharan.
\newblock Graph clustering using effective resistance.
\newblock {\em arXiv preprint arXiv:1711.06530}, 2017.

\bibitem[ACK{\etalchar{+}}16]{AndoniCKQWZ16}
Alexandr Andoni, Jiecao Chen, Robert Krauthgamer, Bo~Qin, David~P. Woodruff,
  and Qin Zhang.
\newblock On sketching quadratic forms.
\newblock {\em Proceedings of the 2016 {ACM} Conference on Innovations in
  Theoretical Computer Science, Cambridge, MA, USA, January 14-16, 2016}, pages
  311--319, 2016.

\bibitem[ADK{\etalchar{+}}16]{dynamicSparsifiers}
Ittai Abraham, David Durfee, Ioannis Koutis, Ssebastian Krinninger, and Richard
  Peng.
\newblock On fully dynamic graph sparsifiers.
\newblock In {\em 57th Annual Symposium on Foundations of Computer Science},
  pages 335--344, 2016.

\bibitem[AGM12a]{ahn2012analyzing}
Kook~Jin Ahn, Sudipto Guha, and Andrew McGregor.
\newblock Analyzing graph structure via linear measurements.
\newblock In {\em Proceedings of the twenty-third annual ACM-SIAM symposium on
  Discrete Algorithms}, pages 459--467. SIAM, 2012.

\bibitem[AGM12b]{ahn2012graph}
Kook~Jin Ahn, Sudipto Guha, and Andrew McGregor.
\newblock Graph sketches: sparsification, spanners, and subgraphs.
\newblock In {\em \PODS{2012}}, pages 5--14, 2012.

\bibitem[AGM13]{ahn2013spectral}
Kook~Jin Ahn, Sudipto Guha, and Andrew McGregor.
\newblock Spectral sparsification in dynamic graph streams.
\newblock In {\em Approximation, Randomization, and Combinatorial Optimization.
  Algorithms and Techniques}, pages 1--10. Springer, 2013.

\bibitem[BBC{\etalchar{+}}17]{blasiok2017}
Jaroslaw Blasiok, Vladimir Braverman, Stephen~R. Chestnut, Robert Krauthgamer,
  and Lin~F. Yang.
\newblock Streaming symmetric norms via measure concentration.
\newblock In {\em 49th Annual Symposium on Theory of Computing}, 2017.

\bibitem[BK96]{BenczurKarger:1996}
Andr\'{a}s~A. Bencz\'{u}r and David~R. Karger.
\newblock Approximating {S-T} minimum cuts in $\tilde{O}(n^2)$ time.
\newblock In {\em 38th Annual Symposium on Theory of Computing}, pages 47--55,
  1996.

\bibitem[BSS12]{BatsonSpielmanSrivastava:2012}
Joshua Batson, Daniel~A. Spielman, and Nikhil Srivastava.
\newblock Twice-{Ra}manujan sparsifiers.
\newblock {\em SIAM Journal on Computing}, 41(6):1704--1721, 2012.

\bibitem[BSST13]{BatsonSurvey}
Joshua Batson, Daniel~A. Spielman, Nikhil Srivastava, and Shang-Hua Teng.
\newblock Spectral sparsification of graphs: Theory and algorithms.
\newblock {\em Communications of the ACM}, 56(8):87--94, 2013.

\bibitem[BZ16]{belazzouguiZhang16}
Djamal Belazzougui and Qin Zhang.
\newblock Edit distance: Sketching, streaming, and document exchange.
\newblock In {\em IEEE 57th Annual Symposium on Foundations of Computer
  Science.}, pages 51--60, 2016.

\bibitem[CCL{\etalchar{+}}15]{ChengCLPT15}
Dehua Cheng, Yu~Cheng, Yan Liu, Richard Peng, and Shang{-}Hua Teng.
\newblock Spectral sparsification of random-walk matrix polynomials.
\newblock {\em CoRR}, abs/1502.03496, 2015.

\bibitem[CGK16]{chakraborty2016}
Diptarka Chakraborty, Elazar Goldenberg, and Michal Kouck\'{y}.
\newblock Streaming algorithms for embedding and computing edit distance in the
  low distance regime.
\newblock In {\em Forty-eighth Annual ACM Symposium on Theory of Computing},
  pages 712--725, 2016.

\bibitem[CGP{\etalchar{+}}18]{ChuGPSSW18}
Timothy Chu, Yu~Gao, Richard Peng, Sushant Sachdeva, Saurabh Sawlani, and
  Junxing Wang.
\newblock Graph sparsification, spectral sketches, and faster resistance
  computation, via short cycle decompositions.
\newblock In {\em 59th {IEEE} Annual Symposium on Foundations of Computer
  Science, {FOCS} 2018, Paris, France, October 7-9, 2018}, pages 361--372,
  2018.

\bibitem[CKM{\etalchar{+}}14]{CohenKMPPRX14}
Michael~B. Cohen, Rasmus Kyng, Gary~L. Miller, Jakub~W. Pachocki, Richard Peng,
  Anup~B. Rao, and Shen~Chen Xu.
\newblock Solving {SDD} linear systems in nearly
  \emph{m}log\({}^{\mbox{1/2}}\)\emph{n} time.
\newblock In {\em Symposium on Theory of Computing, {STOC} 2014, New York, NY,
  USA, May 31 - June 03, 2014}, pages 343--352, 2014.

\bibitem[CM05]{cormode2005improved}
Graham Cormode and Shan Muthukrishnan.
\newblock An improved data stream summary: the count-min sketch and its
  applications.
\newblock {\em Journal of Algorithms}, 55(1):58--75, 2005.

\bibitem[CW79]{carterWegman79}
J.~Lawrence Carter and Mark~N. Wegman.
\newblock Universal classes of hash functions.
\newblock {\em Journal of Computer and System Sciences}, 18(2):143--154, 1979.

\bibitem[DPVR12]{deVidickJournal}
Anandiya De, Christopher Portmann, Thomas Vidick, and Renato Renner.
\newblock Trevisan's extractor in the presence of quantum side information.
\newblock {\em SIAM Journal on Computing}, 41(4):915--940, 2012.

\bibitem[DV10]{DeVidick10}
Anindya De and Thomas Vidick.
\newblock Near-optimal extractors against quantum storage.
\newblock In {\em \STOC{2010}}, pages 161--170, 2010.

\bibitem[FKM{\etalchar{+}}05]{Feigenbaum05}
Joan Feigenbaum, Sampath Kannan, Andrew McGregor, Siddharth Suri, and Jian
  Zhang.
\newblock On graph problems in a semi-streaming model.
\newblock {\em Theor. Comput. Sci.}, 348(2):207--216, 2005.

\bibitem[GI10]{GilbertIndyk:2010}
Anna Gilbert and Piotr Indyk.
\newblock Sparse recovery using sparse matrices.
\newblock {\em Proceedings of the IEEE}, 98:937--947, 2010.

\bibitem[GKP12]{GoelKapralovPost12}
Ashish Goel, Michael Kapralov, and Ian Post.
\newblock Single pass sparsification in the streaming model with edge
  deletions.
\newblock {\em arXiv}, abs/1203.4900, 2012.
\newblock \arXiv{1203.4900}.

\bibitem[Hoe63]{hoeffding1963probability}
Wassily Hoeffding.
\newblock Probability inequalities for sums of bounded random variables.
\newblock {\em Journal of the American statistical association},
  58(301):13--30, 1963.

\bibitem[Ind00]{indyk2000stable}
Piotr Indyk.
\newblock Stable distributions, pseudorandom generators, embeddings and data
  stream computation.
\newblock In {\em Foundations of Computer Science, 2000. Proceedings. 41st
  Annual Symposium on}, pages 189--197. IEEE, 2000.

\bibitem[Ind06]{Indyk06}
Piotr Indyk.
\newblock Stable distributions, pseudorandom generators, embeddings, and data
  stream computation.
\newblock {\em J. ACM}, 53(3):307--323, 2006.
\newblock \pFOCS{2000}.

\bibitem[JKPS17]{JindalKPS17}
Gorav Jindal, Pavel Kolev, Richard Peng, and Saurabh Sawlani.
\newblock Density independent algorithms for sparsifying k-step random walks.
\newblock In {\em Approximation, Randomization, and Combinatorial Optimization.
  Algorithms and Techniques, {APPROX/RANDOM} 2017, August 16-18, 2017,
  Berkeley, CA, {USA}}, pages 14:1--14:17, 2017.

\bibitem[JS18a]{jambulapati2018efficient}
Arun Jambulapati and Aaron Sidford.
\newblock Efficient ${\wt o} (n/\epsilon)$ spectral sketches for the laplacian
  and its pseudoinverse.
\newblock In {\em Proceedings of the Twenty-Ninth Annual ACM-SIAM Symposium on
  Discrete Algorithms}, pages 2487--2503. SIAM, 2018.

\bibitem[JS18b]{JambulapatiS18}
Arun Jambulapati and Aaron Sidford.
\newblock Efficient spectral sketches for the laplacian and its pseudoinverse.
\newblock {\em Proceedings of the Twenty-Ninth Annual {ACM-SIAM} Symposium on
  Discrete Algorithms, {SODA} 2018, New Orleans, LA, USA, January 7-10, 2018},
  pages 2487--2503, 2018.

\bibitem[KL13]{KelnerLevin:2013}
Jonathan~A. Kelner and Alex Levin.
\newblock Spectral sparsification in the semi-streaming setting.
\newblock {\em Theory of Computing Systems}, 53(2):243--262, 2013.

\bibitem[KLM{\etalchar{+}}17]{kapralov2017single}
Michael Kapralov, Yin~Tat Lee, CN~Musco, CP~Musco, and Aaron Sidford.
\newblock Single pass spectral sparsification in dynamic streams.
\newblock {\em SIAM Journal on Computing}, 46(1):456--477, 2017.

\bibitem[KLP16a]{KoutisLP16}
Ioannis Koutis, Alex Levin, and Richard Peng.
\newblock Faster spectral sparsification and numerical algorithms for {SDD}
  matrices.
\newblock {\em {ACM} Trans. Algorithms}, 12(2):17:1--17:16, 2016.

\bibitem[KLP{\etalchar{+}}16b]{KyngLPSS16}
Rasmus Kyng, Yin~Tat Lee, Richard Peng, Sushant Sachdeva, and Daniel~A.
  Spielman.
\newblock Sparsified cholesky and multigrid solvers for connection laplacians.
\newblock In {\em Proceedings of the 48th Annual {ACM} {SIGACT} Symposium on
  Theory of Computing, {STOC} 2016, Cambridge, MA, USA, June 18-21, 2016},
  pages 842--850, 2016.

\bibitem[KM09]{KelnerM09}
Jonathan~A. Kelner and Aleksander Madry.
\newblock Faster generation of random spanning trees.
\newblock In {\em 50th Annual {IEEE} Symposium on Foundations of Computer
  Science, {FOCS} 2009, October 25-27, 2009, Atlanta, Georgia, {USA}}, pages
  13--21. {IEEE} Computer Society, 2009.

\bibitem[KMP10]{KoutisMP10}
Ioannis Koutis, Gary~L. Miller, and Richard Peng.
\newblock Approaching optimality for solving {SDD} linear systems.
\newblock In {\em 51th Annual {IEEE} Symposium on Foundations of Computer
  Science, {FOCS} 2010, October 23-26, 2010, Las Vegas, Nevada, {USA}}, pages
  235--244, 2010.

\bibitem[KMP11]{KoutisMP11}
Ioannis Koutis, Gary~L. Miller, and Richard Peng.
\newblock A nearly-m log n time solver for {SDD} linear systems.
\newblock In {\em {IEEE} 52nd Annual Symposium on Foundations of Computer
  Science, {FOCS} 2011, Palm Springs, CA, USA, October 22-25, 2011}, pages
  590--598, 2011.

\bibitem[KNST19]{KNST19}
Michael Kapralov, Navid Nouri, Aaron Sidford, and Jakab Tardos.
\newblock Dynamic streaming spectral sparsification in nearly linear time and
  space.
\newblock 2019.

\bibitem[KNW10]{KaneNelsonWoodruff10}
Daniel~M. Kane, Jelani Nelson, and David~P. Woodruff.
\newblock On the exact space complexity of sketching and streaming small norms.
\newblock In {\em \SODA{2010}}, pages 1161--1178, 2010.

\bibitem[KP12]{KapralovP12}
Michael Kapralov and Rina Panigrahy.
\newblock Spectral sparsification via random spanners.
\newblock In {\em Innovations in Theoretical Computer Science 2012, Cambridge,
  MA, USA, January 8-10, 2012}, pages 393--398, 2012.

\bibitem[KS16]{KyngS16}
Rasmus Kyng and Sushant Sachdeva.
\newblock Approximate gaussian elimination for laplacians - fast, sparse, and
  simple.
\newblock In {\em {IEEE} 57th Annual Symposium on Foundations of Computer
  Science, {FOCS} 2016, 9-11 October 2016, Hyatt Regency, New Brunswick, New
  Jersey, {USA}}, pages 573--582, 2016.

\bibitem[KW14]{KapralovW14}
Michael Kapralov and David~P. Woodruff.
\newblock Spanners and sparsifiers in dynamic streams.
\newblock {\em {ACM} Symposium on Principles of Distributed Computing, {PODC}
  '14, Paris, France, July 15-18, 2014}, pages 272--281, 2014.

\bibitem[LMP13]{LiMP13}
Mu~Li, Gary~L. Miller, and Richard Peng.
\newblock Iterative row sampling.
\newblock In {\em 54th Annual {IEEE} Symposium on Foundations of Computer
  Science, {FOCS} 2013, 26-29 October, 2013, Berkeley, CA, {USA}}, pages
  127--136. {IEEE} Computer Society, 2013.

\bibitem[LNW14]{LiYiWoodruff14}
Yi~Li, Huy~L. Nguyen, and David~P. Woodruff.
\newblock Turnstile streaming algorithms might as well be linear sketches.
\newblock In {\em \STOC{2014}}, pages 174--183, 2014.

\bibitem[LS14]{LeeS14}
Yin~Tat Lee and Aaron Sidford.
\newblock Path finding methods for linear programming: Solving linear programs
  in {\~{o}}(vrank) iterations and faster algorithms for maximum flow.
\newblock In {\em 55th {IEEE} Annual Symposium on Foundations of Computer
  Science, {FOCS} 2014, Philadelphia, PA, USA, October 18-21, 2014}, pages
  424--433, 2014.

\bibitem[LS15]{LeeSun15}
Yin~Tat Lee and He~Sun.
\newblock Constructing linear-sized spectral sparsification in almost-linear
  time.
\newblock In {\em 56th Annual Symposium on Foundations of Computer Science},
  pages 250--269, 2015.

\bibitem[LS17]{yinTatSDP}
Yin~Tat Lee and He~Sun.
\newblock An sdp-based algorithm for linear-sized spectral sparsification.
\newblock In {\em Proceedings of the 49th Annual Symposium on Theory of
  Computing}, pages 678--687, 2017.

\bibitem[LS18]{LiS18}
Huan Li and Aaron Schild.
\newblock Spectral subspace sparsification.
\newblock In {\em 59th {IEEE} Annual Symposium on Foundations of Computer
  Science, {FOCS} 2018, Paris, France, October 7-9, 2018}, pages 385--396,
  2018.

\bibitem[Lu02]{lu2002hyper}
Chi-Jen Lu.
\newblock Hyper-encryption against space-bounded adversaries from on-line
  strong extractors.
\newblock In {\em Advances in Cryptology-CRYTO 2002}, pages 257--271. Springer,
  2002.

\bibitem[McG14]{McGregor14}
Andrew McGregor.
\newblock Graph stream algorithms: A survey.
\newblock {\em SIGMOD Rec.}, 43(1):9--20, 2014.
\newblock \pICALP{2004}.

\bibitem[MST15]{MadryST15}
Aleksander Madry, Damian Straszak, and Jakub Tarnawski.
\newblock Fast generation of random spanning trees and the effective resistance
  metric.
\newblock In Piotr Indyk, editor, {\em Proceedings of the Twenty-Sixth Annual
  {ACM-SIAM} Symposium on Discrete Algorithms, {SODA} 2015, San Diego, CA, USA,
  January 4-6, 2015}, pages 2019--2036. {SIAM}, 2015.

\bibitem[Mut05]{Muthukrishnan2005}
S.~Muthukrishnan.
\newblock Data streams: Algorithms and applications.
\newblock {\em Found. Trends Theor. Comput. Sci.}, 1(2):117--236, 2005.

\bibitem[Nis92]{nisan1992pseudorandom}
Noam Nisan.
\newblock Pseudorandom generators for space-bounded computation.
\newblock {\em Combinatorica}, 12(4):449--461, 1992.
\newblock \pSTOC{1990}.

\bibitem[NZ96]{NisanZuckerman96}
Noam Nisan and David Zuckerman.
\newblock Randomness is linear in space.
\newblock {\em J. Comput. Syst. Sci.}, 52(1):43--52, 1996.
\newblock \pSTOC{1993}.

\bibitem[PS14]{PengS14}
Richard Peng and Daniel~A. Spielman.
\newblock An efficient parallel solver for {SDD} linear systems.
\newblock In {\em Symposium on Theory of Computing, {STOC} 2014, New York, NY,
  USA, May 31 - June 03, 2014}, pages 333--342, 2014.

\bibitem[RTS00]{Radhakrishnan00boundsfor}
Jaikumar Radhakrishnan and Amnon Ta-Shma.
\newblock Bounds for dispersers, extractors, and depth-two superconcentrators.
\newblock {\em SIAM JOURNAL ON DISCRETE MATHEMATICS}, 13:2000, 2000.

\bibitem[RU10]{rudraUurtamo2010}
Atri Rudra and Steve Uurtamo.
\newblock Data stream algorithms for codeword testing.
\newblock In {\em International Colloquium on Automata, Languages and
  Programming}, pages 629--640, 2010.

\bibitem[Sch18]{Schild18}
Aaron Schild.
\newblock An almost-linear time algorithm for uniform random spanning tree
  generation.
\newblock In Ilias Diakonikolas, David Kempe, and Monika Henzinger, editors,
  {\em Proceedings of the 50th Annual {ACM} {SIGACT} Symposium on Theory of
  Computing, {STOC} 2018, Los Angeles, CA, USA, June 25-29, 2018}, pages
  214--227. {ACM}, 2018.

\bibitem[SS11]{spielman2011graph}
Daniel~A Spielman and Nikhil Srivastava.
\newblock Graph sparsification by effective resistances.
\newblock {\em SIAM Journal on Computing}, 40(6):1913--1926, 2011.

\bibitem[ST04]{SpielmanT04}
Daniel~A. Spielman and Shang{-}Hua Teng.
\newblock Nearly-linear time algorithms for graph partitioning, graph
  sparsification, and solving linear systems.
\newblock In {\em Proceedings of the 36th Annual {ACM} Symposium on Theory of
  Computing, Chicago, IL, USA, June 13-16, 2004}, pages 81--90, 2004.

\bibitem[ST11]{SpielmanTengSparsification}
Daniel~A. Spielman and Shang-Hua Teng.
\newblock Spectral sparsification of graphs.
\newblock {\em SIAM Journal on Computing}, 40(4):981--1025, 2011.

\bibitem[Tre01]{Trevisan01}
Luca Trevisan.
\newblock Extractors and pseudorandom generators.
\newblock {\em J. ACM}, 48(4):860--879, 2001.
\newblock \pSTOC{1999}.

\bibitem[Vad04]{Vadhan04}
Salil~P. Vadhan.
\newblock Constructing locally computable extractors and cryptosystems in the
  bounded-storage model.
\newblock {\em J. Cryptol.}, 17(1):43--77, 2004.

\end{thebibliography}

\newcommand{\etalchar}[1]{$^{#1}$}

\appendix

\section{Partitioning graphs into low diameter clusters in effective resistance metric}\label{sec:partitioning}

\noindent{\em {\bf Theorem~\ref{thm:main-partition}} (Restated)
	For any unweighted graph $G=(V,E)$ that $|V|=n$ and with min-degree at least $n^{0.4}\log^2 n$, set of vertices, $V$, admits a partitioning into $V=C_1\cup \dots\cup C_k$, that for any 
	$$\forall i\in [k], \text{diam}_{\text{eff}}^{\text{Ind}}(C_i)	\le \frac{10}{n^{0.4}}$$
	and $$k \le  c \cdot n\sqrt{\log n}\sqrt{\frac{1}{n^{0.4}\log^2 n}}=c\cdot n^{0.8} \sqrt{\frac{1}{\log n}}=O(n^{0.8})$$
}

\begin{definition}

	For any non-empty set of vertices, we call it a \textbf{singleton cluster} if it has size $1$, otherwise we call it a \textbf{non-singleton cluster}.
\end{definition}
\begin{definition}
	In graph $G=(V,E)$, for any subset of vertices $S$, $\Vol_G(S)$ is defined as follows
	$$\Vol_G(S):=\sum_{u\in S} \text{deg}_G(u)$$
	where $\text{deg}_G(u)$ is the degree of node $u$ in graph $G$.
\end{definition}
\begin{definition}
	In graph $G=(V,E)$, for any set $U\subseteq V$, $\partial_G(U)$ is defined as follows
	$$\partial_G(U):=\left(U\times \left(V\setminus U\right)\right)\cap E$$
\end{definition}
\begin{definition}\label{Def:phi}
In graph $G=(V,E)$, for any set $U\subseteq V$, $\Phi(U)$ is defined as follows
$$\Phi_G(U):=\frac{|\partial_G(U)|}{\min\{\Vol_G(U),\Vol_G(U^c)\}}$$
where $U^c=V\setminus U$.
\end{definition}

This section strongly relies on the clustering results of \cite{alev2017graph}. Specifically, we need the following Lemma.
\begin{lemma}(\textbf{Corollary 2}, \cite{alev2017graph})\label{lem:Shayan}
	Let $G = (V, E)$ be an unweighted graph. If $deg(v) \geq d_{\text{min}}$ for all $v \in V$ and $R_{\text{diam}}\le \text{diam}_{\text{eff}}(G)$, then for
	any $0 < \epsilon < 1/2$, there is a subset of vertices $U \subseteq V$ and a constant $c>0$ such that
	\begin{align}
	\Phi_G(U)\le c\cdot  \frac{\left(\frac{1}{d_{\text{min}}}\right)^{\epsilon}}{\sqrt{R_{\text{diam}} \cdot \epsilon}}\cdot \Vol_G(U)^{\epsilon-1/2}\label{eq:Shayan}
	\end{align}
\end{lemma} 
Now, we plug $\e=\frac{1}{\log n}$ into \eqref{eq:Shayan} as follows
\begin{align}
		\Phi_G(U)&\le c\cdot  \frac{\left(\frac{1}{d_{\text{min}}}\right)^{\epsilon}}{\sqrt{R_{\text{diam}} \cdot \epsilon}}\cdot \Vol_G(U)^{\epsilon-0.5} \Bigg|_{\e=\frac{1}{\log n}} \nonumber\\
		&= c\cdot  \frac{\left(\frac{1}{d_{\text{min}}}\right)^{\frac{1}{\log n}}}{\sqrt{R_{\text{diam}} \cdot 1/\log n}}\cdot \Vol_G(U)^{\frac{1}{\log n}-0.5}\nonumber \\
		&\le c\cdot  \frac{\left(\frac{1}{d_{\text{min}}}\right)^{\frac{1}{\log n}}\left( n^2\right)^{\frac{1}{\log n}}}{\sqrt{R_{\text{diam}} \cdot 1/\log n}}\cdot \Vol_G(U)^{-0.5}&& \text{Since } \Vol_G(U) \le n^2\nonumber \\
		&\le c\cdot \frac{4}{\sqrt{R_{\text{diam}} \cdot 1/\log n}}\cdot \Vol_G(U)^{-0.5} && \text{Since } \frac{1}{d_{\text{min}}}\le 1 \nonumber\\
		&= c'\cdot \sqrt{\frac{\log n }{R_{\text{diam}} \Vol_G(U)}} \label{eq:Shayan-eps}
\end{align}
where $c'=4c$. \\

The following algorithm, given a graph with min-degree at least $10d_{\text{min}}$, returns a partition of the vertices, where the effective resistance diameter of each cluster is at most $R_{\text{diam}}$ and induced degree of each vertex in any cluster is at least $d_{\text{min}}$.
\begin{algorithm}[H]
	\caption{\textsc{Decompose}($H,d_{\text{min}},R_{\text{diam}}$)~~~~~\Comment{Note that we do not run this algorithm, we just need it for the analysis.}}
	\label{alg-Decomposition}
	\hspace*{\algorithmicindent} \textbf{Input} A graph $H$, which doesn't have any vertex with degree less than $10d_{\text{min}}$ \\
	\hspace*{\algorithmicindent} \textbf{Output} A partition of $V(H)$ 
	\begin{algorithmic}[1]
	
		\State  If $H$ contains a vertex with degree less than $d_{min}$ put it in a cluster (singleton) and cut all the edges of that vertex. Repeat this step until no such vertex remains and remove all such vertices from $H$.
		\State Find $u,v \in V(H)$ such that $R_{uv} =\text{diam}_{\text{eff}}(H)$ 
		\State If $R_{uv} \le R_{diam}$ return $V(H)$
		\State Otherwise, find cut $(U, U^c)$ with $\Phi_H(U)\le c'\cdot \sqrt{\frac{\log(n)}{R_{\text{diam}}\Vol_H(U)}}$ and $\Vol_H(U)\le \Vol_H(U^c)$, for some constant $c'$, by invoking Lemma~\ref{lem:Shayan} with $\e=1/\log n$ (See \eqref{eq:Shayan-eps}).\label{line:cut-uiu}

		\State  Call the algorithm recursively on $H[U]$ and $H[U^c]$			
		\State Return the union of the outputs of both recursive calls. 
	\end{algorithmic}
	
\end{algorithm}

We have the following claims about the decomposition algorithm.
\begin{lemma}
	For any unweighted graph $H$, some positive integer $d_{\text{min}}$, and any $R_{\text{diam}}>0$, in the output of \textsc{Decompose}($H,d_{\text{min}},R_{\text{diam}}$) (see Algorithm~\ref{alg-Decomposition}), the number of non-singleton clusters is at most $\frac{n}{d_{\text{min}}}$.
\end{lemma}
\begin{proof}
	Each non-singleton cluster has at least $d_{\text{min}}$ nodes, since the minimum degree in each non-singleton cluster is $d_{\text{min}}$. Thus, we can have at most $\frac{n}{d_{\text{min}}}$ non-singleton clusters.
\end{proof}

\begin{lemma}\label{lem-cut-vol-1}
	For any graph $H=(V, E), |V|=n,$ with min-degree at least $10d_{\text{min}}$, and $R_{\text{diam}}\ge\frac{\log^2n}{d_{\text{min}}}$, when we run \textsc{Decompose}($H,d_{\text{min}},R_{\text{diam}}$) (see Algorithm~\ref{alg-Decomposition}),
in each iteration when line \ref{line:cut-uiu} of the algorithm cuts $(U_i,U_i^c)$, at most 
$c'\sqrt{d_{min}^{H_i}\frac{2}{\log n}}|U_i|$ edges are in the cut. 
\end{lemma}
\begin{proof}
	Using line \ref{line:cut-uiu} of the decomposition algorithm (Algorithm~\ref{alg-Decomposition}) we have,
	\begin{align}
	|\partial_{H_i}(U_i)|&=\Phi_{H_i}(U_i) \cdot \Vol_{H_i}(U_i)&&\text{By Definition \ref{Def:phi}} \nonumber\\
	&\le c' \cdot \sqrt{\frac{\log n}{R_{\text{diam}}}\Vol_{H_i}(U_i)}\label{partial_vol}&&\text{By line \ref{line:cut-uiu} of Algorithm~\ref{alg-Decomposition}}\\
	&\le c' \cdot \sqrt{ \frac{d_{\text{min}}^{H_i}\Vol_{H_i}(U_i)}{\log n}}&&\text{By the fact that } R_{\text{diam}}\ge \frac{\log^2n}{d_{\text{min}}}\nonumber
	\end{align}

	Now we show that $\Vol_{H_i}(U_i)\le 2\cdot |U_i|^2$, which is equivalent to show that $d_{\text{avg}}^{H_i}(U_i)\le 2\cdot |U_i|$, where $d_{\text{avg}}^{H_i}:=\frac{\Vol_{H_i}(U_i)}{|U_i|}$. First, we show that $|\partial_{H_i}(U_i)|\le \frac{c'}{\sqrt{\log n}}\Vol_{H_i}(U_i)$. 

	\begin{align*}
		|\partial_{H_i}(U_i)|&\le c' \cdot \sqrt{\frac{\log n}{R_{\text{diam}}}\Vol_{H_i}(U_i)}&& \text{By inequality \eqref{partial_vol}}\\
		&\le c' \cdot \sqrt{ \frac{d_{\text{min}}^{H_i}\cdot \Vol_{H_i}(U_i)}{\log n}}&&\text{By the fact that } R_{\text{diam}}\ge \frac{\log^2n}{d_{\text{min}}^{H_i}}\\
		&\le \frac{c'}{\sqrt{\log n}} \Vol_{H_i}(U_i)		&&\text{Since } d_\text{min}^{H_i}\le \Vol_{H_i}(U_i)
	\end{align*}
Then,
	\begin{align*}
	\sum_{u\in U_i} \left(\text{deg}_{H_i}(u)-|U_i|\right) &\le |\partial_{H_i}(U_i)| \\
	&\le \frac{c'}{\sqrt{\log n}} \Vol_{H_i}(U_i)\\
	&=\frac{c'}{\sqrt{\log n}}|U_i|d_{\text{avg}}^{H_i}(U_i)
	\end{align*}
	Which implies
	$$\sum_{u\in U_i} \text{deg}_{H_i}(u)-|U_i|^2=|U_i|d_{\text{avg}}^{H_i}(U_i)-|U_i|^2\le \frac{c'}{\sqrt{\log n}}|U_i|d_{\text{avg}}^{H_i}(U_i)$$
	Therefore
	$$d_{\text{avg}}^{H_i}(U_i)\le \frac{1}{1-\frac{c'}{\sqrt{\log n}}} \cdot  |U_i|$$
	For $n\ge 2^{4c'^2}$ we have, 
	$$d_{\text{avg}}^{H_i}(U_i)\le 2 \cdot  |U_i|$$	
	
	So the claim holds. 
\end{proof}

\begin{lemma}\label{lem:number-edges-cut-up}
	For every graph $H=(V, E), |V|=n$ with min-degree at least $10d_{\text{min}}$, if $R_{\text{diam}}\ge \frac{\log^2n}{d_{\text{min}}}$, then the partition output by \textsc{Decompose}($H,d_{\text{min}},R_{\text{diam}}$) (see Algorithm~\ref{alg-Decomposition}), cuts at most $\wt{O}(n\sqrt{d_{\text{min}}})$ edges.
\end{lemma}
\begin{proof}
	Suppose that we have a vector $f\in \R^{V(G)}$, which is set to be equal to all zero vector initially, and in each cut $(U_i,U_i^c)$, we update $f$ as follows.
	\[
	f(v)= 
	\begin{cases}
	f(v)+1,& \text{if      } v\in U_i\\
	f(v),              & \text{otherwise}
	\end{cases}
	\]
	So, we have,
	\begin{align*}
	\sum_i |\partial_{H_i}(U_i)| &\le \sqrt{d_{\text{min}}} \sum_i c'\sqrt{\frac{2}{\log n}}|U_i|&&\text{By Lemma  \ref{lem-cut-vol-1}}\\
	&=c'\sqrt{\frac{2}{\log n}}\sqrt{d_{\text{min}}}\sum_{v\in V} f(v)&& \text{By definition of vector $f$}
	\end{align*}
	
	We also know that $f(v)\le 4\log (\Vol(G))$, 

	We argue this by the binary tree of the decomposition as follows. Whenever we charge one vertex, we know that the volume of corresponding $U_i$ decreased by a factor of 2. So, we charged each vertex at most $2\log(\Vol(G))$ times. Putting these results together we proved that $\sum_i |\partial_{H_i}(U_i)| \le 4n\cdot c'\sqrt{\frac{2}{\log n}}\cdot \sqrt{d_{\text{min}}}\cdot  \log(\Vol(G)) $. 
\end{proof}
\begin{lemma}\label{lem:single-num}
		For any graph $H$ with min-degree at least $10d_{\text{min}}$, for some $d_{\text{min}}$, in the output of Algorithm \textsc{Decompose}($H,d_{\text{min}},R_{\text{diam}}$), for $R_{\text{diam}}\ge\frac{\log^2n}{d_{\text{min}}}$, we have at most $\wt{O}(\frac{n}{\sqrt{d_{\text{min}}}})$ singletons in the output. 
\end{lemma}
\begin{proof}
 Now, we count overall number of cut edges. We just cut edges in two cases: 
	\begin{enumerate}
		\item {When we want to remove vertices with degree less than $d_{\text{min}}$. }
		\item{When we find the sparse cut in line \ref{line:cut-uiu} of the decomposition algorithm. }
	\end{enumerate}
Suppose that we run the decomposition algorithm and we get $k$ singletons after it terminates. First, for the upper-bound on the number of edges cut by the algorithm in total, clearly\footnote{The summation terms, $\sum_i$, in this proof basically sums over all nodes of recursion tree of this algorithm, and $H_i$ is the graph at that node in the recursion tree with low degree vertices removed. }

\begin{align}\label{ineq:cut-up}
\# \text{ of cut edges} \le \sum_i |\partial_{H_i}(U_i)| + k\cdot  d_{\text{min}}
\end{align}

On the other hand for the lower bound, note that in the beginning, every vertex had degree at least $10d_{\text{min}}$ and it becomes singleton if it loses at least $9/10$ of its edges, so
\begin{align}\label{ineq:cut-low}
\frac{9k}{2}d_{\text{min}} \le \# \text{ of cut edges}
\end{align} 
So, by \eqref{ineq:cut-up} and \eqref{ineq:cut-low}, we have
\begin{align}
\frac{7k}{2}d_{\text{min}}\le \sum_i |\partial_{H_i}(U_i)|
\end{align}
Also, by Lemma \ref{lem:number-edges-cut-up}, we know that $\sum_i |\partial_{H_i}(U_i)| \le 4n\cdot c'\sqrt{\frac{2}{\log n}}\sqrt{d_{\text{min}}}\cdot \log(\Vol(G)) $. Hence, 
\begin{align*}
	k\le\frac{8}{7} \cdot n\cdot c'\sqrt{\frac{2}{\log n}}\sqrt{\frac{1}{d_{\text{min}}}}\cdot \log(\Vol(G))
\end{align*}
\end{proof}

\begin{proofof}{Theorem \ref{thm:main-partition}}
Proof easily follows from invoking Lemma \ref{lem:single-num}, when $d_{\text{min}}=\frac{1}{10}n^{0.4}\log^2 n$ (note that it guarantees $n^{0.4}\log^2 n$ min-degree for graph, see Algorithm \ref{alg-Decomposition}) and $R_{\text{diam}}=\frac{\log^2 n}{d_{\text{min}}}$. 
\end{proofof}

\section{Proofs omitted from Section~\ref{sec:algo} and Section \ref{sec:BC}}
\label{app:algo}
We will need Lemma \ref{lem:chain_coarse} that we use in the proof of Lemma \ref{lem:corr-main-SP}.

\begin{lemma}[Chain of Coarse Sparsifiers \cite{LiMP13, kapralov2017single}]
	\label{lem:chain_coarse}
	Consider any PSD matrix $K$ with
	maximum eigenvalue bounded from above by $\lambda_u=2n$ and minimum nonzero eigenvalue bounded from below by $\lambda_\ell=8/n^2$. Let $d = \ceil{\log_\Gamma \frac{\lambda_u}{\lambda_\ell}}$. For $\ell \in \{0, 1, 2, \ldots, d'\}$, that $d'\ge d$, define: $\gamma(\ell)=\frac{\lambda_u}{\Gamma^\ell}$. So $\gamma(d)\leq \lambda_\ell$, and $\gamma(0)=\lambda_u$. Then the chain of $PSD$ matrices, $[K(0), K(1), \ldots, K(d')]$ with $K(\ell)=K+\gamma(\ell)I$ satisfies the following relations:\Navid{-}
	\begin{enumerate}
		\item $K \preceq_r K(d') \preceq_r \Gamma \cdot K $
		\label{itm:last}
		\item $K(\ell) \preceq K(\ell-1) \preceq \Gamma \cdot K(\ell) $ for all $\ell \in {1,\ldots, d'}$\label{itm:mid}
		\item $K(0) \preceq \Gamma\cdot \gamma(0) \cdot I \preceq \Gamma \cdot K(0) $ \label{itm:base}
	\end{enumerate}
\end{lemma}

We will need Lemma \ref{lem:classic_result} that we use in the proof of Lemma \ref{lem:corr-main-SP}. It is well known that by sampling the edges  of $B$ according to their effective resistance, it is possible to obtain a matrix $\wt{B}$ such that $(1-\e)B^\top B \preceq \wt{B}^\top\wt{B} \preceq (1+\e)B^\top B$ with high probability (see Lemma \ref{lem:classic_result}).

\begin{lemma}[Spectral Approximation via Effective Resistance Sampling \cite{spielman2011graph}] 
	\label{lem:classic_result}
	Let $B\in \R^{{{n}\choose{2}}\times n}$, $K=B^\top B$, and $\wt{\tau}$ be a vector of leverage
	score overestimates for $B$'s rows such that $\wt{\tau}_y \geq  \mathbf{b}_y^\top K^{+} \mathbf{b}_y$ for all $y \in [m]$. For $\e \in (0,1)$ and fixed constant $c$, define the sampling probability for row $\mathbf{b}_y$ to be $p_y = \min\{1, c'\e^{-2}\log n \cdot\wt{\tau}_y \}$. Define a diagonal sampling matrix $W$ with $W(y,y) = \frac{1}{p_y}$ with probability $p_y$ and $W(y,y) = 0$ otherwise. With high probability,
	$\wt{K}=B^\top WB\ape K$. Furthermore $W$ has $O(||\wt{\tau}||_1\log n \epsilon^{-2})$ non-zeros with high probability.
\end{lemma}

We will use the following lemma in proof of Lemma~\ref{lem:corr-main-SP}.
\begin{lemma}
\label{lem:main_heavy_sample}

For any integer $i,j,k,\ell$, such that  $0\leq k\leq j$ and $0\leq i\leq j-k$, let node $a$ in the recursion tree (see section \ref{sec:index}) be such that $\textsf{label}(a)=(\textsc{Sparsify},i,\ell)$, and let node $b$ be a child of node $a$ such that $\textsf{label}(b)=(\textsc{HeavyEdges},j-k,\ell+j-k-i)$. Let $G_a$ and $G_b$ denote the graph corresponding to node $a$ and $b$ respectively. Suppose that $G_a=(V,E_a)\cup \frac{\gamma}{n} K_n$, and $G_b=(V,E_b)\cup \frac{\gamma'}{n} K_n$ where $\gamma'=\gamma\cdot\Gamma^{i-j+k}$. For edge $e\in E_b \subseteq E_a$, let $R_e^{G_a}$ and $R_e^{G_b}$ denote the effective resistance of edge $e$ in graph $G_a$ and $G_b$ respectively. Let $p_e^{G_a}=\min \{1,c'\log n \e^{-2}R_e^{G_a}\}$. If $p_e^{G_a}\in [\Gamma^{i-j-1},\Gamma^{i-j}]$, if $e\in G_b$, then $$ R_e^{G_{b}}\geq \frac{1}{500c'\e^{-2} \cdot\Gamma^{1+k} \log n}$$ with probability at least $1-\frac{1}{n^{100}}$. 
\end{lemma}
\begin{proof}

For graph $G=(V,E)$, and vector $\varphi\in \R^{|V|}$, we define $\mathcal{E}(\varphi,E):=\sum_{(u,v)\in E}(\varphi(u)-\varphi(v))^2$, and $\mathcal{E}(\varphi)=\sum_{\{u,v\}\in \binom{V}{2}}(\varphi(u)-\varphi(v))^2$.

Consider graph $G_a=(V,E_a)\cup \frac{\gamma}{n}K_n$. Recall from Section \ref{sec:index} that since $\textsf{label}(a)=(\textsc{Sparsify},i,\ell)$, therefore $E_a$ corresponds to the set of sampled edges, and $\frac{\gamma}{n}K_n$ corresponds to the complete graph where $\gamma=\frac{\lambda_u}{ \Gamma^\ell}$. 

Let $e=(u,v)$, and suppose that in graph $G_a$, we inject $\frac{1}{R_e^{G_a}}$ unit of flow to node $u$ and extract it from $v$. Let vector $\varphi=\frac{1}{R_e^{G}}L^{+}\mathbf{b}_e$ denote the potentials induced at the vertices in graph $G$. By Corollary~\ref{cor:Reff-energy} we have
\begin{align*}
 \frac{1}{R_e^{G_a}}
 &= \mathcal{E}(\varphi,E_a) + \frac{\gamma}{n}\mathcal{E}(\varphi) \\
 &= \left(\varphi(u)-\varphi(v)\right) ^2+ \mathcal{E}(\varphi,E_a \setminus \{e\})+ \frac{\gamma}{n}\mathcal{E}(\varphi)\nonumber\\
 &=1+ \mathcal{E}\left(\varphi,E_a \setminus \{e\}\right) + \frac{\gamma}{n} \mathcal{E}(\varphi) && \text{Since } \varphi(u)-\varphi(v)=1  
\end{align*}

We define $s=\mathcal{E}(\varphi,E_a\setminus \{e\})$, and $t=\frac{\gamma}{n} \mathcal{E}(\varphi)$. Therefore we have
\begin{equation}
\label{eq:1Rst}
\frac{1}{R_e^{G_a}}=1+s+t \text{.}
\end{equation}

Consider graph $G_b=(V,E_b)\cup \frac{\gamma'}{n}K_n$. Recall from Section \ref{sec:index} that, since $\textsf{label}(a)=(\textsc{Sparsify},j-k,\ell+j-k-i)$, therefore $E_b$ corresponds to the set of sampled edges, where edges of $E_b$ are sampled from $E_a$ at rate $\Gamma^{i-{j-k}}$.
Moreover, $\frac{\gamma'}{n}K_n$ corresponds to the complete graph where $\gamma'=\frac{\lambda_u}{ \Gamma^{\ell+j-k-i}}$.

Since $\varphi(u)-\varphi(v)=1$, by Corollary  \ref{cor:Ggamma-eff-pot}, we have
\begin{equation}
\label{eq:mnR_j-R}
\frac{1}{R_e^{G_{b}}} \le \mathcal{E}(\varphi,E_{b})  + \frac{\gamma'}{n} \mathcal{E}(\varphi)
\end{equation}
Therefore, to complete the proof, it's sufficient to find an upper bound for  $$\mathcal{E}(\varphi,E_{b}) + \frac{\gamma'}{n} \mathcal{E}(\varphi).$$ 
Observe that since $e\in E_{b}$ we have
\begin{align}
\mathcal{E}(\varphi,E_{b})  +\frac{\gamma'}{n}  \mathcal{E}(\varphi)  \nonumber 
&= \left(\varphi(u)-\varphi(v)\right)^2+\mathcal{E}(\varphi,E_{b}\setminus \{e\})  +\frac{\Gamma^{i-j}\cdot\gamma}{n}\mathcal{E}(\varphi) && \text{Since } \gamma'=\Gamma^{i-j+k}\cdot \gamma\nonumber \\
&= 1+\mathcal{E}(\varphi,E_{b}\setminus \{e\})  +\Gamma^{i-j+k}\cdot \frac{\gamma}{n}\mathcal{E}(\varphi)\text{.} &&  \text{Since } \varphi(u)-\varphi(v)=1  
\end{align}
We define $S=\mathcal{E}(\varphi, E_{b}\setminus \{e\})$. Recall that $t=\frac{\gamma}{n}\mathcal{E}(\varphi)$. Therefore we have
\begin{equation} \label{eq:Gj-phi-St}
\mathcal{E}(\varphi,E_b) + \frac{\gamma'}{n} \mathcal{E}(\varphi)= 1+S  +\Gamma^{i-j+k}\cdot t \text{.}
\end{equation}
Note that $S$ is a random variable where $\mathbb{E}_{E_b}\left[S  \right]$ is 
\begin{align}
\mathbb{E}_{E_b}\left[S  \right]&=\mathbb{E}_{E_b}\left[\mathcal{E}(\varphi,E_b\setminus \{e\}) \right] \nonumber \\
&=\Gamma^{i-j+k} \cdot\mathcal{E}(\varphi,E_b\setminus \{e\}) \nonumber\\
&= \Gamma^{i-j+k} \cdot s \text{.}\label{eq:expS}
\end{align}
Putting~\eqref{eq:Gj-phi-St} and ~\eqref{eq:expS} together we get
\begin{align} 
\mu&=\mathbb{E}_{E_b}\left[\mathcal{E}(\varphi,E_b)  + \frac{\gamma'}{n} \mathcal{E}(\varphi)  \mid e\in E_b \right] \nonumber \\
&=\mathbb{E}_{E_b}\left[ 1+S+\Gamma^{i-j+k} \cdot t \right] \nonumber \\
&= 1+\Gamma^{i-j+k} \cdot t + \mathbb{E}_{E_b}\left[S  \right]\nonumber \\
&= 1+\Gamma^{i-j+k} \cdot t + \Gamma^{i-j+k}\cdot s \text{.} \label{eq:mu}
\end{align}
Therefore we have
\begin{align}
&\Pr_{E_b}\left[ \left(\mathcal{E}(\varphi,E_b)  + \frac{\gamma'}{n} \mathcal{E}(\varphi) \right) \geq (1+\delta)\mu  \mid e\in E_b \right] \nonumber \\
&= \Pr_{E_b}\left[ 1+S+\Gamma^{i-j+k} \cdot t \geq (1+\delta)(1+\Gamma^{i-j+k} \cdot t + \Gamma^{i-j+k}\cdot s) \right] \nonumber \\
&=\Pr_{E_b}\left[ S \geq (1+\delta)( \Gamma^{i-j+k}\cdot s) +\delta (1+\Gamma^{i-j+k} \cdot t)\right] \nonumber \\
&=\Pr_{E_b}\left[ S \geq ( \Gamma^{i-j+k}\cdot s) \left(1+\delta +\frac{\delta (1+\Gamma^{i-j+k} \cdot t)}{ \Gamma^{i-j+k}\cdot s} \right)\right] \text{.} \label{eq:pr}  \\
\end{align}

We define  
\begin{equation}
\label{eq:delta'}
\delta':=\delta +\frac{\delta (1+\Gamma^{i-j+k} \cdot t)}{ \Gamma^{i-j+k}\cdot s} \text{.}
\end{equation}

Observe that for any $(x,y)\in E_b$, $\left(\varphi(x)-\varphi(y)\right)^2 \in [0,1]$ by Fact \ref{fact:maxphi}, hence we can apply standard multiplicative Chernoff bound to show the the concentration \cite{hoeffding1963probability}.
Thus  we have,
\begin{align}
\Pr_{E_b}{\left[S > (1+\delta') \cdot\mathbb{E}[S] \right]} 
&\leq \text{exp}\left(-\frac{\delta'}{3}\cdot\mathbb{E}[S]\right) \nonumber \\
&=\text{exp}\left(-\frac{\Gamma^{i-j+k}\cdot s}{3}\cdot \left(\delta +\frac{\delta (1+\Gamma^{i-j+k} \cdot t)}{ \Gamma^{i-j+k}\cdot s}\right) \right) &&\text{By~\eqref{eq:delta'} and~\eqref{eq:expS}} \nonumber \\
&=\text{exp}\left(-\frac{\delta}{3} \cdot \left(\Gamma^{i-j+k}\cdot s +1+\Gamma^{i-j+k} \cdot t\right) \right) \nonumber \\
&\leq\text{exp}\left(-\frac{\delta\cdot \Gamma^{i-j+k}}{3} (1+s+t) \right)  && \text{Since }  \Gamma^{i-j+k} \leq 1\nonumber \\
&= \text{exp}\left(-\frac{\delta\cdot \Gamma^{i-j+k}}{3} \cdot \frac{1}{R_e^{G_a}} \right)  && \text{By~\eqref{eq:1Rst}} \label{eq:err}  
\end{align}
Note that $i=j-k$ immediately translates to $G_a=G_{b}$. So, if $i=j-k$ and $p_e^{G_a}=1$, one has $p_e^{G_a}\leq c'R_e^{G_a} \log n \epsilon^{-2}=c'R_e^{G_{b}} \log n \epsilon^{-2}$, hence, 
\begin{equation}
\label{eq:mnR_lwbnd}
R_e^{G_{b}} \geq \frac{1}{c'\log n \e^{-2}}
\end{equation}

However, if $p_e^{G_a} < 1$, we have
\begin{equation}
\label{eq:mnR_upbnd}
R_e^{G_a}= \frac{p_e^{G_a}}{c'\log n \e^{-2}} \leq \frac{\Gamma^{i-j}}{c'\log n \e^{-2}}
\end{equation}
and 
\begin{equation}
\label{eq:mnR_lowbnd}
R_e^{G_a}= \frac{p_e^{G_a}}{c'\log n \e^{-2}} \geq \frac{\Gamma^{i-j-1}}{c'\log n \e^{-2}}
\end{equation}

We set $\delta=300\e^2c'^{-1}$. Putting~\eqref{eq:err} and~\eqref{eq:mnR_upbnd} together we get
\begin{align}
\label{eq:finalerr}
\Pr_{E_b}{\left[S > (1+\delta') \cdot\mathbb{E}[S] \right]} 
&\leq \text{exp}\left(-\frac{\delta\cdot \Gamma^{i-j+k}}{3} \cdot \frac{1}{R_e^{G_a}} \right)   \leq n^{-100}
\end{align}
Therefore putting~\eqref{eq:finalerr} and ~\eqref{eq:pr} together we get 
\begin{align}
&\Pr_{E_b}\left[ \left(\mathcal{E}(\varphi,E_b)+ \frac{\gamma'}{n} \mathcal{E}(\varphi) \right) \geq (1+\delta)\mu  \mid e\in E_b \right] \nonumber\\
&=\Pr_{E_b}{\left[S > (1+\delta') \cdot\mathbb{E}[S] \right]}  \nonumber \leq n^{-100}
\end{align}
Thus with probability at least $1-\frac{1}{n^{100}}$ we have
\begin{align}
\mathcal{E}(\varphi,E_b)  + \frac{\gamma'}{n} \mathcal{E}(\varphi) &\leq (1+\delta)\mu \nonumber \\
&= (1+\delta)\left( 1+\Gamma^{i-j+k} \cdot t + \Gamma^{i-j+k}\cdot s \right) && \text{By~\eqref{eq:mu}} \nonumber\\
&= (1+\delta)\left(1+\Gamma^{i-j+k}\left(\frac{1}{R_e^{G_a}}-1\right)\right) && \text{By~\eqref{eq:1Rst}} \nonumber \\
&\leq (1+300\e^2c'^{-1}) \left(1+\frac{\Gamma^{i-j+k}}{R_e^{G_a}} \right) && \text{Since } \delta=300\e^2c'^{-1} \nonumber \\
&\leq (1+300\e^2c'^{-1}) \left(1+\frac{\Gamma^{i-j+k}}{\frac{\Gamma^{i-j-1}}{c'\log n \e^{-2}}} \right) && \text{By~\eqref{eq:mnR_lowbnd}} \nonumber \\
&=(1+300\e^2c'^{-1})(1+\Gamma^{k+1} c' \e^{-2} \log n ) \nonumber \\
&\leq 500c'\e^{-2}\cdot\Gamma^{k+1}\log n \label{eq:phi_upbnd}.
\end{align}
Putting~\eqref{eq:phi_upbnd}, and ~\eqref{eq:mnR_j-R} together we get with probability at least $1-\frac{1}{n^{100}}$
\begin{align}\label{eq:sdfhio}
\mathcal{E}(\varphi,E_{b}) + \frac{\gamma'}{n} \mathcal{E}(\varphi) \leq 500c'\e^{-2}\cdot\Gamma^{k+1}\log n.
\end{align}
Therefore, in total, by combining \eqref{eq:mnR_lwbnd} and \eqref{eq:sdfhio}, with probability at least $1-\frac{1}{n^{100}}$, we have
$$R_e^{G_{b}}\geq \frac{1}{500c'\e^{-2} \cdot\Gamma^{k+1} \log n} \text{.}$$
\end{proof}
	
The following Lemma quantifies the effect of contracting a subset of vertices on effective resistance metric.

\noindent{\em {\bf Lemma~\ref{lem:additive}} (Restated)
	In graph $G=(V_G,E_G)$, suppose that vertex $u\in V_G$ belongs to a set of vertices $C$, where $\text{diam}_{\text{eff}}(C)\le \beta$. Also assume that $H=G\slash C$ and let $c$ denote the corresponding super-node
	, i.e., $H=\left(\{c\}\cup V_G\setminus C, E_H, w_H\right)$ is the resulting graph after contracting vertices of $C$ in $G$. Then for any $v\notin C$ such that $R_{uv}^G\ge \beta$ one has $$R_{cv}^H\ge R_{uv}^G \left(1-\frac{\beta}{R_{uv}^G}\right)^2.$$
}

\begin{proof}
	For every vertex $x \in V_G\setminus \{u,v\}$, $\varphi(v_i):=\frac{\mathbf{b}_{uv}^\top L^+\mathbf{b}_{xv}}{\mathbf{b}_{uv}^\top L^+\mathbf{b}_{uv}}$ and let $\varphi(u)=1$ and $\varphi(v)=0$. Then, for every $x \in C$ we have:
	\begin{align}
	\varphi(x)&=1-\frac{\mathbf{b}_{uv}^\top L^+\mathbf{b}_{ux}}{\mathbf{b}_{uv}^\top L^+\mathbf{b}_{uv}}\nonumber\\ 
	&\ge 1-\frac{R_{ux}^G}{\mathbf{b}_{uv}^\top L^+\mathbf{b}_{uv}}&&\text{By Fact \ref{fact:maxphi}}\nonumber\\
	&\ge 1-\frac{\beta}{\mathbf{b}_{uv}^\top L^+\mathbf{b}_{uv}}&&\text{Since } u\in C \text{ and }\text{diam}_{\text{eff}}(C)\le \beta \label{eq:eta-beta-1}
	\end{align}
	On the other hand, by Lemma \ref{lem:reff-energy},
	\begin{align}{\label{st2}}
	R_{uv}^G=\frac{1}{\sum_{e=(x,y)\in E_G}\left(\varphi(x)-\varphi(y)\right)^2}
	\end{align}
	Next, we define a potential vector $\varphi^*$ for vertices of graph $H$. For any vertex $x\in V_G\setminus C$, let $\varphi^*(x)=\min\{\frac{\varphi(x)}{\eta},1\}$, where $\eta=1-\frac{\beta}{R_{uv}^G}$, and let $\varphi^*(c)=1$. One should note that since we are contracting vertices of set $C$, then for any edge $e=(c,v')$, for some $v'\in V_G\setminus C$, $w_H(e)=|\left(C\times \{v'\}\right)\cap E_G|$.
	Therefore, we get,
	\begin{align}{\label{st1}}
	\sum_{e=(x,y)\in E_H}w_H(e)\left(\varphi^*(x)-\varphi^*(y)\right)^2 \le \frac{1}{\eta^2}\sum_{e=(x,y)\in E_G}\left(\varphi(x)-\varphi(y)\right)^2
	\end{align}
	Also, note that since $\varphi^*(c)=1$ and $\varphi^*(v)=0$, by Fact \ref{fact:eff-pot}, we get,
	\begin{align*}
	R_{cv}^H&\ge \frac{1}{\sum_{e=(x,y)\in E_H}w_H(e)\left(\varphi^*(x)-\varphi^*(y)\right)^2 }\\
	&\ge \frac{\eta^2}{\sum_{(x,y)\in E_G}\left(\varphi(x)-\varphi(y)\right)^2}&&\text{By } \eqref{st1}\\
	&= R_{uv}^G\cdot \eta^2&&\text{By } \eqref{st2}\\
	&=R_{uv}^G \left(1-\frac{\beta}{R_{uv}^G}\right)^2&&\text{By definition of $\eta$ }
	\end{align*}
\end{proof}
Now, we present the proof of Lemma \ref{lm:decomp} from Section \ref{sec:BC}.

\begin{proofof}{Lemma \ref{lm:decomp}}
	The proof is by (strong) induction on the number of nonzeros in $\sigma$, denoted by $q:=||\sigma||_0$.
	
	The {\bf base case} is provided by $q=0$ and $q=2$. In the former case there is nothing to prove. In the latter case $\sigma$ contains two nonzeros of opposite signs, so there exist $s_1, t_1$ such that $\sigma=\alpha \mathbf{b}_{s_1 t_1}$, where $\alpha=||\sigma||_1/2$, as required.
	
	We now prove the {\bf inductive step}: $\{2,\ldots, q-2, q-1\}\to q$. Let $x\gets \text{argmin}_{u\in V: \sigma_u\neq 0} |\sigma_u|$, and let $y\in V$ be such that $\sigma_x\cdot \sigma_y<0$. Note that $|\sigma_y|\geq |\sigma_x|$. Also, assume that $\sigma_x>0$ (The other case is similar \xxx[NN]{We can add the other case, but it is pretty much the same}). Let $\beta=\sigma_x/(\mathbf{b}_{x y})_x$, so that 
	$$
	(\beta \cdot \mathbf{b}_{x y})_u=\left\lbrace
	\begin{array}{cc}
	\sigma_s&\text{~if~}u=x\\
	-\sigma_s&\text{~if~}u=y\\
	0&\text{o.w.}
	\end{array}
	\right.
	$$
	Note that $\beta>0$. Letting $\sigma'=\sigma-\beta\cdot \mathbf{b}_{x y}$, we get using the equation above and the fact that $|\sigma_x|\leq |\sigma_y|$ and $\sigma_y\cdot \sigma_x<0$ that $||\sigma'||_1=||\sigma||_1+||\beta\cdot \mathbf{b}_{s t}||_1=||\sigma||_1+2|\beta|$. Since $||\sigma'||_0<q=||\sigma||_0$, we get by the inductive hypothesis that there exist pairs $(s_i, t_i), i=1,\ldots, r$ and coefficients $\alpha_i, i=1,\ldots, l$ such that 
	$\sigma'=\sum_{i=1}^l \alpha_i \cdot \mathbf{b}_{s_i t_i}$ and $||\alpha||_1=||\sigma'||_1/2$. Letting $\alpha_{l+1}:=\beta$, $(s_{l+1}, t_{l+1}):=(x, y)$, we get that $\sum_{i=1}^{l+1} \alpha_i \cdot \mathbf{b}_{s_i t_i}=\sigma$ and $||\alpha||_1=||\sigma||_1/2$, as required. This completes the inductive step.
\end{proofof}

\section{Useful primitives}
\label{sec:util}
In this section we explain different sketches and decoding algorithms that we use through the paper. In all of our algorithms in the paper we always use $SB$ as the sketch of matrix $B$, however we implicitly assume that the corresponding decoding algorithm uses the relevant sketch. More precisely, $S$ is a randomly constructed matrix with ${n \choose 2}$ columns that corresponds to the concatenation of the following matrices: The sampling matrix i.e., $\Pi \in \mathbb{R}^{{n \choose 2}\times {n \choose 2}}$ (Section~\ref{sec:index}), the sketch to find the edges with connectivity at most $\lambda$, i.e., $S^f\Sigma \in\mathbb{R} ^{\lambda\cdot\poly(\log n)\times {n \choose 2}}$ (Section~\ref{subsec:find_low}), the \textsc{SparseRecovery} sketch to recover $k$-sparse vectors, i.e., $S^r\in\mathbb{R} ^{k\cdot\poly(\log n)\times {n \choose 2}}$ (Section~\ref{subsec:sp-rec}), and the \textsc{HeavyHitter} sketch to find the edges that are heavy with respect to parameter $\eta$, i.e., $S^h\in\mathbb{R} ^{\eta^{-2}\cdot\poly(\log n)\times {n \choose 2}}$ (Section~\ref{subsec:hv-hit}). 

\subsection{Recovering Low Connectivity Edges}
\label{subsec:find_low}
The goal of this section is to find low connectivity edges. Let $G=(V,E)$ be  an unweighted graph with $n$ vertices. Let $B\in \mathbb{R}^{{n \choose 2} \times n}$ denote the vertex edge incidence matrix of graph $G$. We start by defining edge-connectivity.

\begin{definition}
\label{def:edg-conn}
Let $G$ be an unweighted graph. We define the \textit{edge-connectivity} of an edge $e = (u,v)$ denoted by $\lambda_e$ as the size of the minimum $u-v$ cut in graph $G$.
\end{definition}

We will use the following lemma from ~\cite{ahn2012analyzing} to design an algorithm for finding low connectivity edges.
\begin{lemma}[\cite{ahn2012analyzing}]\label{lem:Low-Conn-E}
Let $B$ denote the vertex edge incidence matrix of a graph $G$. There exists a sketching matrix $S^f$ and a single-pass, $O(n\cdot\text{poly}(\log n))$-space and time algorithm denoted by \textsc{SpanningForest}, for dynamic connectivity, such that Algorithm \textsc{SpanningForest}$(S^fB)$, returns a spanning forest of graph $G$.
\end{lemma}

Let $h_{\lambda}={n \choose 2}\rightarrow \{0,1\}$, be a pairwise independent hash function such that for any $i \in {n \choose 2}$, $$\text{Pr}[h_\lambda(i)=1]=\frac{1}{10\lambda}\text{.}$$ 
Let $B_\lambda$ be $B$ with all rows except those with $h_\lambda(e)=0$ zeroed out. So $B_\lambda$ is $B$ with rows sampled independently at rate $\frac{1}{10\lambda}$. We build a diagonal matrix $\Sigma\in \R^{\binom{n}{2}\times \binom{n}{2}}$, based on hash functions $h_\lambda$ that serves as a sampling matrix as follows.
$$
\Sigma(e,e):= h_\lambda(e) 
$$
Then clearly $B_\lambda=\Sigma  B$.

Then Algorithm~\ref{alg:low-connectivity-edges}, given $S^{f}\Sigma B$ returns edges with connectivity at most $\lambda$ with high probability.

\begin{algorithm}
	\caption{\textsc{FindLowConnectivityEdges}: recovers edges with connectivity at most $\lambda$}\label{alg:low-connectivity-edges}{\label{spanning-forest-ALG}}
	\begin{algorithmic}[1]
	\Procedure{FindLowConnectivityEdges($S^{f}\Sigma B,\lambda$)}{}
		\State \Comment{$B_\lambda$ is $B$ with rows sampled independently at rate $\frac{1}{10\lambda}$}
		\State $E' \leftarrow \emptyset$
		\For {$1\le i\le 200 \,\lambda \log n$} \label{ln:loop}
			\State $E'\gets E'\cup \textsc{SpanningForest}\left(S^{f}\Sigma B \right)$. \Comment \text{see Lemma~\ref{lem:Low-Conn-E}} \label{ln:sp-for}
		\EndFor
		\Return $E'$
		\EndProcedure
	\end{algorithmic}
\end{algorithm}
\begin{lemma}
\label{lem:recover-low-connectivity}
Let $G$ be an unweighted graph, and $B$ denote the vertex edge incidence matrix of a graph $G$. Then Algorithm~\ref{alg:low-connectivity-edges} i.e., $\textsc{FindLowConnectivityEdges}(S^fB_\lambda,\lambda)$ returns all the edges with edge-connectivity at most $\lambda$ with probability at least $1-n^{-10}$. Algorithm~\ref{alg:low-connectivity-edges} runs in $O(\lambda n \cdot\poly(\log n))$-space and time.
\end{lemma}
\begin{proof}
Consider edge $e_1=(u,v)$ with edge-connectivity at most $\lambda$. Therefore, edge $e_1$ belongs to a cut of size at most $\lambda$. Let $e_2,e_3,\ldots,e_k$ denote all other edges belong to the same cut as $e_1$. Since $\lambda_e\leq \lambda$, thus $k\leq \lambda$. Therefore
$$
\text{Pr}[e_1\in B_\lambda]=\frac{1}{10\lambda} \text{.}
$$
Moreover, since $h_\lambda$ is pairwise independent, for all $2\leq i \leq k$ we have
$$
\text{Pr}[e_i \in B_\lambda | e_1\in B_\lambda]=\frac{1}{10\lambda} 
$$
Therefore, since $k\leq \lambda$, by union bound over all edges $e_2,e_3, \ldots, e_k$ we have
\begin{align*}
\text{Pr}[\forall i\in \{2,\ldots,k\}\text{ } e_i \notin B_\lambda | e_1\in B_\lambda]&=1-\frac{k-1}{10\lambda} \leq \frac{9}{10}  \text{.}
\end{align*}
Therefore, with probability at least $\frac{9}{10}\cdot \frac{1}{10\lambda}$, edge $e_1$ is the unique edge sampled from $u-v$ cut. Hence, since edge $e_1$ is the unique edge of that cut, then  by Lemma~\ref{lem:Low-Conn-E}, edge $e_1$ is recovered in the spanning forest constructed by $\textsc{SpanningForest}\left(S^f B_\lambda \right)$ in line~\ref{ln:sp-for} of Algorithm~\ref{alg:low-connectivity-edges}. 

Observe that as per line~\ref{ln:loop} of algorithm we repeat this process $200 \lambda \log n$ times independently. Hence, edge $e_1$ is the unique edge of the cut in at least one of the iterations, with probability at least 
$$1-\left(1-\frac{9}{100\lambda}\right)^{200\lambda\log n} \geq 1-n^{-18}\text{.}$$
Therefore by union bound over all edges with connectivity at most $\lambda$, we have that $E'$ contains all the edges with edge-connectivity at most $\lambda$ with probability at least $1-n^{-10}$. 

Note that we run $O(\lambda \log n)$ independent copies of Algorithm $\textsc{SpanningForest}$ where its space and runtime is given by $O(n\cdot\poly(\log n))$ by Lemma~\ref{lem:Low-Conn-E}. Therefore, overall the space and runtime of Algorithm \textsc{FindLowConnectivityEdges} is at most $O(\lambda n \cdot\poly(\log n))$.
Also note that $|E'|=O(\lambda n\log n)$.
\end{proof}

\subsection{Sparse Recovery}\label{sec:Sparse-Rec}
\label{subsec:sp-rec}
In this section we explain how to recover set of neighbors of a vertex in a graph. To that end we use the following lemma that is rather standard (see, e.g. \cite{cormode2005improved}).
\begin{lemma}
	\label{lem:sparse-recovery}
	There exists a distribution over $m\times n$ matrices $\mathcal{S}$, $m = O\left(k \cdot \poly({\log n})\right)$, and an algorithm denoted by \textsc{SparseRecovery}, such that for any $k$-sparse vector $x\in \mathbb{R}^n$, given $S^r x$, \textsc{SparseRecovery}($S^rx,k$)  returns $x$ with probability at least $1-\frac{1}{\poly(n)}$. Time to recover $x$ given $S^rx$ is $O(k \cdot \poly(\log n))$, and time to update $S^rx$ after incrementing one of the coordinates of $x$ is $\poly(\log n)$.
\end{lemma}
Let $G=(V,E)$ be an unweighted graph with $n$ vertices. Recall that $B\in \mathbb{R}^{{n \choose 2} \times n}$ is the vertex edge incidence matrix of graph $G$. Let vector $\mathbf{b}_v$ denote a column of matrix $B$ that corresponds to vertex $v$. Note that $|\text{supp}(\mathbf{b}_v)|=\text{deg}(v)$. Therefore, by application of Lemma~\ref{lem:sparse-recovery}, by setting $x=\mathbf{b}_v$ and $k=\text{deg}(v)$, we get that $\textsc{SparseRecovery}(S^r\mathbf{b}_v,\text{deg}(v))$ recovers  neighbors of vertex $v$ with high probability.
\subsection{Heavy Hitters}
\label{subsec:hv-hit}
\begin{lemma}\label{lem:HH}[$\ell_2$ Heavy Hitters] For any $\eta > 0$, there is a decoding algorithm denoted by \textsc{HeavyHitter} and a distribution on matrices $S^h$ in $\R^{O(\eta^{-2} \polylog(N))\times N}$ such that, for any $x \in \R^N$, given $S^h x$, the algorithm $\textsc{HeavyHitter}(S^h x, \eta)$ returns a vector $w$ such that $w$ has $O(\eta^{-2} \polylog(N))$ non-zeros and satisfies
	$$||x-w||_\infty \leq \eta ||x||_2 $$
	with probability $1 - \frac{1}{\poly(N)}$ over the choice of $S^h$. The sketch $S^hx$ can be maintained and decoded in $O(\eta^{-2} \polylog(N))$ time and space.
\end{lemma}
This procedure allows us to distinguish from a sketch whether or not a specified entry in $x$ has value $> 2\eta ||x||_2$.

\end{document}